\documentclass[11pt,a4paper]{article}
\pdfoutput=1
\usepackage{jheppub}
\usepackage{epsfig}   
\usepackage{fancyhdr}
\usepackage{graphicx}
\usepackage{pifont}
\usepackage{amsmath}
\usepackage{multirow}
\usepackage{subfigure}
\usepackage{array}
\usepackage{slashed}
\usepackage{amssymb}

\title{High-energy suppression of the Higgsstrahlung cross section in the Minimal Composite Higgs Model}

\author{Katy Hartling\footnote{Previously Katy Hally.}}
\author{and Heather E. Logan}
\affiliation{Ottawa-Carleton Institute for Physics, Carleton University, Ottawa, Ontario K1S 5B6 Canada}
\emailAdd{khally@physics.carleton.ca}
\emailAdd{logan@physics.carleton.ca}

\abstract{If the Higgs boson is composite, signs of this compositeness should appear via a formfactor-like suppression of Higgs scattering cross sections at momentum transfers above the compositeness scale.  We explore this by computing the cross section for $e^+e^- \rightarrow ZH$ (Higgsstrahlung) in a warped five-dimensional gauge-Higgs unification model known as the Minimal Composite Higgs Model (MCHM). We observe that the Higgsstrahlung cross section in the MCHM is strongly suppressed compared to that in the Standard Model at center-of-mass energies above the scale of the first Kaluza-Klein excitations, due to cancellations among the contributions of successive $Z$ boson Kaluza-Klein modes. We also show that the magnitude and sign of the coupling of the first Kaluza-Klein mode can be measured at a future electron-positron collider such as the proposed International Linear Collider or Compact Linear Collider.}

\arxivnumber{}
\begin{document}
\maketitle

 \section{Introduction \label{chap_intro}}

Over the past several decades, the Standard Model (SM) has provided a consistent and elegant description of particle physics and has withstood many experimental tests. In the SM, the quarks, charged leptons, and weak gauge bosons acquire masses through their interactions with a single elementary scalar field called the Higgs field. The Higgs field is an isospin doublet with nonzero hypercharge, which initiates spontaneous breaking of the electroweak symmetry, ${\rm SU}(2)_L \times {\rm U}(1)_Y\rightarrow {\rm U}(1)_{EM}$, due to its non-zero vacuum expectation value (vev). The physical SM Higgs boson is constrained to be heavier than 114.4~GeV by direct experimental searches at the CERN Large Electron-Positron Collider (LEP)~\cite{LEP}. A SM Higgs has further been constrained to be lighter than about 152~GeV from electroweak precision constraints (EWPCs)~\cite{LEPEWWG}, and has very recently been excluded for masses in the ranges 110--122~GeV and 127--600~GeV by direct searches at the CERN Large Hadron Collider (LHC)~\cite{LHCJuly4}.  These searches have now revealed a new particle with mass around 126~GeV and properties consistent with the SM Higgs boson~\cite{LHCJuly4}.

Despite its simplicity and consistency with experiment, the SM Higgs mechanism has a number of deficiencies. The most important of these is the hierarchy problem: radiative corrections to the Higgs mass-squared parameter yield additive contributions proportional to the square of the cutoff scale of the SM.  For a cutoff at the Planck scale $M_P \sim 10^{18}$~GeV, this requires a cancellation against the high-scale Higgs mass-squared parameter fine-tuned to better than a part in $10^{30}$ in order to yield the low-scale Higgs mass-squared parameter of order $M_{EW}^2 \sim (100~{\rm GeV})^2$. This has prompted the development of many extended models of electroweak symmetry breaking (EWSB) that address the hierarchy problem, including supersymmetry~\cite{SUSY1,SUSY2,SUSY3,SUSY4,SUSY5,SUSY6},  little Higgs models~\cite{LittleHiggs1,LittleHiggs2,LittleHiggs3,LittleHiggs4}, composite Higgs models~\cite{TQC,TC,FG,CS}, and technicolor~\cite{Technicolor,Technicolor2,Technicolor3,ExTech,WTech,Technicolor4}. 

One such solution to the hierarchy problem is provided by models with a warped extra spatial dimension,  based on ideas first proposed by Randall and Sundrum in 1999~\cite{RS1}. These models are constructed in a five-dimensional (5D) anti--de~Sitter (AdS) spacetime bounded by two four-dimensional (4D) Minkowski boundaries.  From a 4D perspective, each 5D particle corresponds to a tower of Kaluza-Klein (KK) ``particle-in-a-box'' excitations. If a 5D field is subject to appropriate boundary conditions at both boundaries, the corresponding KK tower will contain a zero mode that can be identified as a SM particle.  The warping in the fifth dimension can be chosen such that a high energy cutoff $\Lambda \sim M_P$ on one boundary (the UV or Planck brane) is redshifted down to an exponentially lower energy scale on the other boundary (the IR or TeV brane).  Localizing the Higgs boson on~\cite{RS1} or near~\cite{0508279,0611358} the TeV boundary thereby provides a solution to the hierarchy problem by cutting off the quadratically-divergent contributions to the Higgs mass near the TeV scale.

These warped extra-dimensional models have a dual interpretation, via the AdS/CFT correspondence~\cite{Maldacena}, as 4D conformal field theories (CFTs).  In this dual description, states localized near the Planck brane correspond to fundamental degrees of freedom in the CFT, while states localized near the TeV brane (including the Higgs and the KK particles) correspond to bound states of the CFT.  This duality allows calculations in a weakly-coupled 5D theory to be used to model the spectrum and interactions of a strongly-coupled conformal theory of a composite Higgs.  Composite Higgs theories solve the hierarchy problem because the Higgs is not a fundamental scalar and instead appears as a bound state only below the compositeness scale, which cuts off quadratically-divergent contributions to its mass.

This leads us to ask the question: can one make a direct experimental probe of the compositeness of the Higgs boson?  Compositeness has historically been demonstrated by the appearance of non-pointlike behavior when the composite object is probed at a length scale comparable to the size of the object---i.e., the pointlike interaction is replaced by a form factor that encodes the suppression of the pointlike scattering cross section above the compositeness energy scale.  In this paper we explore this question for a composite Higgs boson by studying a process that effectively hits the Higgs with a short-wavelength probe.  The simplest such process is Higgsstrahlung, $f \bar f \to Z H$, in which the short-wavelength probe is the off-shell $s$-channel $Z$ boson.\footnote{In the model we study, $s$-channel exchange of the KK excitations of the $Z$ boson plays a critical role in the formfactor-like behavior.}  For simplicity we take the initial state to be $e^+e^-$; the underlying physics is largely unchanged for a light $q \bar q$ initial state.

In this paper we compute the Higgsstrahlung cross section in 
a warped 5D theory with the Higgs localized near the TeV brane.  For concreteness, we 
use
the Minimal Composite Higgs Model (MCHM), which was proposed by Agashe, Contino and Pomarol in 2005~\cite{MCHM}.  The MCHM has an ${\rm SO}(5) \times {\rm U}(1)_{B-L}$ gauge symmetry in the 5D bulk, which is broken at the boundaries by boundary conditions.  This large gauge group allows the preservation of both an ${\rm SU}(2)_L \times {\rm SU}(2)_R \approx {\rm SO}(4)$ and an $O(3)$ custodial symmetry which prevent large corrections to electroweak precision observables~\cite{0308036,Zbb,0612048}. For concreteness we will adopt the fermion embedding of Medina, Shah and Wagner~\cite{MSW}, which is consistent with electroweak precision tests~\cite{MSW,0612048,MCHMEWPT}.  Further constraints on the model parameters have been studied in Refs.~\cite{EWCs,LightKK,CMSW,Neutrinos}.  

In the MCHM, the Higgs doublet arises as the zero modes of the fifth components of 5D gauge fields belonging to the ${\rm SO}(5)/{\rm SO}(4)$ coset, referred to as gauge-Higgs unification~\cite{GHU}.  In the 4D dual description, this corresponds to the Higgs boson being a pseudo-Nambu-Goldstone boson arising from spontaneous breaking of a global SO(5) symmetry of the 4D strong dynamics down to SO(4), leading to a natural little hierarchy between the Higgs mass and the KK scale.
As a pseudo-Nambu-Goldstone boson, the Higgs has no potential at tree level; instead an appropriate potential is generated at one loop via the Coleman-Weinberg mechanism~\cite{Coleman:1973jx} yielding a Higgs mass in the range 114--160~GeV~\cite{MSW}.  
Crucially, the radiatively-induced Higgs potential is finite and calculable~\cite{Hosotani,Falkowski,MSW,HiggsPGB}---i.e., quadratic divergences are absent and the hierarchy problem is thus solved.  This is accomplished because the one-loop integrals that contribute to the Higgs potential are exponentially suppressed at momenta above the KK scale due to the warped 5D propagators; in the dual theory this is interpreted as a consequence of the compositeness of the Higgs.  We will show that the formfactor-like suppression of the $e^+e^- \to ZH$ cross section arises in a similar, but not identical, way.  

We emphasize that we expect this formfactor-like suppression to be a general feature of warped 5D models that are dual to 4D conformal composite Higgs models.  In all such models, the process $f \bar f \to ZH$ involves a 5D $Z$ propagator ``stretched'' across the extra dimension from the light fermions (localized near the Planck brane) to the Higgs (localized near or on the TeV brane).  At energies above the KK scale, the process becomes non-local and the propagator is generically suppressed (see, e.g., Ref.~\cite{0302001}; similar suppression effects from a nonzero impact parameter in extra dimensions were discussed in Ref.~\cite{9909411}).  This can also be thought of as an exclusion of short-wavelength modes from the IR region of the warped space.  Our goals in this paper are to study the details of the implementation of this suppression in the MCHM and to explore how they can be probed in future high-energy collider experiments.

The paper is organized as follows. In Sec.~\ref{sec:model} we review the MCHM with the Medina-Shah-Wagner fermion embedding.
In Sec.~\ref{sec:xsec} we compute the Higgsstrahlung cross section and illustrate the formfactor-like suppression at high center-of-mass energies.
In Sec.~\ref{sec:coupling} we study the prospects at future linear colliders for measuring the couplings of the first and second KK excitations of the $Z$ boson that are responsible for this suppression.  Section~\ref{chapConc} contains our conclusions.  Some technical details are collected in the appendices.

\section{The Minimal Composite Higgs Model \label{sec:model}}

\subsection{Metric and gauge structure}

The Minimal Composite Higgs model is defined in a 5D AdS spacetime with metric~\cite{MCHM,WarpedReview},
\begin{equation}
	ds^2 = \frac{1}{(kz)^2}\left[\eta_{\mu\nu} \, dx^\mu \, dx^\nu - (dz)^2\right] 
	\equiv g_{MN}\, dx^M \, dx^N,
	\label{metric}
\end{equation}
where $M,N = 0...3, 5 $ are Lorentz indices in the full 5D space, $k$ is the curvature of the fifth dimension, and $z \equiv x^5$ is the warped-space coordinate of the fifth dimension.  The warped-space coordiate $z$ can be re-expressed in terms of a flat-space coordinate $y$ according to $z = e^{ky}/k$.  The fifth dimension is bounded by two branes so that $L_0 \leq z \leq L_1$, with $L_0 = 1/k \sim \mathcal{O}(1/M_P)$ and $L_1 = 1/M_{KK} \sim 1/{\rm TeV}$ (to set the scale, the first gauge boson KK modes will appear at about $2.5 \, M_{KK}$, and subsequent modes will be separated by about $3.1 \, M_{KK}$).  The warp factor embedded in the 5D metric causes the energy scale to decrease along the fifth dimension such that if the boundary at $L_0$ (known as the UV or Planck boundary) has energies up to the Planck
scale $M_P$, the one at $L_1$ (known as the Weak, TeV, or IR boundary) will have energies only up to the TeV scale, thereby solving the hierarchy problem. The Higgs boson will be localized near the TeV boundary, so that any large contribution to the Higgs mass parameter will be ``warped down'' to the weak scale.

The fifth dimension is orbifolded by imposing an $S^1/Z_2$ symmetry, where $S^1$ is a circle parametrized by the flat-space coordinate $y$ and $Z_2$ is the transformation $y \rightarrow -y$. The $Z_2$ symmetry allows for the 5D fermion fields---which are inherently non-chiral Dirac spinors---to transform as ${\Psi}(-y)=\mp\gamma_5 \Psi(y)$, allowing the identification of right- and left-handed Weyl spinors, ${\Psi}_{L,R}=\pm\gamma_5 \Psi_{L,R}$ with opposite boundary conditions. The periodicity of the $S^1$ will allow the 5D fields to be expressed as a superposition of 4D KK modes with a profile describing their location along the fifth dimension.

The bulk of the 5D space contains an ${\rm SU}(3)_c \times {\rm SO}(5) \times {\rm U}(1)_{B-L}$ gauge symmetry and a collection of fermion fields (we will discuss the fermion sector in Sec.~\ref{sec:fermions}).  We will denote the bosons corresponding to the ${\rm U}(1)_{B-L}$ and ${\rm SO}(5)$ gauge groups by $U_M$ and $A^a_M$, respectively (recall that $M$ is the 5D Lorentz index).  The electroweak and fermionic sectors of the model are described by the bulk Lagrangian~\cite{MCHM,RandallSchwartz,WarpedReview},
\begin{eqnarray}
	\mathcal{L} &=&-\frac{\sqrt{\bar{g}}}{4} g^{MN} g^{RS} 
	\left( F^a_{MR} F^a_{NS} + U_{MR} U_{NS} \right) 
	+ \sqrt{\bar{g}}\, \mathcal{L}_{GF} 
	\nonumber \\
	&& + \sqrt{\bar{g}} \left[\frac{i}{2}\bar{\Psi} e^M_A \Gamma^A \mathcal{D}_M \Psi 
	- \frac{i}{2}\left(\mathcal{D}_M\Psi\right)^\dagger \Gamma^0 e^M_A \Gamma^A \Psi 
	- M_\Psi \bar{\Psi}\Psi \right],
	\label{L}
\end{eqnarray}
where $\bar{g} = (kz)^{-10}$ is the determinant of the metric $g_{MN}$, $e^M_A=kz \delta_A^M$ is the vielbein, and the five-dimensional Dirac matrices are $\Gamma^M=\left\{\gamma^\mu,-i\gamma^5\right\}$. $\mathcal{L}_{GF}$ is the gauge-fixing term~\cite{HiggsPGB,RandallSchwartz},
\begin{equation}
	\sqrt{\bar{g}} \, \mathcal{L}_{GF} = -\frac{1}{2\epsilon k z}\left[\partial^\mu A^a_\mu 
	- \epsilon z \partial_z \left(\frac{1}{z}A^a_5\right)\right]^2,
	\label{LGF}
\end{equation} 
with gauge-fixing parameter $\epsilon$, and the gauge field strength tensors are defined as 
\begin{eqnarray}
	F^a_{MN} &=& \partial_M A^a_N - \partial_N A^a_M + g_5\, A^{a}_{M N}, \nonumber \\
	U_{MN} &=& \partial_M U_N - \partial_N U_M,
	\label{F}
\end{eqnarray}
where $A^a_{MN}$ is defined as the coefficient of the appropriate piece of the matrix $A^{a}_{MN} T^{a} = - i \, [T^b,T^c] A^b_M A^c_N$. The generators $T^a$ of ${\rm SO}(5)$ are collected in Appendix~\ref{generators} for convenience.  The 5D gauge coupling for ${\rm SO}(5)$ is given in terms of the usual 4D ${\rm SU}(2)_L$ gauge coupling as 
\begin{equation}
	g_5 = g \sqrt{\ln(kL_1)/k}.
	\label{eq:g5}
\end{equation}
Finally, the covariant derivative acting on the fermions is
\begin{equation}
	\mathcal{D}_M = \partial_M + \frac{1}{8}\omega_{MAB}\left[\Gamma^A, \Gamma^B\right]
	- i g_5 A_M - i g_5^{\prime\prime} Q_{B-L} U_M,
	\label{covder}
\end{equation}
where $\omega_{MAB}$ is the spin connection (this term cancels in a diagonal metric such as we use here), $A_M \equiv T^a A^a_M$, $Q_{B-L}$ is the $B-L$ (baryon number minus lepton number) charge of the fermion in question, and $g_5^{\prime\prime}$ is the 5D ${\rm U}(1)_{B-L}$ coupling, which we will fix in terms of $\tan\theta_W$ below Eq.~(\ref{URot}).

The first line of Eq.~(\ref{L}) can be expanded by expressing the five-component gauge bosons $A^a_M$, $U_M$ in terms of a (5D) four-component vector gauge boson $A^a_\mu$, $U_\mu$ and a (5D) scalar fifth component $A^a_5$, $U_5$.  These 5D vector and scalar fields---as well as the chiral fermion fields $\Psi_{L,R}$---can be further decomposed into a tower of 4D Kaluza-Klein fields~\cite{RandallSchwartz,Houches,FifthDim},
\begin{equation}
	A(x^{\mu},z) = \sum_{n = 0,1}^\infty f(m_n,z)\,A(m_n,x^{\mu}).
\end{equation}
Here $A(x^{\mu},z)$ represents any 5D field; the sum starts at $n=0$ if the boundary conditions of the field are such that it has a zero mode, and at $n=1$ otherwise.  The periodicity of the fifth dimension allows this field to be decomposed into a tower of 4D fields (KK modes) $A(m_n,x^{\mu}) \equiv A^{(n)}(x)$ with masses $m_n$, each of which has a fixed profile $f(m_n,z) \equiv f^{(n)}(z)$ in the fifth dimension.  The profiles and masses are determined via separation of variables, by solving the equation of motion in the fifth dimension and applying the appropriate boundary conditions.  

The profiles of the 4D gauge bosons (and their KK modes) satisfy the gauge boson equation of motion which can be derived from the first line of Eq.~(\ref{L}), 
\begin{equation}
	\left[p^2 - \frac{1}{z} \partial_z + \partial_z ^2 \right]f_G(p,z)=0,
	\label{ProfileEq}
\end{equation}
where $p \equiv \sqrt{p^2}$ will be equal to the KK mode mass $m_n$~\cite{RandallSchwartz,WarpedReview,Houches}.  The gauge boson profiles will be normalized according to~\cite{WarpedReview}
\begin{equation}
	\int_{L_0}^{L_1} \frac{dz}{kz}\,f_G^{(n)}(z)\,f_G^{(m)}(z)=\,\delta_{mn}.
	\label{gaugenormb}
\end{equation}

It is convenient to define even and odd solutions $C_A(m_n,z)$ and $S_A(m_n,z)$ with Neumann and Dirichlet boundary conditions, respectively, on the Planck brane:
\begin{equation}
	\left. \partial_z C_A(m_n,z) \right|_{z = L_0} = 0 \quad {\rm (Neumann)}, \qquad
	\left. S_A(m_n, z) \right|_{z = L_0} = 0 \quad {\rm (Dirichlet)}.
\end{equation}
The solutions are conventionally normalized so that
\begin{equation}
	\left. C_A(m_n,z) \right|_{z = L_0} = 1, \qquad 
	\left. \partial_z S_A(m_n,z) \right|_{z = L_0} = m_n.
\end{equation} 
These solutions are given explicitly by~\cite{Davoudiasl:1999tf,MSW,Falkowski},
\begin{eqnarray}
	C_A(m_n,z) &=& \frac{\pi\,m_n}{2}\,z\,\left[J_1(m_n z)\,Y_0(m_n \, L_0) 
	- J_0(m_n \, L_0)\,Y_1(m_n z)\right], \label{Cgaugeb} \\
	S_A(m_n,z) &=& \frac{\pi\,m_n}{2}\,z\,\left[J_1(m_n \, L_0)\,Y_1(m_n z) 
	- J_1(m_n z)\,Y_1(m_n \, L_0)\right], \label{Sgaugeb}
\end{eqnarray}
where $J_n(x)$ and $Y_n(x)$ are the $n$th-order Bessel functions of the first and second kind (some useful identities involving these functions are collected in Appendix~\ref{AppA}).  
The KK mode mass eigenvalues $m_n$ are then determined by imposing the appropriate boundary conditions at the TeV brane.

Note in particular that the gauge boson zero mode profile can be determined by solving Eq.~(\ref{ProfileEq}) directly with $p=m_n=0$, yielding a constant profile independent of $z$.  Imposing the normalization condition gives,
\begin{equation}
	f_{G}^{(0)}(z)=\sqrt{\frac{k}{\ln(k L_1)}}\,.
	\label{gauge0b}
\end{equation}
This solution is consistent only with Neumann boundary conditions on both the Planck and TeV boundaries.  Therefore, gauge bosons with a Dirichlet boundary condition at either or both of the boundaries will not have a gauge boson zero mode.

The bulk gauge symmetry is broken to different subgroups on the Planck and TeV branes.  This is achieved by applying appropriate boundary conditions to the four-component gauge fields $A^a_\mu$.  A Neumann boundary condition for the four-component gauge field preserves the gauge symmetry on the brane, while a Dirichlet boundary condition breaks it~\cite{MCHM,WarpedReview}.  The corresponding scalar fifth components automatically obtain opposite boundary conditions to the corresponding vector gauge field~\cite{HiggsPGB,MCHM}.

In the MCHM we break the bulk ${\rm SO}(5)\times {\rm U}(1)_{B-L}$ gauge symmetry down to the SM ${\rm SU}(2)_L \times {\rm U}(1)_Y$ on the Planck brane, and to the larger group ${\rm SO}(4) \times {\rm U}(1)_{B-L}$ on the TeV brane~\cite{MCHM,HiggsPGB}.  The ${\rm SO}(4)$ group preserved on the TeV brane consists of ${\rm SU}(2)_L \times {\rm SU}(2)_R$.  The SM hypercharge interaction ${\rm U}(1)_Y$ is a linear combination of ${\rm U}(1)_{B-L}$ and the third generator of ${\rm SU}(2)_R$, $Y = Q_{B-L} + T^{3_R}$.
There are ten ${\rm SO}(5)$ gauge bosons $A^a_M$; we will denote the six corresponding to the ${\rm SU}(2)_L \times {\rm SU}(2)_R$ symmetry with the index $a =a_{L,R}=1_{L,R},2_{L,R},3_{L,R}$, and the four corresponding to the remaining (broken) ${\rm SO}(5)/{\rm SO}(4)$ coset with the index $a=\hat{a}=\hat{1},\hat{2},\hat{3},\hat{4}$. As the $A^{a_{L}}_\mu$ bosons will correspond to the usual SM ${\rm SU}(2)_L$ bosons, we will rename the $A^{a_{L,R}}_\mu$ bosons using the more familiar notation $W^{a_{L,R}}_\mu$.

 To preserve an unbroken ${\rm SU}(2)_L \times {\rm SU}(2)_R \times {\rm U}(1)_{B-L}$ symmetry on the TeV boundary, the $W^{a_{L,R}}_\mu$ and $U_\mu$ bosons must have Neumann boundary conditions at $z=L_1$, while the $A^{\hat{a}}_\mu$ bosons must have Dirichlet boundary conditions.
Similarly, at the Planck boundary the $W^{a_{L}}_\mu$ must have Neumann boundary conditions to preserve the ${\rm SU}(2)_L$ gauge symmetry. However, to preserve the SM hypercharge ${\rm U}(1)_Y$, we must apply the Neumann boundary condition to the linear combination of $U_\mu$ and $W^{3_{R}}_\mu$ that corresponds to the SM hypercharge boson $B_\mu$.  The orthogonal linear combination, $X_\mu$, will have a Dirichlet boundary condition at the Planck boundary, but a Neumann boundary condition at the TeV boundary (since it is part of ${\rm SO}(4) \times {\rm U}(1)_{B-L}$). We define a rotation
\begin{eqnarray}
	W_\mu^{3_R} &=& \cos\theta_H B_\mu -\sin\theta_H X_\mu, \label{RRot} \\
	U_\mu &=& \sin\theta_H B_\mu + \cos\theta_H X_\mu, \label{URot}
\end{eqnarray}
where the mixing angle $\theta_H$ is defined by 
\begin{equation}
	\cos\theta_H = \frac{g_5^{\prime\prime}}{\sqrt{g_5^2+g_5^{\prime \prime 2}}},
	\qquad 
	\sin\theta_H = \frac{g_5}{\sqrt{g_5^2+g_5^{\prime\prime 2}}},
\end{equation} 
and is related to the Weinberg angle through $\cos\theta_H = \tan\theta_W$.\footnote{This relation is determined by requiring that the coupling of the photon to two $A^{\hat 4}_5$ scalars (which will be the physical Higgs boson) is zero.  In this case the coupling of the photon to two $Z_{\mu}$ or two $X_{\mu}$ bosons is also zero, as it should be.} The remaining gauge bosons, $A^{1_R}_\mu$, $A^{2_R}_\mu$ and $A^{\hat{a}}_\mu$, must have Dirichlet boundary conditions at the Planck boundary.

By performing the usual SM gauge field rotations (we denote the photon by $V_\mu$ to avoid confusion with our notation for the generic ${\rm SO}(5)$ vector gauge bosons $A_\mu$),
\begin{eqnarray}
	W^\pm_{L\,\mu}  &=& \frac{1}{\sqrt{2}}\left(W_{\mu}^{1_L} \mp i\,W_\mu^{2_L}\right),
	\nonumber \\
	Z_\mu &=& \cos\theta_W\,W^3_\mu - \sin\theta_W\,B_\mu, \qquad
	V_\mu  = \sin\theta_W\,W^3_\mu + \cos\theta_W\,B_\mu,
	\label{Rot}
\end{eqnarray}
as well as the analogous rotations to the charge basis
\begin{equation}
	W^\pm_{R\,\mu} = \frac{1}{\sqrt{2}}\left(W_{\mu}^{1_R} \mp i\,W_\mu^{2_R}\right), \qquad
	A^{\widehat \pm}_\mu = \frac{1}{\sqrt{2}}\left(A_\mu^{\hat{1}} \mp i\,A_\mu^{\hat{2}}\right)\,,
	\label{Rot2}
\end{equation}
we obtain the MCHM spectrum of physical gauge states.

From the 4D perspective, then, the MCHM contains a tower of KK modes with a zero mode---due to the Neumann-Neumann boundary conditions---for each of the SM electroweak gauge bosons: $W^\pm_{L\,\mu}$, $Z_\mu$, and the photon $V_\mu$.  It also contains seven extra gauge KK towers without zero modes: $W^\pm_{R\,\mu}$, $X_\mu$, $A^{\widehat \pm}_\mu$, $A^{\hat{3}}_\mu$ and $A^{\hat{4}}_\mu$. 

Each of these towers of vector bosons has an accompanying tower of scalars arising from the corresponding $A^a_5$.  Almost all of these scalars can be eliminated by performing a 5D gauge transformation; they are thus Goldstone bosons which are eaten by the corresponding massive 4D gauge KK mode to become its third polarization degree of freedom.  The exception is the zero-mode scalars, which cannot be gauged away.  The boundary conditions for the fifth component of the gauge field are opposite those of the four vector components; thus only the $A^{\hat a}_5$ scalar KK towers have zero modes.  These four massless scalars transform as a $\bf 4$ of ${\rm SO}(4)$ and are identified with the Higgs doublet.  As in the SM, three of these scalars will be eaten by the zero modes of the $W^\pm_{L\,\mu}$ and $Z_\mu$ towers after EWSB.  Only one physical scalar boson is left in the spectrum; it becomes the Higgs boson. We will choose the Higgs to be $H_5(x^{\mu},z)=A^{\hat{4}}_5$.\footnote{Here the $5$ subscript on $H_5$ indicates that this is the 5D Higgs field.}  This mechanism, by which the Higgs arises naturally out of the gauge structure of the model, is known as gauge-Higgs unification. In unitary gauge we are then left with the spectrum of gauge bosons and the single scalar Higgs boson outlined in Table~\ref{MCHMboson}.

\begin{table}
\begin{center}
\begin{tabular}{c c c}
\hline \hline
\multicolumn{2}{c}{Boundary condition} & \\
Planck brane & TeV brane & Particles \\
\hline
Neumann & Neumann & $W^{\pm}_{L\,\mu}$,  $Z_{\mu}$, $V_{\mu}$, $H_5$ \\
\hline
Dirichlet & Neumann & $W^{\pm}_{R\,\mu}$, $X_\mu$ \\
\hline
Dirichlet & Dirichlet & $A^{\pm}_\mu$, $A^{\hat{3}}_\mu$, $A^{\hat{4}}_\mu$ \\
\hline \hline
\end{tabular}
\end{center}
\caption{Boundary conditions for the gauge bosons of the MCHM.  (We denote the photon by $V_{\mu}$.)}
\label{MCHMboson}
\end{table}

The profile of the Higgs in the fifth dimension is fixed by the requirement that zero-mode particles be massless before EWSB.  This leads to a Higgs profile linear in $z$~\cite{MCHM},
\begin{equation}
	f_H(z)=z\sqrt{\frac{2k}{L_1^2-L_0^2}},
	\label{fh}
\end{equation}
where $H_5(x^{\mu},z) = f_H(z) H(x^{\mu})$.  Note that although this profile is linear in the warped coordinate $z$, in terms of the flat-space coordinate $y$ the profile is exponentially peaked toward the TeV brane.

\subsection{Fermion sector \label{sec:fermions}}

To incorporate fermions into the 5D model, they must be embedded in an appropriate representation of ${\rm SO}(5) \times {\rm U}(1)_{B-L}$. The choice of embedding strongly affects the TeV-scale physics~\cite{MCHMEWPT,LightKK,EWCs}, though it will have little direct effect on our ultimate conclusions. We adopt the Medina-Shah-Wagner (MSW) embedding~\cite{MSW}, which has been shown to satisfy electroweak precision constraints (EWPCs)~\cite{EWCs,MSW}.\footnote{Other common fermion embeddings are the Hosotani-Oda-Ohnuma-Sakamura (HOOS) embedding~\cite{HOOS1,HOOS2}, and the original MCHM spinorial embedding known as MCHM4~\cite{MCHM}. In the former, however, the $ZZH$ coupling, which is key to our calculation, does not exist, and the latter is difficult to reconcile with EWPCs~\cite{MCHMEWPT,EWCs}.} In the MSW embedding, each generation of quarks is embedded into two ${\bf 5}$'s and one ${\bf 10}$ of ${\rm SO}(5)$ as follows. The left-handed quark doublet and the right-handed up-type quark singlet are each embedded in a different ${\bf 5}_{2/3}$ representation, while the down-type quark singlet is embedded in a ${\bf 10}_{2/3}$ representation (here the 2/3 subscript denotes the ${\rm U}(1)_{B-L}$ charge). The quark fields of generation $i$ can be written explicitly in terms of their ${\rm SU}(2)_L \times {\rm SU}(2)_R$ transformation properties as,
\begin{eqnarray}
\xi^{q_i}_{1_L} &=& Q^{i}_{1_L} \oplus \hat{t}^{i}_{1_L} =
	\left(
	\begin{array}{c c}
	\chi^i_{1_L} (-,+)_{\frac{5}{3}} & t^i_{1_L} (+,+)_{\frac{2}{3}} \\
	\tilde{t}^i_{1_L} (-,+)_{\frac{2}{3}} & b^i_{1_L} (+,+)_{-\frac{1}{3}} \\
	\end{array}
	\right)\oplus \hat{t}^{i}_{1_L} (-,+)_{\frac{2}{3}} 
	\nonumber \\ 
\xi^{q_i}_{2_R} &=& Q^{i}_{2_R} \oplus \hat{t}^{i}_{2_R} =
	\left(
	\begin{array}{c c}
	\chi^i_{2_R} (-,+)_{\frac{5}{3}} & t^i_{2_R} (-,+)_{\frac{2}{3}} \\
	\tilde{t}^i_{2_R} (-,+)_{\frac{2}{3}} & b^i_{2_R} (-,+)_{-\frac{1}{3}} \\\end{array}
	\right)\oplus \hat{t}^{i}_{2_R} (+,+)_{\frac{2}{3}} 
	\nonumber \\
\xi^{q_i}_{3_R} &=& T^{i}_{1_R} \oplus  T^{i}_{2_R} \oplus Q^{i}_{3_R} \nonumber\\
	&=&
	\left(\!\!
	\begin{array}{c}
	\Xi^{i}_{3_R}(-,+)_{\frac{5}{3}} \\
	T_{3_R}^{i}(-,+)_{\frac{2}{3}} \\
	B_{3_R}^{i}(-,+)_{-\frac{1}{3}} \\
	\end{array}
	\!\!\right)
	\!\oplus\!
	\left(\!\!
	\begin{array}{c}
	\Xi^{'i}_{3_R} (-,+)_{\frac{5}{3}} \\
	T_{3_R}^{\prime i} (-,+)_{\frac{2}{3}} \\
	B_{3_R}^{\prime i} (+,+)_{-\frac{1}{3}} \\
	\end{array}
	\!\!\right)
	\!\oplus\!
	\left(\!\!
	\begin{array}{c c}
	\chi^i_{3_R} (-,+)_{\frac{5}{3}} \!\!& t^i_{3_R} (-,+)_{\frac{2}{3}} \\
	\tilde{t}^i_{3_R} (-,+)_{\frac{2}{3}} \!\!& b^i_{3_R} (-,+)_{-\frac{1}{3}} \\
	\end{array}
	\!\!\right),
	\label{quark123}
\end{eqnarray}
where $\hat{t}$ denotes an ${\rm SU}(2)_L \times {\rm SU}(2)_R$ singlet, $T_1$ and $T_2$ transform as ($\bf 3,1$) and ($\bf 1,3$) of ${\rm SU}(2)_L \times {\rm SU}(2)_R$, and $Q$ denotes a bidoublet of ${\rm SU}(2)_L \times {\rm SU}(2)_R$ (where ${\rm SU}(2)_L$ acts vertically and ${\rm SU}(2)_R$ acts horizontally)~\cite{MSW,CMSW}. The final subscripts on each field on the right-hand side denote the electromagnetic charges. The plus and minus signs in parentheses denote even and odd boundary conditions, respectively; the first entry corresponds to the Planck brane boundary condition, while the second corresponds to the TeV brane boundary condition. For fermions, an odd boundary condition is the usual Dirichlet condition, but an even boundary condition is a superposition of Dirichlet and Neumann boundary conditions. The four quarks of each generation with even boundary conditions $(+,+)$ on both boundaries correspond to SM particles. In particular, $t_{1_L}^i$ and $b_{1_L}^i$ together correspond to the left-handed SM doublet of generation $i$. Similarly, $\hat{t}_{2_R}^i$ and $B^{\prime i}_{3_R}$ correspond to the right-handed up- and down-type quark singlets.\footnote{Note that 5D fermion fields are inherently non-chiral Dirac spinors. Chiral 4D spinors can be obtained from these 5D Dirac fermions because of the $Z_2$ orbifold symmetry that is imposed on the fifth dimension~\cite{FifthDim}, allowing the identification of right- and left-handed chiral states, ${\Psi}_{L,R}=\mp\gamma_5 \Psi_{L,R}$.  The boundary conditions given in Eq.~(\ref{quark123}) above are applied to the specified chiral state; the opposite chiral state automatically receives opposite boundary conditions.}

The lepton embedding takes a similar form.  The left-handed lepton doublet and right-handed neutrino of each generation are embedded in ${\bf 5_{0}}$ representations, while the right-handed charged leptons are embedded in ${\bf 10_{0}}$ representations~\cite{Neutrinos},
\begin{eqnarray}
\xi^{\ell_i}_{1_L} &=& Q^{\ell_i}_{1_L} \oplus \hat{n}^{i}_{1_L}  =
	\left(
	\begin{array}{c c}
	\kappa_{1_L}^{i} (-,+)_{1} & n_{1_L}^{i} (+,+)_{0} \\
	\tilde{n}_{1_L}^{i} (-,+)_{0} & \ell_{1_L}^{i} (+,+)_{-1} \\
	\end{array}
	\right)\oplus \hat{n}^{i}_{1_L} (-,+)_{0}
	\nonumber \\
\xi^{\ell_i}_{2_R} &=& Q^{\ell_i}_{2_R} \oplus \hat{n}^{i}_{2_R} =
	\left(
	\begin{array}{c c}
	\kappa_{2_R}^{i} (-,+)_{1} & n^{i}_{2_R} (-,+)_{0} \\
	\tilde{n}_{2_R}^{i} (-,+)_{0} & \ell^{i}_{2_R} (-,+)_{-1} \\
	\end{array}
	\right)\oplus \hat{n}^{i}_{2_R} (+,+)_{0}
	\nonumber \\
\xi^{\ell_i}_{3_R} &=& T^{\ell_i}_{1_R} \oplus  T^{\ell_i}_{2_R} \oplus Q^{\ell_i}_{3_R} \nonumber\\
	&=&
	\left(
	\begin{array}{c}
	K^{i}_{3_R}(-,+)_{1} \\
	N^{i}_{3_R}(-,+)_{0} \\
	L^{i}_{3_R}(-,+)_{-1} \\
	\end{array}
	\right)
	\!\!\oplus\!\!
	\left(
	\begin{array}{c}
	K^{\prime i}_{3_R}(-,+)_{1} \\
	N^{\prime i}_{3_R}(-,+)_{0} \\
	L^{\prime i}_{3_R}(+,+)_{-1} \\\end{array}
	\right)
	\!\!\oplus\!\!
	\left(
	\begin{array}{c c}
	\kappa_{3_R}^{i}(-,+)_{1}\! & n^{i}_{3_R}(-,+)_{0} \\
	\tilde{n}_{3_R}^{i}(-,+)_{0}\! & \ell^{i}_{3_R}(-,+)_{-1} \\
	\end{array}
	\right),
	\label{lep123}
\end{eqnarray}
where $\hat{n}$ denotes an ${\rm SU}(2)_L \times {\rm SU}(2)_R$ singlet, while $T_i$ and $Q_i$ transform as before. Similarly to the quark case, $\ell^i_{1_L}$, $n^i_{1_L}$ ($L^{\prime i}_{3_R}$, $\hat{n}^i_{2_R}$) are the left-handed (right-handed) SM lepton and its associated neutrino of generation $i$.\footnote{The right-handed neutrinos were used to construct a realistic neutrino mass model in Ref.~\cite{Neutrinos}.}

The fermion dynamics (before EWSB) are described by the second line of Eq.~(\ref{L}), which can be reduced to 
\begin{equation}
	\mathcal{L}_f=\frac{1}{(k z)^4}\bar{\Psi}\left[\slashed{p} + \gamma^5 \partial_5  - \frac{1}{kz}\left(2k \gamma^5 +  M_\Psi \right)\right]\Psi\,.
\label{Lfer}
\end{equation}
Making use of  $\gamma_5\,\Psi_{L,R} =\mp \Psi_{L,R}$ for the chiral components of $\Psi= \Psi_L + \Psi_R$, it can be shown from Eq.~(\ref{Lfer}) that the chiral components are related through
\begin{eqnarray}
	\left[\partial_z - \frac{1}{z}(2-c)\right]\Psi_L &=& - m_n \Psi_R, 
	\label{ChiralRelationLRb} \\
	\left[\partial_z - \frac{1}{z}(2+c)\right]\Psi_R &=& m_n \Psi_L,
	\label{ChiralRelationRLb}
\end{eqnarray}
where we have defined $c \equiv M_\Psi/k$.  Combining these, we obtain an equation that must be satisfied by the fifth-dimensional profiles of the fermion KK modes,
\begin{equation}
	\left[\partial_z - \frac{1}{z}(2 \pm c)\right] \left[\partial_z - \frac{1}{z}(2 \mp c)\right] 
	f_{L,R}^{(n)}(z) = -m_n^2 \, f^{(n)}_{L,R}(z).
	\label{FProfileEq}
\end{equation}
The parameter $c$ determines the location of the fermion along the fifth dimension. Note that there are three $c$ values ($c_1^i$, $c_2^i$, $c_3^i$) for each generation of quarks and leptons, one for each multiplet $\xi_1^i$, $\xi_2^i$, $\xi_3^i$.
It is convenient to define the following solution~\cite{MSW,CMSW} to Eq.~(\ref{FProfileEq}):
\begin{equation}
	S^{\pm}_{c} (m_n,z) = \frac{\pi m_n}{2 k}(kz)^{\frac{5}{2}}
	\left[J_{\pm c+\frac{1}{2}}(m_n L_0) Y_{\pm c+\frac{1}{2}}(m_n z)
	- J_{\pm c+\frac{1}{2}}(m_n z) Y_{\pm c+\frac{1}{2}}(m_n L_0) \right].
	\label{Sfermion}
\end{equation}
Up to a normalization constant, $S^+_{c}(m_n,z)$ ($S^-_{c}(m_n,z)$) is the profile of a left-handed (right-handed) fermion with a Dirichlet boundary condition on the Planck brane. Its chiral partner has an even boundary condition on the Planck brane (for fermions, this is a mixture of Dirichlet and Neumann boundary conditions), with profiles given by~\cite{WarpedReview,MSW,CMSW}
\begin{equation}
	\dot{S}^{\pm}_{c}(m_n,z) = \mp\frac{1}{m_n}\left[\partial_z - \frac{1}{z}(2\mp c)\right]
	S^{\pm}_{c}(m_n,z).
	\label{dotSFermion}
\end{equation}
Again, the fermion mass eigenvalues $m_n$ are determined by imposing the TeV boundary condition upon the profiles above. The fermion profiles are normalized according to~\cite{WarpedReview},
\begin{equation}
	\int_{L_0}^{L_1} \frac{dz}{kz} \, \frac{f_f^{(n)}(z)}{(kz)^{3/2}} \, 
	\frac{f_f^{(m)}(z)}{(kz)^{3/2}} = \delta_{mn},
	\label{fermnorm}
\end{equation}
where we have grouped the metric factors $kz$ for later convenience.

The fermion zero-mode profile is found by solving Eqs.~(\ref{ChiralRelationLRb}-\ref{ChiralRelationRLb}) with $m_n=0$ and imposing the normalization condition, yielding
\begin{eqnarray}
	f_L^{(0)}(z) &=& \sqrt{\frac{(1- 2c) k}{\left(k L_1\right)^{1 - 2c}-1}}\left(kz\right)^{2 - c} 
	\qquad {\rm for} \ c \neq 1/2, \nonumber \\
	f_L^{(0)}(z) &=& \sqrt{k \ln(kL_1)} \, (kz)^{3/2} \qquad {\rm for} \ c = 1/2,
	\label{f0}
\end{eqnarray}
and equivalent expressions with $c \to -c$ for $f_R^{(0)}$.
Note in particular that the shape of the zero-mode fermion profile depends significantly on the parameter $c$.  For a left-handed fermion, $c = 1/2$ yields a ``flat'' fermion profile---that is, flat when written in terms of the flat-space coordinate $y$ [this can also be seen after the metric factors are taken into account---compare Eq.~(\ref{fermnorm})].
Similarly, for a left-handed fermion with $c > 1/2$ ($c < 1/2$), the fermion's profile is peaked toward the Planck (TeV) brane.  Zero-mode fermion masses will be generated through their couplings to the Higgs after EWSB, which depend on the overlap between the fermion profiles and the Higgs profile; as such, light fermions require $c>1/2$, while the heavier top and bottom quarks are assigned $c<1/2$.\footnote{Fermions are also subject to an ${\rm SU}(2)_L \times {\rm SU}(2)_R \times {\rm U}(1)_{B-L}$ invariant boundary mass Lagrangian on the TeV brane, which marries the ${\rm SU}(2)_L \times {\rm SU}(2)_R$ bidoublets and singlets via Dirac boundary mass terms and thereby introduces mixing among fermion states of the same electric charge~\cite{MCHM}. The quark boundary mass Lagrangian takes the form, 
\begin{equation}
	\mathcal{L}_{\rm bound.} = -2 \delta(z-L_1) 
	\left[\bar{u}'_L M_{q_1} u_R +\bar{Q}_{1_L} M_{q_2} Q_{3_R} + \rm{h.c.} \right],
\end{equation}
where $M_{q_{1,2}}$ are dimensionless matrices of Dirac mass terms~\cite{Neutrinos}.  The lepton mass Lagrangian is analogous. As shown by Ref.~\cite{MSW}, such a boundary term involving mass parameter $M$ and two fields $\bar{\Psi}_L^1$ and ${\Psi}_R^2$ with profiles $g_L(z)$ and $h_R(z)$, respectively, will lead (via the equations of motion) to the boundary conditions
\begin{eqnarray}
	\lim_{\epsilon\rightarrow 0} g_R(L_1-\epsilon) &=& - M h_R(L_1), \nonumber \\
	\lim_{\epsilon\rightarrow 0} h_L(L_1-\epsilon) &=&   M g_L(L_1). \label{FermBC}
\end{eqnarray}
One may also introduce terms for both quarks and leptons that mix $\bar{Q}_{1_L}$ and $Q_{2_R}$, as well as Majorana mass terms for the right-handed neutrino~\cite{MSW,CMSW,Neutrinos}. For simplicity, we will set all of these boundary mass terms to zero. For our purposes, the Majorana mass effects are only relevant to the KK gauge boson decay widths; we will discuss the resulting model dependence in Sec.~\ref{CrossSectionComparison}.}

Although the MCHM contains many more fermion fields than the Standard Model, only the SM fermions have zero modes. As in the gauge sector, the new degrees of freedom appear only as KK modes.  Their main effect on our calculation is through their contributions to the KK gauge boson decay widths.

\subsection{Mixing effects from electroweak symmetry breaking \label{Mixing}}

Before EWSB, all the zero modes (including the Higgs itself) are massless.  The Higgs acquires a Coleman-Weinberg potential at one loop which triggers EWSB.  As well as giving masses to the zero-mode fermions and weak gauge bosons, the Higgs vev induces mixing among the gauge boson and fermion states, leading to mass eigenstates that are superpositions of particles with different gauge transformation identities.  These EWSB-induced mixing effects are small, being generically suppressed by $\mathcal{O}(v^2/M_{KK}^2)$, where $v \simeq 246$~GeV is the Higgs vev.  They will have only a very small effect on the couplings relevant to the process $e^+e^- \to ZH$; however, the mixing does have a significant effect on the widths of the higher $Z$ boson KK modes because it opens new decay channels that were previously forbidden due to the absence of the relevant couplings.

\begin{figure}
\begin{center}
\includegraphics[scale=0.3]{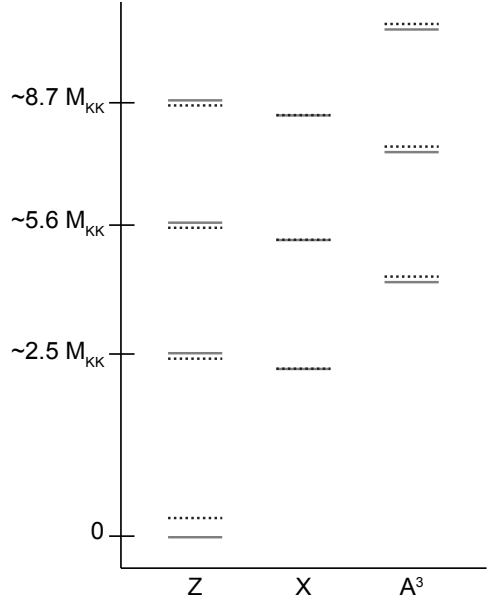}
\caption{A sketch of the KK mode spectrum for the neutral $Z$, $X$ and $A^{\hat{3}}$ bosons before (solid lines) and after (dotted lines) EWSB.  The masses of the $X$ bosons are not shifted by EWSB.  Numbers for a particular parameter set are given in Table~\ref{gaugemass}. Not to scale.}
\label{MCHMmass}
\end{center}
\end{figure}

EWSB induces mixing among the three neutral gauge bosons $Z$, $X$, and $A^{\hat 3}$, and between the three charged gauge bosons $W^{\pm_L}$, $W^{\pm_R}$, and $A^{\widehat{\pm}}$.  The photon and the neutral 4D gauge partner $A^{\hat 4}_{\mu}$ of the 5D Higgs do not participate in the mixing.  This mixing shifts the masses of $Z$, $A^{\hat 3}$, $W^{\pm_L}$, and $A^{\widehat \pm}$, while leaving the masses of $X$ and $W^{\pm_R}$ unaffected.  This is sketched for the neutral gauge sector in Fig.~\ref{MCHMmass}.  Similarly, EWSB induces mixing among the fermions with a common electric charge.

The mixing is implemented as follows.  Because the Higgs in the MCHM arises from the 5D gauge sector, the mixed gauge boson profiles $f_\alpha(m,z;v)$ for arbitrary Higgs vev $v$ can be related to the pre-EWSB profiles $f_\alpha(m,z;0)$ via a 5D gauge transformation~\cite{Falkowski,MSW},
\begin{equation}
	f_\alpha(m,z;v) \, T^\alpha = \Omega^{-1}(z,v)\,f_\alpha(m,z;0)\,T^\alpha \, \Omega(z,v)
	\label{gaugetrans}
\end{equation}
where $\alpha=a_{L,R},\hat{a}$ is the gauge index, $T^\alpha$ is the corresponding generator, and the gauge transformation is
\begin{equation}
	\Omega(z,v)=\exp\left[-i g_5 v \int_{L_0}^z dz^{\prime} \,f_H(z^{\prime})\right] 
	= \exp\left[-i\sqrt{2} \, \theta_G(z,v) T^{\hat{4}} \right].
	\label{Omega}
\end{equation}
Here $f_H(z^{\prime})$ is the 5D Higgs profile from Eq.~(\ref{fh}).
The function $\theta_G(z,v)$ is obtained by integrating, yielding
\begin{equation}
	\theta_G(z,v)= g_5 v \sqrt{\frac{k}{L_1^2-L_0^2}} \frac{(z^2-L_0^2)}{2}.
	\label{Theta}
\end{equation}
This gauge transformation ``turns on'' a non-zero vev $v$ for the Higgs. This can be seen by applying the gauge transformation Eq.~(\ref{Omega}) to the Higgs $H_5 = H f_H$ with zero vev,
\begin{eqnarray}
	H_5 T^{\hat{4}} &\rightarrow& \Omega \, H_5 T^{\hat{4}} \, \Omega^{\dag} 
	+ \frac{i}{g_5} \Omega\,\partial_z \Omega^{\dag} 
	= \Omega \,H f_H T^{\hat{4}} \,\Omega^{\dag} + \frac{i}{g_5} \Omega
	\left(-i g_5 v f_H \right) \Omega^{\dag} \nonumber \\
	&=& \Omega \,(H + v) f_H T^{\hat{4}} \,\Omega^{\dag}=(H + v) f_H T^{\hat{4}},
\end{eqnarray}
where we used the fact that $\Omega$ commutes with $T^{\hat 4}$ and $\Omega \, \Omega^{\dagger} = 1$.  It is also interesting to note that $\theta_G(L_0)=0$ and therefore $\Omega(L_0,v)=1$ on the Planck brane, so that the mixing does not affect the implementation of the Planck-brane boundary conditions. 

The EWSB-induced mixing is implemented by applying the gauge transformation to the gauge boson and fermion profiles, imposing the TeV-brane boundary conditions upon the post-EWSB profiles, and solving the resulting set of equations for the normalization coefficients and mass eigenvalue corresponding to each KK mode.  The gauge-transformed profiles, mass conditions, and solutions for the normalization coefficients for the gauge and fermion sectors are collected in Appendix~\ref{MixProfiles}.

The main effect of EWSB-induced mixing on our calculation is through its effect on particle couplings.  All interactions in the MCHM, including those of the Higgs, arise from the gauge structure of the theory.  The Feynman rules before EWSB are collected for convenience in Appendix~\ref{sec:vertices}.  After EWSB, the profiles of the gauge KK mass eigenstates are in general no longer factorizable from the associated generators; likewise, the fermions become mixtures of states with different gauge transformation properties.  Interaction vertices among 4D KK states are then computed by summing over the components of the mixed states before performing the integrations over the fifth dimension.


\subsection{The Coleman-Weinberg potential}

Because the Higgs boson arises from the gauge sector in the MCHM, it has no potential at tree level.  Instead, the Higgs potential arises from loop contributions, primarily from the $W$ and $Z$ gauge bosons and the top and bottom quarks (we neglect the contributions of the light SM fermions in comparison with these).  These loop contributions are the same effects that lead to the quadratically-divergent corrections to the Higgs mass in the SM. 

The Higgs potential is given at one-loop order by the Coleman-Weinberg potential, which can be written as~\cite{Falkowski,MSW}
\begin{eqnarray}
	V_{CW}(v) &=&  \sum_r \pm \frac{N_r}{(4\pi)^2} \int dp\,p^3 \ln \left[ \rho_r(-p^2) \right] 
	\nonumber \\
	&=& \frac{1}{(4\pi)^2}\int_0^\infty dp\,p^3  \left\{ 6\ln \left[ \rho_W(-p^2) \right] 
	+ 3\ln \left[ \rho_Z(-p^2) \right] \right. 
	\nonumber \\
	&& \left. - 12 \ln \left[ \rho_t(-p^2) \right] - 12 \ln \left[ \rho_b(-p^2) \right] \right\},
	\label{CW}
\end{eqnarray}
where in the first line $N_r$ is the number of degrees of freedom, the plus sign applies to bosons, and the minus sign applies to fermions. The spectral functions $\rho_r$ are obtained from the post-EWSB mass conditions for the $W$, $Z$, bottom, and top quark (see Appendix~\ref{MixProfiles}) normalized to 1 in the absence of EWSB by dividing out the term independent of $\sin\theta_G$.  They are given by
\begin{eqnarray}
	\rho_i(m^2) &=& 1+F_i(m^2)\sin^2\theta_G(L_1,v), \qquad i=W,Z,b, \nonumber \\
	\rho_t(m^2) &=& 1+\frac{F_{t1}(m^2)}{2 F_0(m^2)} \sin^2\theta_G(L_1,v)
	+ \frac{F_{t2}(m^2)}{2 F_0(m^2)}\sin^4\theta_G(L_1,v),
\end{eqnarray}
with the functions $F_i(m^2)$ defined as
\begin{eqnarray}
	F_Z(m^2) &=& \sec^2\theta_W\, F_W(m^2) = \frac{k L_1 m \sec^2\theta_W}
	{2 C'_A(m,L_1) S_A(m,L_1)}, \nonumber \\
	F_b(m^2) &=& -\frac{ (k L_1)^4 M_2^2 \dot{S}_{c_1}^{-}}
	{2 S_{c_3}^{+} (M_2^2 S_{c_3}^{-} \dot{S}_{c_1}^{-} + S_{c_1}^{-} \dot{S}_{c_3}^{-} )}, \nonumber \\
	F_0(m^2) &=& M_1^2 S^+_{c_1} \dot{S}^-_{c_2} \dot{S}^+_{c_2}\left(M_2^2 S^-_{c_3} \dot{S}^-_{c_1} + S^-_{c_1} \dot{S}^-_{c_3}\right)+ S^+_{c_2} \dot{S}^-_{c_2} \left(M_2^2 \dot{S}^-_{c_1} \dot{S}^+_{c_1} S^-_{c_3} + S^-_{c_1}\dot{S}^+_{c_1} \dot{S}^-_{c_3}\right),
\nonumber \\
	F_{t1}(m^2) &=& M_2^2 S^+_{c_2} S^-_{c_3} \dot{S}^-_{c_2} + M_1^2 \left(2 M_2^2 S^+_{c_1} S^-_{c_3} \dot{S}^-_{c_1} + 2 S^+_{c_1} S^-_{c_1} \dot{S}^-_{c_3} -   \dot{S}^+_{c_2} \dot{S}^-_{c_2} \dot{S}^-_{c_3}\right),
\nonumber \\
	F_{t2}(m^2) &=& - (k L_1)^8 M_1^2 \dot{S}^{-}_{c_3},
\end{eqnarray}
where the arguments of $S_c(m,L_1;v)$ have been suppressed for compactness.

The Higgs vev is determined by minimizing the Coleman-Weinberg potential, while its mass is determined by evaluating the second derivative at the vev~\cite{MSW}. 
Crucially, unlike in the SM, the 5D potential of Eq.~(\ref{CW}) is finite and calculable; the integrand is exponentially suppressed with momentum for momenta above the KK scale, creating an effective cut-off for the loop integrals and thereby avoiding the hierarchy problem. 

This exponential suppression can be demonstrated analytically using the asymptotic properties of Bessel functions (Appendix~\ref{AppA}). Consider for example the $Z$ boson loop.  This involves the function $F_Z(m^2)$, which is inversely proportional to $C'_A(m,L_1) S_A(m,L_1)$. Assuming that $p L_1 \gg 1$, $p L_0 \ll 1$, and using Eqs.~(\ref{ItoJ}-\ref{asympK}), one can show that, after Wick rotating (here $\gamma$ is the Euler-Mascheroni constant and $m = p \equiv \sqrt{p^2}$),
\begin{equation}
	 C'_A(m,L_1) 
	\stackrel{\rm Wick\, rot.}{\simeq} i m^2 L_1 \left\{ \frac{\exp(m L_1)}{\sqrt{2\pi m L_1}}
	\left[ \ln \left(\frac{m L_0}{2}\right) + \gamma \right] 
	+ \sqrt{\frac{\pi}{2 m L_1}}\exp(-m L_1) \right\}.
\end{equation}
For $m L_1 \gg 1$, the first term dominates, and $C'_A(m,L_1)$ is approximately proportional to $\exp(m L_1)$. $S_A(m_n,L_1)$ can be shown to have a similar exponential dependence, and thus the function $F_Z(m^2)$ is exponentially suppressed with increasing $p$.  We may then Taylor-expand the logarithm in Eq.~(\ref{CW}) to show that the integrand itself is proportional to $F_Z(m^2)$, and therefore exponentially suppressed as well.  

We implement the Coleman-Weinberg potential numerically and extract values for $v$ and $M_H$ for a particular set of input parameters, given next.


\subsection{Input parameters and particle masses \label{Parameters}}

The MCHM depends on the parameters $g$, $\theta_W$, $M_Z^{(0)}$, $\ln(k L_1)$, and the set of $c_i$ and $M_i$ values necessary to specify the fermion profiles.  We set the first three of these equal to the usual SM parameters: $g=0.649$, $\theta_W = 28.75^{\circ}$ (or $\sin^2\theta_W = 0.2314$), and $M_Z^{(0)}=91.1876$~GeV~\cite{PDG}. The fourth parameter is usually taken to be $\ln(k L_1)\sim 30$~\cite{MSW}; we choose $\ln(kL_1) = 30$.  

Once the fermion parameters of the third quark generation have been chosen, the Coleman-Weinberg potential can be minimized to determine the value of $\theta_G(L_1,v)$~\cite{MSW}.\footnote{Note that the minimization condition can be written entirely in terms of the SM input parameters and $\theta_G(L_1,v)$, and independently of $k$, by transforming the integration variable to $p L_0$.} Using this value, $k$ is determined by solving the $Z$ boson mass condition such that $m=M_Z^{(0)}$. With $k$ in hand, $L_0$ is given by $L_0 = 1/k$ and $L_1$ is obtained from $\ln(kL_1)$.  This also fixes $g_5$ via Eq.~(\ref{eq:g5}).

At this point the Higgs vev can be obtained by rearranging the equation for $\theta_G(L_1,v)$, 
\begin{equation}
	v=\frac{2\theta_G(L_1,v)}{g_5\sqrt{k(L_1^2-L_0^2)}}\,,
\end{equation}
and, finally, the Higgs mass can be determined by evaluating the second derivative of the Coleman-Weinberg potential at the vev, $M_H^2 = V_{CW}^{\prime\prime}(v)$~\cite{MSW}.

The choice of parameters is constrained by electroweak precision measurements. In the MCHM with the MSW fermion embedding, if the light fermions are placed close to the Planck brane---as is required to obtain realistic masses---then the KK mass scale $M_{\rm KK} = 1/L_1 \gtrsim 1.4$~TeV~\cite{EWCs,LightKK}. The parameters of the third quark generation are constrained to the following regions of parameter space: $0\leq |c_1| \leq 0.3$, $0.35\leq |c_2| \leq 0.45$, $0.55\leq |c_3| \leq 0.6$, $M_1 \geq 1$, and $M_2 < M_1$; within these requirements, $c_1>0$ and $c_2<-0.4$ are favored by electroweak precision constraints~\cite{MSW,CMSW}. Furthermore, while light fermions generally require $c$ values above $0.5$ to obtain the right masses, electroweak precision constraints require that $c<0.75$~\cite{Neutrinos}.

In what follows we use the input parameters given in Table~\ref{parameters}.  These satisfy the electroweak precision constraints with a KK mass scale just above the lower bound, so that the KK gauge boson masses are as small as possible.  Parts of this parameter set have been used for other purposes in the literature~\cite{CMSW}, which allowed for cross-checks of our work.
As we will be neglecting zero-mode fermion masses, for simplicity we will not distinguish among the different light fermions; we will use the same parameters for all leptons, neutrinos, and the first two quark generations. However, it should be noted that different choices of fermion parameters will yield extremely different spectra of KK fermions.

\begin{table}
\begin{center}
\begin{tabular}{ c  c  c  c  c  c  c }
\hline \hline
$\ln(kL_1)$ & $c^{(q_3)}_{1}$ & $c^{(q_3)}_{2}$ & $c^{(q_3)}_{3}$ & $c_{\rm light}$ & $M^{(q_3)}_1$ & $M^{(q_3)}_2$ \\
\hline
 $30$ & $0.24$ & $-0.41$ & $-0.58$ & $0.70$ & $2.3$ & $0.5$ \\
\hline \hline
\end{tabular}
\caption{Input parameters used to solve the Coleman-Weinberg potential. The parameters with superscript $q_3$ refer to those of the third quark generation, while $c_{\rm light}$ applies to all other fermions.}
\label{parameters}
\end{center}
\end{table}

The resulting Higgs mass and vev and the positions of the branes are given in Table~\ref{predictions},\footnote{We chose our parameters before the LHC Higgs discovery~\cite{LHCJuly4}.  The Higgs mass can be lowered to the preferred experimental value by slightly varying $\ln(kL_1)$.  This change will have little effect on our conclusions.}
\begin{table}
\begin{center}
\begin{tabular}{cccc}
\hline \hline
$v$ & $M_H$ &  $k = 1/L_0$ & $M_{\rm KK} = 1/L_1$ \\
\hline
$250.218$~GeV & 131.6~GeV &   $1.497\times 10^{16}$~GeV & 1401~GeV \\
\hline \hline
\end{tabular}
\caption{Predicted values resulting from the input parameters of Table~\ref{parameters}.}
\label{predictions}
\end{center}
\end{table}
and the spectrum of masses for the $Z$, $X$, and $A^{\hat{3}}$ bosons is given in Table~\ref{gaugemass}.  The lightest KK modes have masses of roughly $2.5 M_{KK} \simeq 3.5$~TeV.  As sketched in Fig.~\ref{MCHMmass}, the $Z^{(0)}$ mass is shifted upwards by EWSB to its SM value, while the higher $Z$ KK-mode masses are shifted slightly downwards, but remain heavier than the $X$ KK-mode masses of the same KK order.  Similarly, the $A^{\hat{3}}$ masses are shifted slightly upward after EWSB.  The $X$ boson masses are unaffected by EWSB.

\begin{table}
\begin{center}
\begin{tabular}{ccccccc}
\hline \hline
\multirow{2}{51bp}{KK order} & \multicolumn{3}{c}{Mass before  EWSB} & \multicolumn{3}{c}{ Mass after  EWSB} \\
 & \multicolumn{1}{c}{$Z$ [TeV]} & \multicolumn{1}{c}{$X$ [TeV]} & \multicolumn{1}{c}{$A^{\hat{3}}$ [TeV]} & \multicolumn{1}{c}{$Z$ [TeV]} & \multicolumn{1}{c}{$X$ [TeV]} & \multicolumn{1}{c}{$A^{\hat{3}}$ [TeV]} \\
\hline\hline
0 & 0 & -- & -- & 0.09119 & -- & -- \\
\hline
1 & 3.442 & 3.368 & 5.367  & 3.437 & 3.368 & 5.372 \\
\hline
2 & 7.809 & 7.732 & 9.826 & 7.804 & 7.732 & 9.831\\
\hline
3 & 12.199 & 12.121 & 14.249 & 12.194 & 12.121 &  14.254\\
\hline
4 & 16.595 & 16.515 & 18.661 & 16.590 & 16.515 & 18.667\\
\hline \hline
\end{tabular}
\caption{Masses of the $Z$, $X$, and $A^{\hat{3}}$ bosons before and after EWSB. \label{gaugemass}}
\end{center}
\end{table}

\section{The Higgsstrahlung cross section \label{sec:xsec}}

\begin{figure}
\begin{center}
\resizebox{\columnwidth}{!}{
\includegraphics{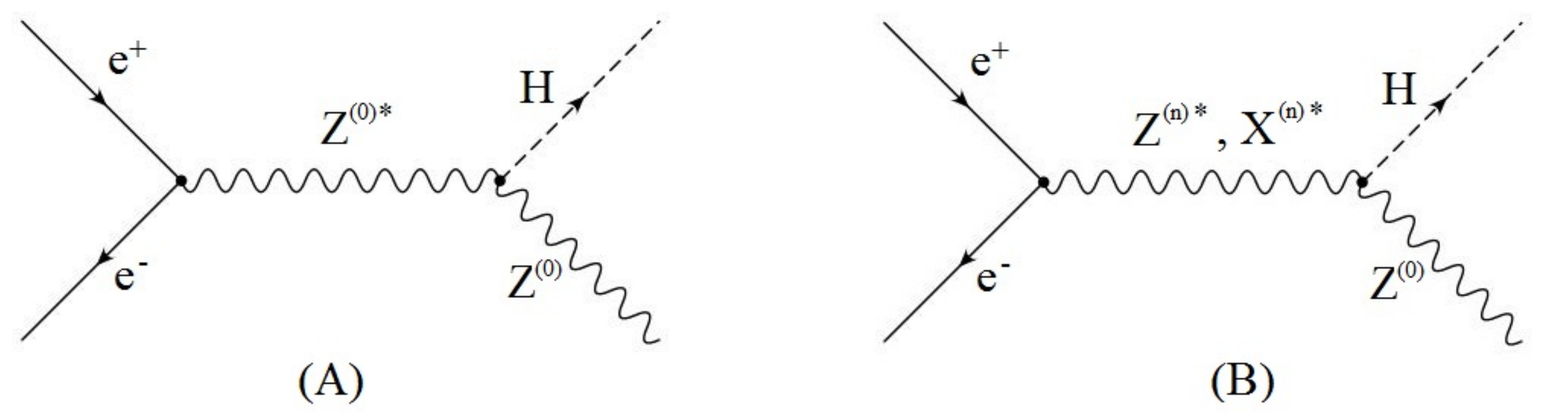}
}

\caption{(A) The SM Higgsstrahlung interaction, $e^+ e^- \rightarrow Z \rightarrow Z H$. (B) The MCHM Higgsstrahlung interaction, $e^+ e^- \rightarrow Z^{(n)}, X^{(n)} \rightarrow Z^{(0)} H$.}
 \label{Higgsstrahlung}
 \end{center}
\end{figure}

We now consider the Higgsstrahlung process, $e^+e^- \rightarrow Z^{(0)}H$, in which a Higgs boson is produced in association with a zero-mode $Z$ boson (see Fig.~\ref{Higgsstrahlung}).  In the SM, this process is mediated by $s$-channel exchange of a $Z$ boson.  In the MCHM, the KK excitations $Z_\mu^{(n)}$ also contribute, as do the $X_\mu^{(n)}$ bosons (though the $X$ contributions are numerically small).  Before EWSB, the $A^{\hat{3}}_\mu$ KK modes do not contribute because their coupling to $e^+e^-$ is zero (see Table~\ref{FermionVertices2}); after EWSB, mixing with the $Z$ and $X$ states leads to a small contribution from $A^{\hat 3}$.  The $A^{\hat{4}}_\mu$ and photon KK modes do not contribute because they do not couple to $Z^{(0)}H$.

The unpolarized SM cross section for this interaction is given by
\begin{equation}
	\sigma_{\rm SM} = 
	\frac{g_Z^2}{96\pi s^2 M_Z^{(0)\,2}} 
	\lambda^{1/2}(s, M_Z^{(0) 2}, M_H^2)
	\left(C_L^2+C_R^2\right)
	\frac{12 s M_Z ^{(0) 2} + \lambda(s, M_Z^{(0) 2}, M_H^2)}
	{(s-M_Z ^{(0) 2})^2 + \Gamma_{Z^{(0)}}^2 M_Z ^{(0) 2}},
	\label{SMHiggsstrahlung}
\end{equation}
where $s \equiv q^2$ is the square of the center-of-mass energy,
\begin{equation}
	\lambda(x,y,z) = x^2 + y^2 + z^2 - 2xy - 2xz - 2yz,
	\label{eq:lambda}
\end{equation}
and 
\begin{equation}
	g_Z=\frac{g}{\cos\theta_W}.
	\label{eq:gZ}
\end{equation}
Here $g$ is the SU(2)$_L$ gauge coupling constant and $\theta_W$ is the Weinberg angle.
We take $g=0.64939$ and $\sin^2\theta_W=0.23135$. 
We also define the left- and right-handed fermion couplings,
\begin{eqnarray}
  C_L &=& T^{3_L} - Q \sin^2 \theta_W, \qquad 
	C_R = - Q \sin^2 \theta_W.
	\label{RLcoupling} 
\end{eqnarray}

Note that at high energies $s \gg M_Z^{(0) 2}$, the kinematic function $\lambda(s, M_Z^{(0) 2}, M_H^2) \simeq s^2$, and therefore the Higgsstrahlung cross section falls like $1/s$. We will show that the corresponding cross section in the MCHM is strongly suppressed compared to this SM cross section for $\sqrt{s}$ above the scale of the first KK gauge excitations.

In what follows, we compute the Higgsstrahlung cross section in the MCHM using the 4D formulation of the theory in terms of KK modes, including the gauge KK modes up to $n = 6$.  We demonstrate the effect of including the mixing induced by EWSB and the decay widths of the KK gauge bosons.  As a cross-check we also compute the cross section in the 5D theory using the full 5D gauge propagator.

\subsection{4D calculation}

To compute the Higgsstrahlung cross section in the MCHM, we first obtain the 4D theory by integrating the 5D Lagrangian over $z$.  The full cross-section calculation formally involves a sum over the infinite tower of KK gauge modes propagating in the $s$-channel in Fig.~\ref{Higgsstrahlung}; by truncating the sum we obtain an approximate cross section.
Calculating in terms of the KK modes has several advantages, notably that it is straightforward to include zero-mode masses, KK gauge boson widths, and the effects of particle mixing induced by EWSB. 

The Higgsstrahlung cross section for left- or right-handed initial-state fermions is given by
\begin{eqnarray}
	\sigma_{L,R} &=&  \frac{1}{96\pi s^2 M_Z ^{(0) 2} }
	\lambda^{1/2}(s, M_Z^{(0) 2}, M_H^2)
	\left[12 s M_Z ^{(0) 2} + \lambda(s, M_Z^{(0) 2}, M_H^2)\right] \nonumber \\ 
	&& \times
	\left|\sum_{n=0}^\infty
	\frac{C_{Zn}^{L,R}}{s-M_Z ^{(n) 2} + i \Gamma_Z^{(n)} M_Z ^{(n)}} +
	\sum_{n=0}^\infty
	\frac{C_{Xn}^{L,R}}{s-M_X ^{(n) 2} + i \Gamma_X^{(n)} M_X ^{(n)}}
	\right|^2,
	\label{4Dcrosstot}
\end{eqnarray}
where $\Gamma_{Z,X}^{(n)}$ is the total decay width of each particle propagating in the $s$-channel, $C^{L,R}_{Zn}$ and $C^{L,R}_{Xn}$ are the appropriate products of couplings, and $L,R$ refer to the polarizations of the initial-state fermions.  The unpolarized cross section corresponding to Eq.~(\ref{SMHiggsstrahlung}) is obtained by averaging over the initial-state fermion polarizations,
\begin{equation}
	\sigma_{\rm tot} = \frac{1}{4}\left(\sigma_L + \sigma_R \right).
\end{equation}

As a first pass, we neglect the particle mixing caused by EWSB.  The products of couplings are then given by $C_{Gn}^{L,R} = C^{(n)}_{Gff_{L,R}} C^{(n)}_{GZH}$, with $G = Z$ or $X$, where
\begin{equation}
	C^{(n)}_{Gff_{L,R}} = g_G \, c_{Gff}^{L,R} Z_{Gff_{L,R}}^{(n)}, \qquad 
	C^{(n)}_{GZH} = \frac{v}{2} g_Z \, g_G \, Z_{GZH}^{(n,0)}.
	\label{overallcoef}
\end{equation}
The fermion couplings constants $c_{Gff}^{L,R}$ are defined as
\begin{eqnarray}
	\!\!\!\!\!\!\! c_{Zff}^{L,R} &=& C_{L,R}, \qquad
	c_{Xff}^{L} = \frac{(Q-T^3_L)\sin^2\theta_W}{\cos{2\theta_W}}, \qquad
	c_{Xff}^{R} = \frac{Q\sin^2\theta_W - T^3_R \cos^2\theta_W}{\cos{2\theta_W}}\,,
\label{coupcoef}
\end{eqnarray}
where $C_L$ and $C_R$ are defined in Eq.~(\ref{RLcoupling}). The gauge coupling of $X$ is given analogously to Eq.~(\ref{eq:gZ}), 
\begin{equation}
	g_X = \frac{g \sqrt{\cos{2\theta_W}}}{\cos\theta_W},
	\label{eq:gX}
\end{equation}
Finally, the coefficients $Z_{Gff_{L,R}}^{(n)}$ and $Z_{GZH}^{(n,0)}$ are the integrals over $z$ of the profiles of the particles involved in each interaction vertex,
\begin{eqnarray}
	Z_{Gff_{L,R}}^{(n)} &=& \int_{L_0}^{L_1} dz \, \frac{1}{(k z)^4} \frac{g_5}{g} f_G^{(n)}(m_n,z) 
	\left[ f_{f_{L,R}}^{(0)}(m_f^{(0)},z) \right]^2, \label{Zgff} \nonumber \\
	Z_{GZH}^{(n,0)} &=& \int_{L_0}^{L_1} dz \, \frac{1}{k z} \left(\frac{g_5}{g}\right)^2 \
	f_G^{(n)}(m_n,z) f_Z^{(0)}(M_Z^{(0)},z) \left[ f_H(z) \right]^2. 
	\label{Zgzh}
\end{eqnarray}

When EWSB-induced mixing is included, the vertex factors $C^{(n)}_{Gff_{L,R}}$ and $C^{(n)}_{GZH}$ incorporate all possible couplings involving the components of the mixed particles.  In this case the gauge generators cannot be factored out from the $z$ integrals and must be evaluated before integration. In both cases, we compute the integrals numerically using {\tt Maple}~\cite{Maple}.

\subsection{5D calculation}

The Higgsstrahlung cross section can be computed directly in the 5D theory by assembling the amplitude in terms of the 5D Feynman rules, multiplying by the external zero-mode profiles, and integrating over the fifth dimension. This method has the advantage of including all of the propagating KK modes automatically; however, the inclusion of widths, mixing effects, and zero-mode masses becomes difficult.  We therefore use this as a check of our 4D calculation.

The cross section is given by
\begin{eqnarray}
	\sigma_{L,R} &=& \frac{g_Z^2 v^2}{384\pi s^2 M_Z ^{(0) 2} s^2 } 
	\lambda^{1/2}(s, M_Z^{(0) 2}, M_H^2)
	\left[12 s M_Z ^{(0) 2} + \lambda(s, M_Z^{(0) 2}, M_H^2) \right] \nonumber \\ 
	&& \times
	\left[g_Z^2 c_{Zff}^{L,R} Z_{Z}(q,c) + g_X^2 c_{Xff}^{L,R} Z_X(q,c) \right]^2,
\label{5Dcrosstot}
\end{eqnarray}
where the constants $c_{Gff}^{L,R}$ were defined in Eq.~(\ref{coupcoef}), $q$ is the center-of-mass four-momentum, and $Z_{G}(q)$ is the integral of the 5D components over the positions $z$ and $z^{\prime}$ of the two vertices in the fifth dimension:
\begin{eqnarray}
	Z_{G}(q,c) &=& \int_{L_0}^{L_1} dz \, \frac{1}{(k z)^4} \left(\frac{g_5}{g}\right)^3 
	\left[ f_{f}^{(0)}(z) \right]^2 \left[ 
	\int_{L_0}^{z} dz^{\prime} \,\frac{1}{k z^{\prime}} 
	G_G(z^{\prime},z;q) f_{Z}^{(0)}(z^{\prime}) \left[ f_H(z^{\prime}) \right]^2 \right.
	\nonumber \\
	&& \left. + \int_{z}^{L_1} dz^{\prime} \,\frac{1}{k z^{\prime}} 
	G_G(z,z^{\prime};q) f_Z^{(0)}(z^{\prime}) \left[ f_H(z^{\prime}) \right]^2 \right],
\label{Z}
\end{eqnarray}
where the dependence on the fermion parameter $c$ appears from the fermion profile.
The function $G(u,v;q)$ arises from the 5D propagator, which is given in unitarity gauge by\footnote{Note that only the $\eta^{\mu\nu}$ term will contribute to our process because we neglect the tiny initial-state fermion masses.}
\begin{equation}
	-i G_G(z,z^{\prime};p) \left(\eta^{\mu\nu}-\frac{p^\mu p^\nu}{p^2}\right) 
	- i G_G(z,z^{\prime};0)\left(\frac{p^\mu p^\nu}{p^2}\right).
\end{equation}
The function $G(u,v;q)$ is defined as the Green's function of the gauge boson equation of motion, 
\begin{equation}
	\left[p^2 - \frac{1}{z} \partial_z + \partial_z ^2 \right] G_G(z,z^{\prime};p)
	= k z \delta(z-z^{\prime}).
	\label{Gdef}
\end{equation}
 The solution is given by
\begin{equation}
	G(u,v;p) = \frac{\pi}{2}\frac{k u v}{AD-BC}
	\left[A J_1 (p u) + B Y_1 (p u)\right]
	\left[C J_1 (p v) + D Y_1 (p v)\right],
	\label{Gpropb}
\end{equation}
where $u={\rm min}(z,z^{\prime})$, $v={\rm max}(z,z^{\prime})$, the coefficients $A, B, C, D$ are determined by applying the boundary conditions for the gauge boson in question, and $J$ and $Y$ are Bessel functions.  This 5D propagator contains all the poles corresponding to all the KK modes of the corresponding gauge boson.
The integration is performed in two pieces to account for the fact that the propagator is defined with $u < v$.  We compute the integrals numerically using {\tt Maple}~\cite{Maple}.

The 5D calculation can be shown explicitly to be equivalent to the 4D calculation (neglecting the gauge KK mode widths, mixing induced by EWSB, and zero-mode masses) using the fact that the 5D gauge boson propagator obeys the identity~\cite{WarpedReview}
\begin{equation}
	G_G(z,z^{\prime};p) = \sum_{n=0}^\infty \frac{f_G^{(n)}(z) f_G^{(n)}(z')}{p^2-M_G^{(n) 2}}.
	\label{Gtoprof}
\end{equation}
Inserting this into Eq.~(\ref{Z}), the double integral can be separated into a sum over gauge KK modes of the gauge propagator times separate integrals over $z$ and $z^{\prime}$, which reduce to the integrals for the couplings in Eq.~(\ref{Zgzh}).
Therefore, the cross section obtained in Eq.~(\ref{5Dcrosstot}) is completely equivalent to that in Eq.~(\ref{4Dcrosstot}) in the limit that $\Gamma_G^{(n)} = 0$ and EWSB-induced mixing is neglected (this also implies that $M_Z^{(0)} = 0$).

\subsection{Numerical results \label{CrossSectionComparison}}

In order to illustrate the origins of the features of the cross section, we present our numerical results in three stages.  First we compare the 4D and 5D calculations, neglecting EWSB-induced particle mixing and gauge boson widths.  We then show the effect of the EWSB-induced mixing.  Finally we include the gauge boson widths, which have a dramatic effect on the behavior of the cross section above the first $Z$ KK mode, especially after EWSB-induced mixing.  In all cases we use the parameters in Table~\ref{parameters}.  We also choose $M_1 = 1.5$ and $M_2=0.627$ for all fermions other than the third-generation quarks. The choice of $M_2$ was made by requiring that the down-type fermion mass condition yield the electron mass, as this is the most important light fermion for the process we consider.

\begin{figure}
\begin{center}
\resizebox{\columnwidth}{!}{
\includegraphics{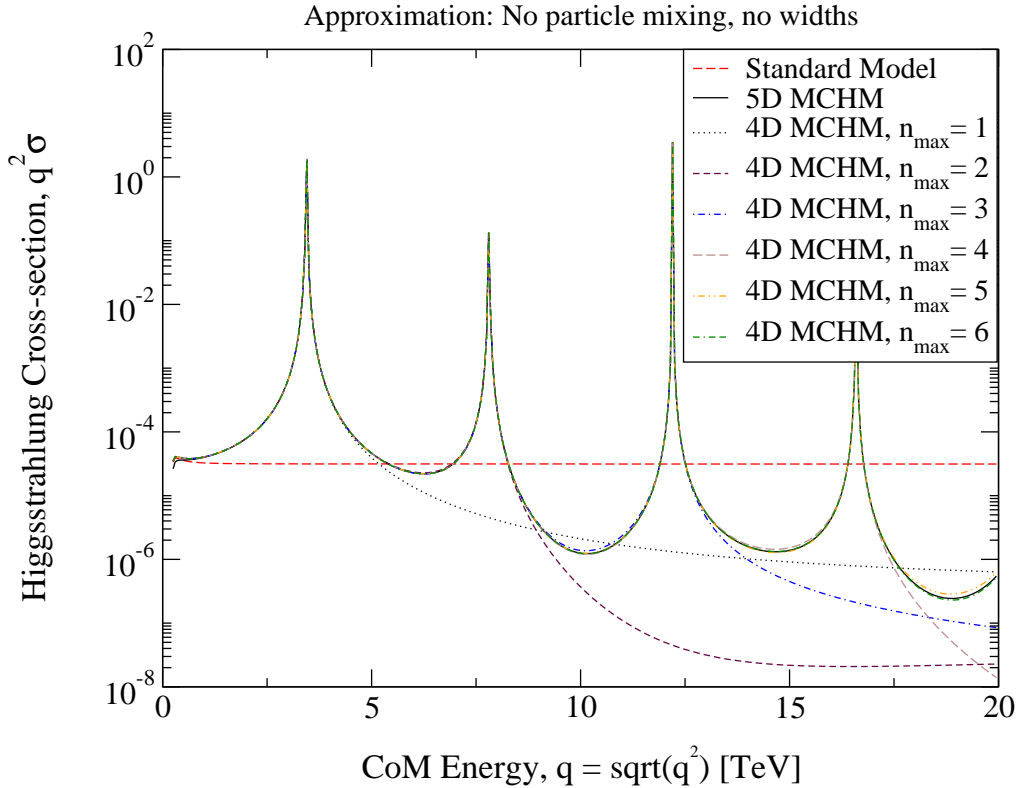}
}
\caption{The unpolarized Higgsstrahlung cross section multiplied by the square of the center-of-mass energy $q^2$.  Shown are the SM (red dashed line), the MCHM 5D calculation (black solid line), and the MCHM 4D calculation with the sum over KK modes truncated at $n = 1$ through 6.  EWSB-induced mixing and gauge KK-mode decay widths are neglected.  (Note that $q^2 \sigma_{\rm tot}$ is dimensionless in natural units.)}
\label{IncKK}
\end{center}
\end{figure}

We begin by ignoring EWSB-induced mixing and setting all gauge boson widths to zero.  We compute the total unpolarized cross section for $e^+e^- \to Z^{(0)}H$ as a function of the center-of-mass energy $\sqrt{s} = \sqrt{q^2} \equiv q$.  Results are shown in Fig.~\ref{IncKK}.  Because the SM cross section is proportional to $1/q^2$ in the high-energy limit, we plot $q^2 \sigma_{\rm tot}$ (this quantity is dimensionless in natural units).  The asymptotic behavior of the SM cross section results in a constant high-energy value for $q^2 \sigma_{\rm tot}$, allowing us to illustrate more clearly the suppression of the cross section in the MCHM.

In addition to the SM cross section, in Fig.~\ref{IncKK} we plot (i) the cross section from the 5D calculation and (ii) the cross section from the 4D calculation including a successively increasing number of gauge boson KK modes.  The resonances clearly visible in the cross section are those of the $Z$ boson KK modes.  The contribution of the $X$ KK modes is numerically negligible for our choice of $c=0.7$ for the electron (this remains true unless the electron $c$ parameter becomes quite close to $0.5$).  This is because the electron's profile (like that of all light fermions) is peaked toward the Planck brane, while the $X$ KK mode profiles are zero on the Planck brane due to the Dirichlet boundary condition. 

Figure~\ref{IncKK} illustrates the excellent agreement between the full 5D cross section calculation and the 4D calculation truncated at a finite KK number.  This agreement holds up to a center-of-mass energy just above the mass of the heaviest KK mode included in the 4D calculation.\footnote{The small discrepancy between the 4D and 5D calculations below 1000~GeV is due to the fact that the 5D calculation uses $M_Z^{(0)} = 0$ since EWSB is not taken into account; we use the physical $Z$ boson mass for $M_Z^{(0)}$ in the 4D calculation.}
We thus learn that we can safely neglect the contribution of KK modes with masses much higher than the center-of-mass energies of interest.  Figure~\ref{IncKK} also provides a first illustration of the suppression of the Higgsstrahlung cross section above the energy scale of the first gauge KK modes, visible at center-of-mass energies away from the gauge KK resonances.

\begin{figure}
\begin{center}
\resizebox{\columnwidth}{!}{
\includegraphics{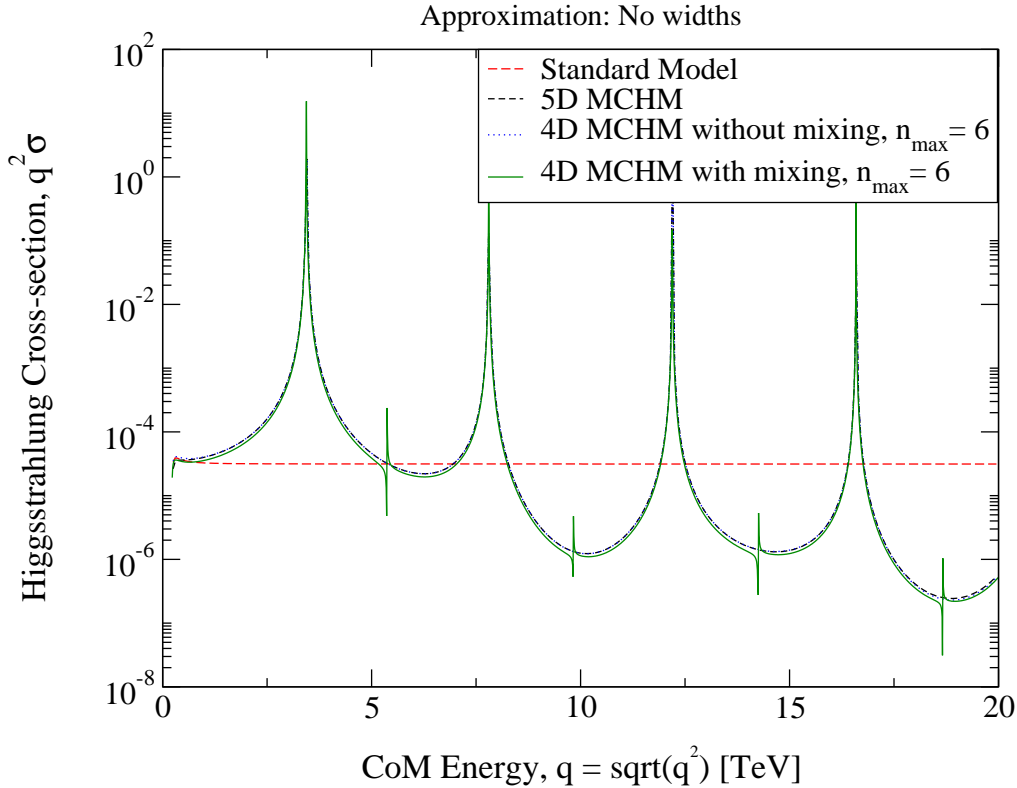}
}
\caption{As in Fig.~\ref{IncKK} but now illustrating the effect of using the KK masses, profiles, and couplings derived including EWSB-induced mixing (solid green line).  Gauge KK-mode widths are neglected.}
\label{NoWidthsMixed}
\end{center}
\end{figure}

We next implement the EWSB-induced particle mixing into the 4D calculation.  This mixing does not substantially change the magnitudes of the couplings relevant to the Higgsstrahlung interaction.  We show this in Fig.~\ref{NoWidthsMixed} by plotting the result of the 4D calculation with and without EWSB-induced mixing, including KK modes with $n \leq 6$.  We again ignore the gauge KK mode widths and plot the result of the 5D (unmixed) calculation for comparison.  The main new feature is the appearance of the $A^{\hat 3}$ resonances at masses in between those of the $Z$ KK modes.  The $A^{\hat 3}$ gauge generators do not permit a coupling to $e^+e^-$ before EWSB; after EWSB, the $A^{\hat 3}$ resonances contribute to Higgsstrahlung only via the small admixture of the $Z$ KK modes into the corresponding mass eigenstates.

A second new feature of the calculation including EWSB is that the Higgsstrahlung cross section  is slightly suppressed at all center-of-mass energies.  This effect is due to slight changes in the normalization of the $Z^{(n)}$ couplings.  Both the $Z\bar{f}f$ and $ZZ^{(0)}H$ couplings are reduced by about 3\% after EWSB-induced mixing, which leads to an overall suppression of roughly 7\% in the cross section.  The implications of this coupling shift for Higgs production and decay at energies below the KK scale have been studied in detail in Ref.~\cite{Espinosa:2010vn}.  The $X$ KK-mode couplings are also modified by EWSB-induced mixing, but the effect is small (the product of couplings is shifted by less than 1\%).  The contributions of the $X$ KK modes to the Higgsstrahlung cross section thus remains numerically negligible.

We finally incorporate the gauge KK-mode decay widths into the 4D cross-section calculation.  We compute the widths including two-body decays to all kinematically accessible Higgs and gauge bosons, including KK modes, and to the SM fermions.  Expressions for the widths are collected in Appendix~\ref{AppDecay}.  For simplicity, we omit decays to fermion KK modes; this lets us avoid the substantial model-dependence of the fermion KK spectrum without significantly changing our conclusions (decays to KK fermions will be discussed below).  

\begin{figure}
\begin{center}
\resizebox{\columnwidth}{!}{
\includegraphics{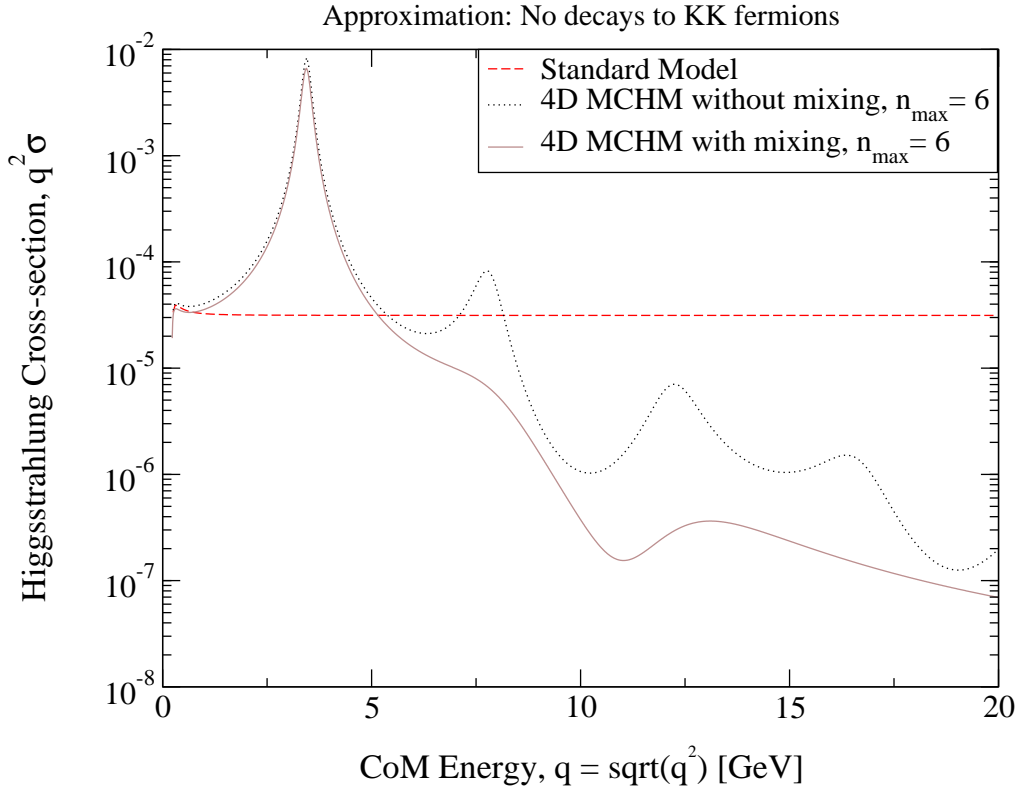}
}
\caption{The unpolarized Higgsstrahlung cross section times $q^2$ computed in the 4D theory with $n \leq 6$, including gauge KK-mode widths (see text for details).  Results are shown with (solid line) and without (dotted line) EWSB-induced mixing.}
\label{WidthsMix}
\end{center}
\end{figure}

Including the gauge KK-mode decay widths has a dramatic effect on the Higgsstrahlung cross section at center-of-mass energies above the $Z^{(1)}$ KK resonance.  We show this in Fig.~\ref{WidthsMix} by plotting the result of the 4D calculation with gauge KK-mode decay widths, with and without EWSB-induced mixing, including KK modes with $n \leq 6$.
Even before EWSB, the widths of the $Z$ KK modes with $n \geq 2$ are quite large.  This is mainly due to the rapid growth of the decay width of a gauge KK mode to two lighter gauge bosons with increasing mass of the KK mode (see Appendix~\ref{AppDecay}), together with the proliferation of kinematically-accessible final states.  
After EWSB, the widths of the $Z$ KK modes with $n \geq 2$ become even larger.  This is due to the appearance of decays involving an $A^{\hat a}$ gauge KK mode in the final state, which are accessible only via the $A^{\hat 3}$ admixture in the $Z$ KK modes after EWSB.  The small mixing is compensated by the large couplings among these gauge KK states.  
These large decay widths flatten the resonance structure of the Higgsstrahlung cross section at center-of-mass energies above the first $Z$ KK mode, yielding a formfactor-like behavior that we interpret as the hallmark of the composite Higgs.

The decay width of the first $Z$ KK mode is relatively modest, around 7\% of its mass both before and after EWSB.  The width of this mode is not significantly altered by EWSB because it is too light to decay into other gauge KK modes.  Its couplings to the kinematically-accessible zero-modes are only slightly modified by EWSB-induced mixing effects.  Up to now we have omitted the contribution to the width from decays into fermion KK modes.  For our choice of parameters, only the first top-quark KK mode is lighter than half the $Z^{(1)}$ mass; including decays to these states increases the $Z^{(1)}$ width to about 10\% of its mass.  Raising the light-fermion parameter $M_1$ to $2.2$ lowers the masses of the first neutrino, up-quark, and charm-quark KK modes so that they can also appear in the final states of $Z^{(1)}$ decays; for this parameter set, the $Z^{(1)}$ width becomes about 13\% of its mass.  Other parameter sets (see Refs.~\cite{MSW,CMSW}) can also result in low masses for the first fermion KK excitations.

The decay widths of the second and higher $Z$ KK modes already reach 30\%--50\% of their masses excluding decays to KK fermions.  The very large multiplicity of accessible final states involving KK fermions will increase these widths further---recall that the MSW embedding contains 20 quarks and 20 leptons per generation, including the exotic fermions that have no zero modes.  The resulting large gauge KK-mode widths begin to call into question the perturbativity of the theory.  However, their effect on the Higgsstrahlung cross section will be only to further flatten the gauge KK resonances, leading to a smoother fall-off of the cross section with increasing center-of-mass energy.

\subsection{Source of the cross-section suppression \label{sec:kkcoup}}

The formfactor-like suppression of the Higgsstrahlung cross section at center-of-mass energies above the first $Z$ KK resonance arises due to progressive cancellations among the KK-mode contributions to the cross section.  To illustrate the cancellation, consider the contribution of the first $n$ KK modes at a center-of-mass energy much higher than the masses of the considered KK modes.\footnote{Because there are an infinite number of KK modes, we are necessarily neglecting the contributions of KK modes with masses near the center-of-mass energy.  Nevertheless, this approximation allows us to illustrate the dominant source of the cross-section suppression at center-of-mass energies up to about an order of magnitude above the mass of the first $Z$ KK mode.}  In this limit, $s \gg M^2, M\Gamma$ in the propagators and we obtain the approximate result
\begin{equation}
	\frac{\sigma_{L,R}^{{\rm MCHM}, n}}{\sigma_{L,R}^{\rm SM}} \approx 
	\left(\sum^{n}_{i=0} \frac{C^{(i)}_{Zff_{L,R}} C^{(i)}_{ZZH} 
	+ C^{(i)}_{Xff_{L,R}} C^{(i)}_{XZH}}
	{C^{(\rm SM)}_{Zff_{L,R}} C^{(\rm SM)}_{ZZH}}\right)^2, \qquad 
	{\rm if} \ \sqrt{s} \gg M^{(n)},
\label{coupsum}
\end{equation}
where $C^{(\rm SM)}_{Zff_{L,R}} = g_Z c^{L,R}_{Zff}$ and $C^{(\rm SM)}_{ZZH} = g_Z^2 v_{\rm SM}/2$ [see Eqs.~(\ref{overallcoef})--(\ref{Zgzh})].  We ignore EWSB-induced mixing, which has very little effect on the couplings involving zero modes.  The product of the $Z^{(n)}$ couplings is then given by,
\begin{equation}
	C^{(n)}_{Zff_{L,R}} C^{(n)}_{ZZH} = g_Z c^{L,R}_{Zff} Z^{(n)}_{Zff_{L,R}} 
	\times g_Z^2 \frac{v}{2} Z^{(n,0)}_{ZZH}.
\end{equation}
In particular, the product of couplings differs from that of the SM $Z$ boson only through the presence of the integrals of the 5D profiles and through the small shift in the Higgs vev (we use the MCHM value from Table~\ref{predictions}).  Because of this, if we neglect the (small) contributions of the $X$ KK modes, we obtain the simple relation,
\begin{equation}
	\frac{\sigma_{L,R}^{{\rm MCHM}, n}}{\sigma_{L,R}^{\rm SM}} \approx 
	\frac{v^2}{v_{\rm SM}^2} 
	\left(\sum^{n}_{i=0} Z_{Zff_{L,R}}^{(i)} Z_{ZZH}^{(i,0)} \right)^2,
 	\qquad {\rm if} \ \sqrt{s} \gg M^{(n)}.
 \label{coupapprox}
\end{equation}

\begin{figure}
\begin{center}
\resizebox{\columnwidth}{!}{
\includegraphics{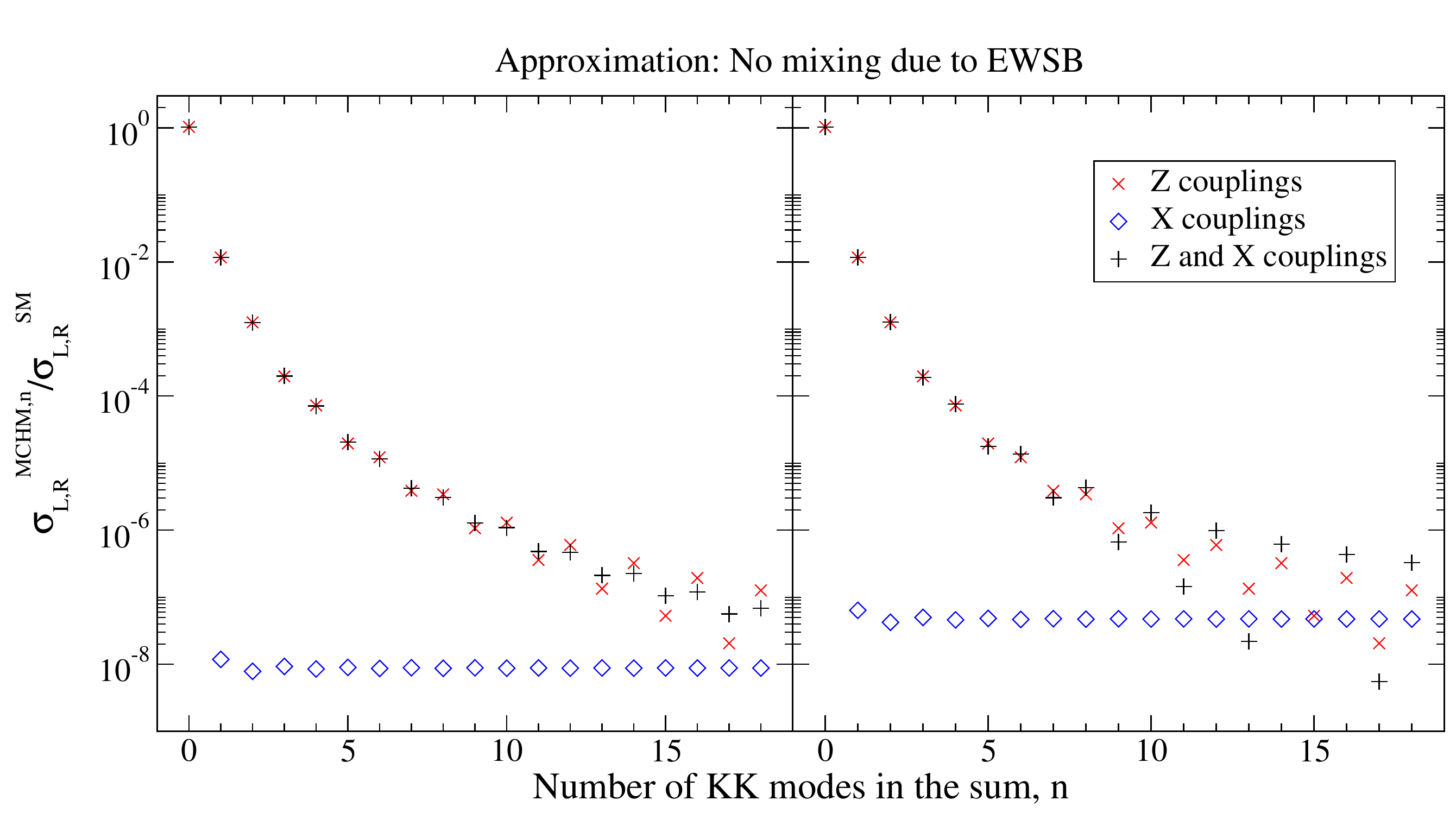}
}
\caption{The ratio of the squared sum of the Higgsstrahlung interaction couplings with respect to the SM Higgsstrahlung couplings from Eq.~(\ref{coupsum}) plotted versus the number of included KK modes.  The left (right) panel shows the ratio for left-handed (right-handed) initial-state electrons. }
\label{SqCoupLR}
\end{center}
\end{figure}

We plot the ratio in Eq.~(\ref{coupsum}) in Fig.~\ref{SqCoupLR}, for both left- and right-handed initial-state electrons.  We show separately the contributions from the $Z$ and $X$ KK modes, together with the complete expression.  The contribution from the $Z$ KK modes overwhelmingly dominates up to $n \sim 10$, and obeys an approximate power-law behavior $\sim n^{-5.5}$.  
This behavior comes from the value of the integrals of the 5D profiles in Eq.~(\ref{coupapprox}), which for the first few $Z$ KK modes reads,
\begin{equation}
	\sum^{n}_{i=0} Z_{Zff_{L,R}}^{(i)} Z_{ZZH}^{(i,0)} 
	\simeq 1.0 - 1.1 + 0.1 - \ldots.
	\label{eq:Zsum}
\end{equation}	
In particular, the product of couplings of the first $Z$ KK mode is about 10\% larger in magnitude and opposite in sign compared to that of the zero-mode $Z$ boson.  The first $Z$ KK mode thus cancels the entire SM Higgsstrahlung amplitude with about 10\% overshoot.  The product of couplings of the second $Z$ KK mode is about 10\% of that of the zero-mode $Z$ boson, with the same sign; this cancels most of the remaining amplitude.  The sum alternates in sign with steadily decreasing magnitude as $n$ increases.

At large $n \gtrsim 10$, the ratio in Eq.~(\ref{coupsum}) deviates from a power law as the $Z$ KK mode contributions asymptote to a finite value and the $X$ KK mode contributions begin to become significant.  At this point the relevance of the truncated KK-mode sum breaks down, because contributions from KK modes with $M^{(n)} \sim \sqrt{s}$ can no longer be neglected.  

We expect that an analytic understanding of the cross-section suppression could be obtained from the 5D gauge propagator including the absorptive part of the one-loop radiative corrections (equivalent to including the gauge KK-mode widths) to push the poles off the real $q^2$ axis.  Indeed, Ref.~\cite{McDonald:2010fe} used this approach to understand the high-energy behavior of scattering mediated by gauge bosons in a warped hidden sector, with the SM confined to the UV brane.  Their approach involved matching the 5D propagator onto a theory in which the IR brane is taken to infinity (equivalent to the IR scale being taken to zero).  This eliminates the poles in the 5D propagator, leading to a smooth cross section for UV-to-UV processes that falls as a power law with collision energy.  However, it is not clear how the details of this approach can be applied to our set-up, since our process involves the Higgs which is localized near the IR brane.  Computation of the one-loop propagator in the 5D theory is beyond the scope of this paper.

\section{Coupling extraction at an electron-positron collider \label{sec:coupling}}

We now consider the prospects for experimentally testing the progressive cancellation of the $Z$-boson KK-mode contributions to the Higgsstrahlung cross section by measuring the relevant product of couplings.  This can be done using measurements of the cross section for $e^+e^- \to Z^{(0)} H$ at more than one center-of-mass energy.  We consider measurement prospects at the proposed International Linear Collider (ILC)~\cite{ILC2} and Compact Linear Collider (CLIC)~\cite{CLIC}.  The expected collision energies of these two machines lead us to consider extraction of the couplings of only the first and second $Z$ KK modes.  For simplicity we use unpolarized cross sections; separating a fixed-luminosity data sample into equal left- and right-polarized samples offers no advantage in this analysis.

We compute the Higgsstrahlung cross section at center-of-mass energies of 500~GeV and 1~TeV (ILC) and 3 and 5~TeV (CLIC).  We use the parameters of Table~\ref{parameters} together with $M_1 = 1.5$ for the light fermions.  We include EWSB-induced mixing and compute the gauge KK-mode widths including decays to all kinematically-accessible boson pairs as well as SM (zero-mode) fermion pairs.  For the $Z^{(1)}$ width we also include decays to all kinematically-accessible fermion KK modes.  We include gauge KK modes with $n \leq 6$.

For each center-of-mass energy, we compute the statistical uncertainty on the Higgsstrahlung cross section assuming 500~fb$^{-1}$ of integrated luminosity at that energy.  We assume that all $e^+e^- \to Z^{(0)} H$ events are detected, and ignore backgrounds and systematic uncertainties; this gives us a best-case estimate of the coupling sensitivity.  The resulting cross sections, numbers of events, and statistical uncertainties are summarized in Table~\ref{ExtractionDataMix}.  Recall that our benchmark parameters yield masses for the first and second $Z$ KK modes of 3.44 and 7.80~TeV, respectively.

\begin{table}
\begin{center}
\begin{tabular}{ccccc}
\hline \hline
$\sqrt{s}$ (TeV) & $\sigma_{\rm tot}$ (fb) & $N_S$ [500~fb$^{-1}$] & $\sqrt{N_S}$ & $\Delta \sigma_{\rm tot}$ (fb) \\
\hline
0.5  & 52.91 & 26,455 & 163 & 0.3253  \\
1 & 13.90  & 6,947 & 83 & 0.1667 \\
3 & 21.51 & 10,752 & 104 & 0.2074 \\
5 & 0.5819  & 291 & 17 & 0.03412  \\
\hline \hline
\end{tabular}
\end{center}
\caption{Unpolarized Higgsstrahlung cross sections, numbers of signal events in 500~fb$^{-1}$, and statistical uncertainties on the number of events and cross section for various $e^+e^-$ center-of-mass energies. \label{ExtractionDataMix}}
\end{table}

We proceed to study how well the relevant products of couplings can be extracted in two scenarios: (i) extraction of the $Z^{(0)}$ and $Z^{(1)}$ couplings, ignoring the presence of higher KK modes; and (ii) extraction of the $Z^{(1)}$ and $Z^{(2)}$ couplings, assuming that the $Z^{(0)} Z^{(0)} H$ coupling has been precisely measured elsewhere (e.g., in $Z^{(0)} H$ production near threshold or in Higgs decays).  In each case we construct a $\Delta \chi^2$ observable,
\begin{equation}
	\Delta\chi^2 = \sum_{i=1}^n 
	\frac{\left(\sigma^{i}_{\rm test}-\sigma^{i}\right)^2}{\left( \Delta\sigma^i \right)^2},
\label{chisq}
\end{equation}
where $\sigma^i$ and $\Delta \sigma^i$ are the measured cross section and its uncertainty from Table~\ref{ExtractionDataMix} and $\sigma^i_{\rm test}$ is a test function that depends on the two unknown products of couplings that we wish to extract.  In each case we will plot 95\% confidence regions ($\Delta \chi^2 = 5.99$) including measurements at two, three, or four different center-of-mass energies.

We first consider extraction of the relevant products of couplings of $Z^{(0)}$ and $Z^{(1)}$.  Because the $Z^{(0)}$ couplings to $e^+e^-$ have already been precisely measured at LEP, the former amounts to a measurement of the $Z^{(0)}Z^{(0)}H$ coupling.  We assume that the $Z^{(1)}$ mass and width will have already been measured, e.g., at the LHC, and ignore their uncertainties.  We construct the test function according to,
\begin{equation}
	\sigma_{\rm test}(s,C_Z^{(0)},C_Z^{(1)}) = {\rm Coef}[s]
	\left|\frac{C_Z^{(0)}}{s - M_Z^{(0) 2}+i\Gamma_Z^{(0)}M_Z^{(0)}} 
	+ \frac{C_Z^{(1)}}{s - M_Z^{(1) 2}+i\Gamma_Z^{(1)}M_Z^{(1)}} \right|^2,
	\label{eq:testfunc1}
\end{equation}
where $C_Z^{(0)}$ and $C_Z^{(1)}$ are the products of couplings for the $Z^{(0)}$ and $Z^{(1)}$ normalized to the corresponding product of SM $Z$ boson couplings (this normalized product of couplings is the same for left-handed and right-handed initial-state electrons).  If we were to neglect EWSB-induced mixing, these normalized products of couplings would correspond to the products of 5D profile integrals times $v/v_{\rm SM}$ as given in Eq.~(\ref{coupapprox}).
Here ${\rm Coef}[s]$ denotes the usual SM coefficients of the $e^+ e^-\rightarrow Z^{(0)} H$ cross section.  

\begin{figure}
\begin{center}
\resizebox{0.5\textwidth}{!}{
\includegraphics{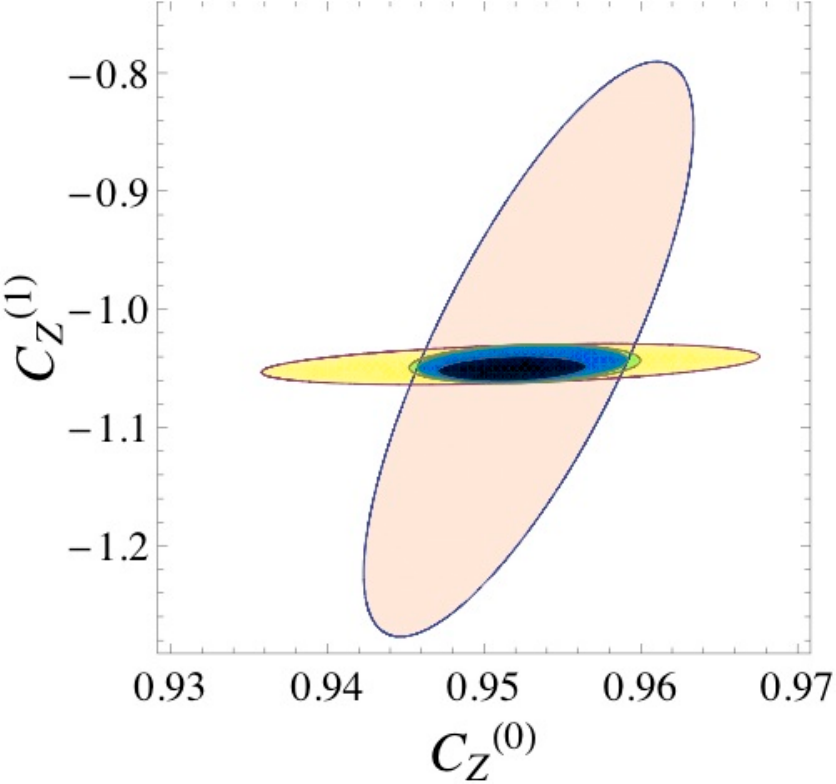}
}\resizebox{0.5\textwidth}{!}{
\includegraphics{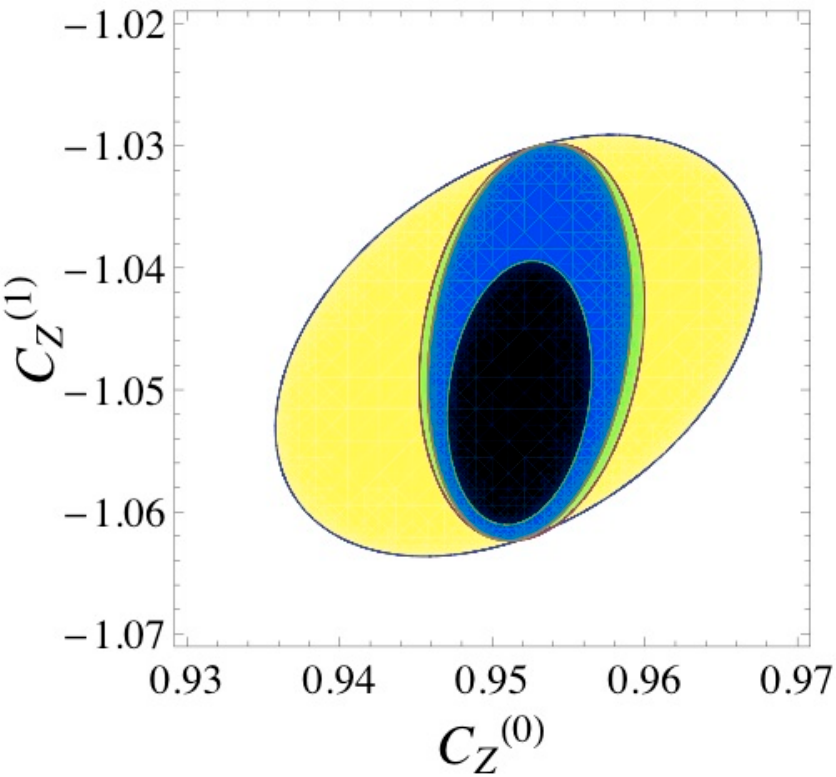}
}
\caption{95\% confidence level regions for $C_Z^{(0)}$ and $C_Z^{(1)}$ from ILC and CLIC cross-section measurements at various center-of-mass energies.  The largest ellipse in the left panel corresponds to cross-section measurements at 0.5 and 1~TeV from the ILC alone.  The right panel shows a blow-up of the region near $C^{(1)}_Z = -1.05$.  From largest to smallest, the ellipses in the right panel correspond to cross-section measurements from both the ILC and CLIC, at center-of-mass energies of 1 and 3~TeV; 0.5 and 3~TeV; 0.5, 1, and 3~TeV; and 0.5, 1, 3, and 5~TeV.  The actual normalized products of couplings are $C_Z^{(0)}=0.953$ and $C_Z^{(1)}=-1.052$.}
\label{Z0Z1Mix}
\end{center}
\end{figure}

The resulting 95\% confidence regions for $C^{(0)}_Z$ and $C^{(1)}_Z$ are shown in Fig.~\ref{Z0Z1Mix}.  The largest ellipse in the left panel of Fig.~\ref{Z0Z1Mix} shows the constraint from cross section measurements at the ILC alone, at 0.5 and 1~TeV.  Even though these collision energies are well below the mass of the first $Z$ KK mode, the ILC is able to clearly detect its influence.  ILC measurements are also enough to determine that the normalized product of $Z^{(1)}$ couplings is opposite in sign and similar in magnitude (to within about $\pm 25\%$ at 95\% confidence level) to that of $Z^{(0)}$, as needed for the cancellation that is responsible for the cross-section suppression.
Furthermore, ILC measurements would clearly differentiate $C_Z^{(0)}$ from the SM expectation $C_Z^{\rm SM} = 1$; this difference is a well-known feature of the $Z^{(0)}Z^{(0)}H$ coupling in the MCHM (see, e.g., Ref.~\cite{Espinosa:2010vn}).

The right panel of Fig.~\ref{Z0Z1Mix} shows the constraints from cross-section measurements at the ILC and CLIC.  A CLIC cross-section measurement at 3~TeV---on the lower-energy flank of the $Z^{(1)}$ resonance---allows $C^{(1)}_Z$ to be extracted to within about $\pm 2\%$ at 95\% confidence level, giving clear evidence of the ``overshoot'' in the coupling cancellation, $|C^{(1)}_Z| > |C^{(0)}_Z|$.  Interestingly, an ILC measurement at 0.5~TeV combined with the 3~TeV CLIC measurement provides a better determination of $C^{(0)}_Z$ than does an ILC measurement at 1~TeV.  This is due to the better statistics from the larger Higgsstrahlung cross section at 0.5~TeV.

We now consider extraction of the relevant products of couplings of $Z^{(1)}$ and $Z^{(2)}$, assuming that the $Z^{(0)}$ coupling to the Higgs has already been measured.  We assume that masses and widths of both $Z^{(1)}$ and $Z^{(2)}$ are known, e.g., from LHC measurements, and neglect their uncertainties.  We construct a new test function analogous to Eq.~(\ref{eq:testfunc1}),
\begin{equation}
	\sigma_{\rm test}(s,C_Z^{(1)},C_Z^{(2)}) = {\rm Coef}[s]
	\left| \sum_{n=0}^2 \frac{C_Z^{(n)}}{s - M_Z^{(n)2} + i \Gamma_Z^{(n)} M_Z^{(n)}}\right|^2,
\end{equation}
where our fit parameters $C^{(1)}_Z$ and $C^{(2)}_Z$ are again the products of couplings for $Z^{(1)}$ and $Z^{(2)}$ normalized to the corresponding product of SM $Z$ boson couplings.  

\begin{figure}
\begin{center}
\resizebox{0.5\textwidth}{!}{
\includegraphics{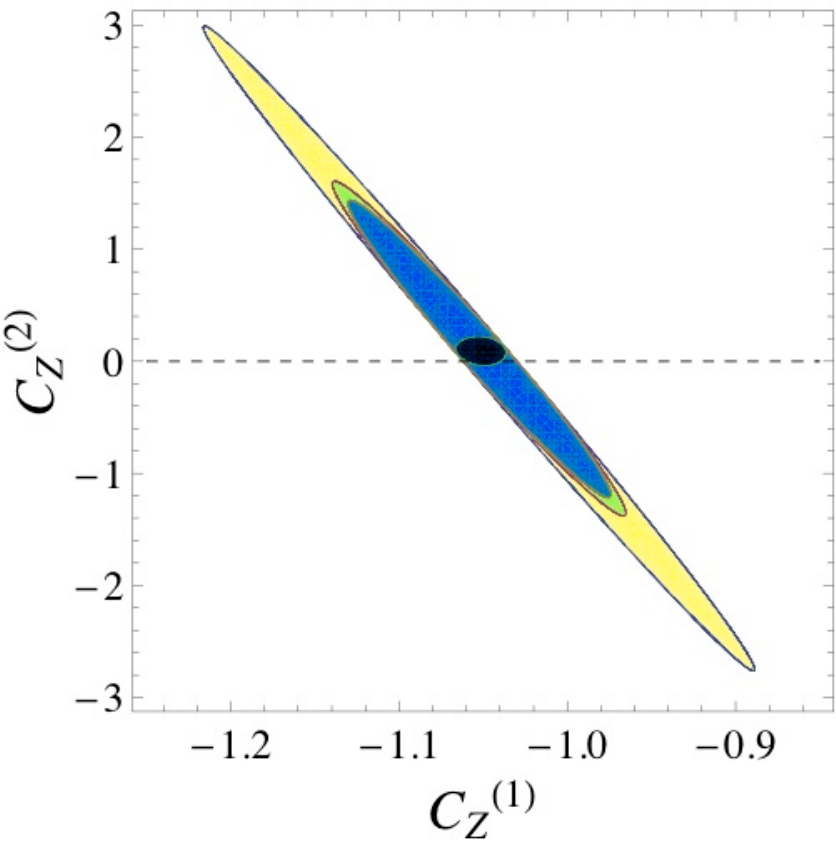}
}\resizebox{0.52\textwidth}{!}{
\includegraphics{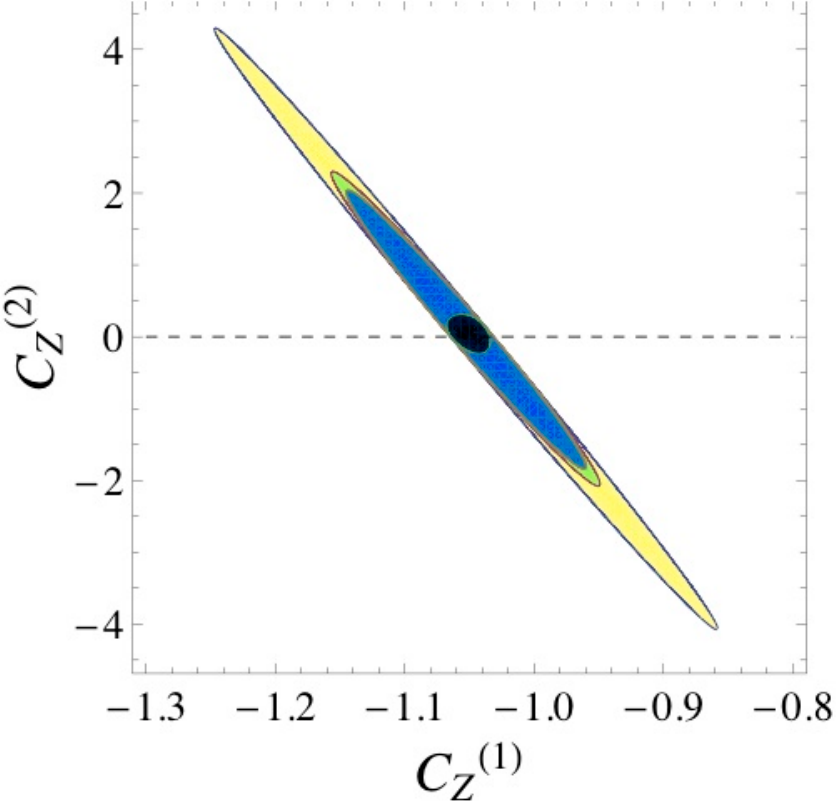}
}
\caption{95\% confidence level regions for $C^{(1)}_Z$ and $C^{(2)}_Z$ from ILC and CLIC cross-section measurements at various center-of-mass energies, assuming that $C^{(0)}_Z$ is already known.  In both plots, from largest to smallest, the ellipses correspond to cross-section measurements from both the ILC and CLIC, at center-of-mass energies of 0.5 and 3~TeV; 1 and 3~TeV; 0.5, 1, and 3~TeV; and 0.5, 1, 3, and 5~TeV.
The left panel is for our benchmark parameter set with $M_1 = 1.5$ for light fermions.  The right panel is for an alternate parameter set with $M_1 = 2.2$ for light fermions (this increases the $Z^{(1)}$ width to about 13\% of its mass) and the $Z^{(2)}$ width set equal to its mass.  The actual normalized products of couplings are $C_Z^{(1)}=-1.052$ and $C_Z^{(2)}=0.134$. }
\label{Z1Z2Mix}
\end{center}
\end{figure}

The resulting 95\% confidence regions for $C^{(1)}_Z$ and $C^{(2)}_Z$ are shown in the left panel of Fig.~\ref{Z1Z2Mix}.  The three largest ellipses show the constraint from cross-section measurements at the ILC (from largest to smallest, at 0.5~TeV, 1~TeV, and including both measurements) combined with a measurement at the 3~TeV CLIC.  Note that, with only a 3~TeV measurement from CLIC, adding the new $C^{(2)}_Z$ parameter to the fit approximately doubles the uncertainty on the extracted value of $C^{(1)}_Z$ by introducing a strong correlation between the two parameters.  Adding a 5~TeV measurement from CLIC (innermost ellipse in the left panel of Fig.~\ref{Z1Z2Mix}) lifts the degeneracy by providing a cross-section measurement above the $Z^{(1)}$ resonance but below the $Z^{(2)}$ resonance.  However, even at this highest center-of-mass energy, $C^{(2)}$ is consistent with zero at the 95\% confidence level; at best, we can determine that its favored value is positive and smaller in magnitude than the first KK mode by at least a factor of four, i.e., $|C_Z^{(2)}| \lesssim 0.25 |C_Z^{(1)}|$ at 95\% confidence level.

We finally recall that the decay width of $Z^{(2)}$ that we use in the cross-section computation does not include decays to KK fermions.  To determine how larger decay widths affect the coupling extraction, in the right panel of Fig.~\ref{Z1Z2Mix} we set the width of $Z^{(2)}$ equal to its mass and also take $M_1 = 2.2$ for light fermions (this increases the $Z^{(1)}$ width to about 13\% of its mass).  Including ILC and CLIC measurements at all four center-of-mass energies, we retain the upper bound $|C_Z^{(2)}| \lesssim 0.25 |C_Z^{(1)}|$ at 95\% confidence level, but we lose the preference for a positive value of $C_Z^{(2)}$.

\section{Conclusions \label{chapConc}}

Our goal in this paper has been to explore the manifestation of Higgs compositeness in a process that can in principle be probed experimentally.  To this end, we computed the cross section for $e^+e^- \to ZH$ in the Minimal Composite Higgs Model.  We examined the dependence of the cross section on the collision energy for energies spanning the masses of the first several gauge KK excitations.  We observed a dramatic suppression of the cross section compared to that in the SM, starting at collision energies about 1.5 times the mass of the first $Z$-boson KK excitation.  The immediate cause of this suppression is a progressive cancellation among the $Z$-boson KK-mode contributions to the amplitude.  

We interpret the suppression as implementing an effective form factor for a process in which the composite Higgs interacts with an $s$-channel probe with wavelength shorter than the compositeness scale.  This behavior is reminiscent of the mechanism by which the MCHM solves the hierarchy problem; radiative contributions to the Higgs mass parameter from gauge boson loops are finite due to an exponential suppression of the integrand in the loop momentum integral.  This suppression arises in a simple way from the form of the 5D gauge boson propagator in the warped extra-dimensional theory.  An analogous understanding of the suppression of the Higgsstrahlung cross section is hindered by the presence of the KK-mode resonance poles on the real $q^2$ axis in the tree-level gauge propagator.  Inclusion of the gauge KK-mode widths shifts the poles off the real axis and allows a realistic computation of the cross section as a function of $q^2$; however, we have implemented this only in the 4D picture.  We expect that a deeper understanding of the cross-section suppression could be obtained from the one-loop 5D gauge propagator, in which the absorptive parts of the gauge boson self-energy (corresponding to the KK-mode decay widths) will shift the poles away from the real axis.  We leave the computation of the one-loop 5D propagator to future work.

We also examined the prospects for extracting the relevant products of couplings of the first two $Z$-boson KK modes at the ILC and CLIC.  We found that ILC measurements at 0.5 and 1~TeV are sufficient to detect the influence of the first $Z$ KK excitation, and to determine that the relevant product of its couplings is opposite in sign and approximately equal in magnitude to that of the zero-mode $Z$ boson.  Adding CLIC measurements at 3 and 5~TeV (the latter energy being above the mass of the first $Z$-boson KK excitation) would show that the first $Z$ KK mode coupling is in fact larger in magnitude than that of the zero-mode $Z$ boson; however, it would also put only an upper bound on the magnitude of the coupling of the second $Z$ KK mode consistent with the progressive cancellation mechanism.

\section*{Acknowledgments}

We thank Y.~Bai and B.~Coleppa for helpful conversations. This work
was supported by the Natural Sciences and Engineering Research Council
of Canada.  K.H.\ was also supported by the Government of Ontario
Graduate Scholarship program.

\appendix

\section{Gauge generators \label{generators}}

In the {\bf 5}-representation, the generators of SO(4) $\approx$ SU(2)$_L$$ \times$ SU(2)$_R$ are
\begin{equation}
	T^{a_{L,R}}_{ij}=-\frac{i}{2}
	\left[\frac{1}{2}\epsilon^{abc}\left(\delta_i^b\,\delta_j^c - \delta_j^b \,\delta_i^c\right) 
	\pm \left(\delta_i^a\,\delta_j^4-\delta_j^a\,\delta_i^4\right)\right],
\end{equation}
while the generators for SO(5)/SO(4) are 
\begin{equation}
	T^{\hat{a}}_{ij}=-\frac{1}{\sqrt{2}}
	\left(\delta_i^{\hat{a}}\delta_j^5-\delta_j^{\hat{a}}\delta_5^i\right),
\end{equation}
where $a_{L,R}=1,2,3$, $\hat{a}=1...4$ and $i,j=1...5$~\cite{MCHM}. 
Explicitly, the generators are given by
\begin{eqnarray}
T^{1_{L,R}} &=& -\frac{i}{2}\left(
\begin{array}{ c c c c c }
0 & 0 & 0 & \pm 1 & 0 \\
0 & 0 & 1 & 0 & 0 \\
0 & -1 & 0 & 0 & 0 \\
\mp 1 & 0 & 0 & 0 & 0 \\
0 & 0 & 0 & 0 & 0 \\
\end{array}
\right),
\quad\quad 
T^{2_{L,R}}=-\frac{i}{2}\left(
\begin{array}{ c c c c c }
0 & 0 & -1 & 0 & 0 \\
0 & 0 & 0 & \pm 1 & 0 \\
1 & 0 & 0 & 0 & 0 \\
0 & \mp 1 & 0 & 0 & 0 \\
0 & 0 & 0 & 0 & 0 \\
\end{array}
\right),
\nonumber \\
T^{3_{L,R}} &=& -\frac{i}{2}\left(
\begin{array}{ c c c c c }
0 & 1 & 0 & 0 & 0 \\
-1 & 0 & 0 & 0 & 0 \\
0 & 0 & 0 & \pm 1 & 0 \\
0 & 0 & \mp 1 & 0 & 0 \\
0 & 0 & 0 & 0 & 0 \\
\end{array}
\right),
\quad\quad
T^{\hat{1}}=-\frac{i}{\sqrt{2}}\left(
\begin{array}{ c c c c c }
0 & 0 & 0 & 0 & 1 \\
0 & 0 & 0 & 0 & 0 \\
0 & 0 & 0 & 0 & 0 \\
0 & 0 & 0 & 0 & 0 \\
-1 & 0 & 0 & 0 & 0 \\
\end{array}
\right),
\\
T^{\hat{2}} &=& -\frac{i}{\sqrt{2}}\left(
\begin{array}{ c c c c c }
0 & 0 & 0 & 0 & 0 \\
0 & 0 & 0 & 0 & 1 \\
0 & 0 & 0 & 0 & 0 \\
0 & 0 & 0 & 0 & 0 \\
0 & -1 & 0 & 0 & 0 \\
\end{array}
\right),
\quad
T^{\hat{3}}=-\frac{i}{\sqrt{2}}\left(
\begin{array}{ c c c c c }
0 & 0 & 0 & 0 & 0 \\
0 & 0 & 0 & 0 & 0 \\
0 & 0 & 0 & 0 & 1 \\
0 & 0 & 0 & 0 & 0 \\
0 & 0 & -1 & 0 & 0 \\
\end{array}
\right),
\quad
T^{\hat{4}}=-\frac{i}{\sqrt{2}}\left(
\begin{array}{ c c c c c }
0 & 0 & 0 & 0 & 0 \\
0 & 0 & 0 & 0 & 0 \\
0 & 0 & 0 & 0 & 0 \\
0 & 0 & 0 & 0 & 1 \\
0 & 0 & 0 & -1 & 0 \\
\end{array}
\right). \nonumber
\end{eqnarray}

In the {\bf 10}-representation, or adjoint representation, $T^{\hat{4}}$, $T^{\hat{3}}$ and $T^{3_{L,R}}$ are given by
\begin{eqnarray}
T_{10}^{\hat{4}} &=& \frac{i}{2} \left(
\begin{array}{cccccccccc}
   0  &  0  &  0  &  0  &  1  &  0  &  0  &  -1  &  0  &  0   \\
   0  &  0  &  0  &  0  &  0  &  1  &  0  &  0  &  -1  &  0   \\
   0  &  0  &  0  &  0  &  0  &  0  &  1  &  0  &  0  &  -1   \\
   0  &  0  &  0  &  0  &  0  &  0  &  0  &  0  &  0  &  0   \\
   -1  &  0  &  0  &  0  &  0  &  0  &  0  &  0  &  0  &  0   \\
   0  &  -1  &  0  &  0  &  0  &  0  &  0  &  0  &  0  &  0   \\
   0  &  0  &  -1  &  0  &  0  &  0  &  0  &  0  &  0  &  0   \\
   1  &  0  &  0  &  0  &  0  &  0  &  0  &  0  &  0  &  0   \\
   0  &  1  &  0  &  0  &  0  &  0  &  0  &  0  &  0  &  0   \\
   0  &  0  &  1  &  0  &  0  &  0  &  0  &  0  &  0  &  0   \\
\end{array}
\right),
\quad
T_{10}^{\hat{3}} = \frac{i}{2} \left(
\begin{array}{ c c c c c c c c c c }
  0  &  0  &  0  &  0  &  0  &  1  &  0  &  0  &  1  &  0   \\
  0  &  0  &  0  &  0  &  -1  &  0  &  0  &  -1  &  0  &  0   \\
  0  &  0  &  0  &  0  &  0  &  0  &  0  &  0  &  0  &  0   \\
  0  &  0  &  0  &  0  &  0  &  0  &  1  &  0  &  0  &  -1   \\
  0  &  1  &  0  &  0  &  0  &  0  &  0  &  0  &  0  &  0   \\
  -1  &  0  &  0  &  0  &  0  &  0  &  0  &  0  &  0  &  0   \\
  0  &  0  &  0  &  -1  &  0  &  0  &  0  &  0  &  0  &  0   \\
  0  &  1  &  0  &  0  &  0  &  0  &  0  &  0  &  0  &  0   \\
  -1  &  0  &  0  &  0  &  0  &  0  &  0  &  0  &  0  &  0   \\
  0  &  0  &  0  &  1  &  0  &  0  &  0  &  0  &  0  &  0   \\
\end{array}
\right),
\nonumber \\
T_{10}^{3_L} &=& \frac{i}{2}\left(
\begin{array}{ c c c c c c c c c c }
  0  &  -1  &  0  &  0  &  0  &  0  &  0  &  0  &  0  &  0  \\
  1  &  0  &  0  &  0  &  0  &  0  &  0  &  0  &  0  &  0  \\
  0  &  0  &  0  &  -1  &  0  &  0  &  0  &  0  &  0  &  0  \\
  0  &  0  &  1  &  0  &  0  &  0  &  0  &  0  &  0  &  0  \\
  0  &  0  &  0  &  0  &  0  &  -2  &  0  &  0  &  0  &  0  \\
  0  &  0  &  0  &  0  &  2  &  0  &  0  &  0  &  0  &  0  \\
  0  &  0  &  0  &  0  &  0  &  0  &  0  &  0  &  0  &  0  \\
  0  &  0  &  0  &  0  &  0  &  0  &  0  &  0  &  0  &  0  \\
  0  &  0  &  0  &  0  &  0  &  0  &  0  &  0  &  0  &  0  \\
  0  &  0  &  0  &  0  &  0  &  0  &  0  &  0  &  0  &  0  \\
\end{array}
\right),
\quad
T_{10}^{3_R} = \frac{i}{2}\left(
\begin{array}{ c c c c c c c c c c }
  0  &  -1  &  0  &  0  &  0  &  0  &  0  &  0  &  0  &  0  \\
  1  &  0  &  0  &  0  &  0  &  0  &  0  &  0  &  0  &  0  \\
  0  &  0  &  0  &  1  &  0  &  0  &  0  &  0  &  0  &  0  \\
  0  &  0  &  -1  &  0  &  0  &  0  &  0  &  0  &  0  &  0  \\
  0  &  0  &  0  &  0  &  0  &  0  &  0  &  0  &  0  &  0  \\
  0  &  0  &  0  &  0  &  0  &  0  &  0  &  0  &  0  &  0  \\
  0  &  0  &  0  &  0  &  0  &  0  &  0  &  0  &  0  &  0  \\
  0  &  0  &  0  &  0  &  0  &  0  &  0  &  0  &  -2  &  0  \\
  0  &  0  &  0  &  0  &  0  &  0  &  0  &  2  &  0  &  0  \\
  0  &  0  &  0  &  0  &  0  &  0  &  0  &  0  &  0  &  0  \\
\end{array}
\right).
\end{eqnarray}

The generators obey the following commutation relations:
\begin{eqnarray}
	[T^{a_{L,R}},T^{b_{L,R}}] &=&  i \epsilon^{abc} T^{c_{L,R}},  \nonumber \\ 
	{[T^{a_L},T^{b_R}]} &=&  0, \nonumber \\
	{[T^{\hat{a}},T^{\hat{b}}]} &=& \frac{1}{2}\left[\delta_i^{\hat{a}} \delta_j^{\hat{b}} 
	- \delta_j^{\hat{a}} \delta_i^{\hat{b}}\right] 
	= \frac{i}{2}\left[ \epsilon^{abc}\left(T^{c_L}+T^{c_R}\right)
	+ \delta_b^4\left(T^{a_L}-T^{a_R}\right)\right], \nonumber \\
	{[T^{a_{L,R}},T^{\hat{b}}]} &=&  \frac{i}{2}
	\left(\epsilon^{abc}T^{\hat{c}} \mp T^{\hat{a}} \delta_{b}^4 \pm T^{\hat{4}} \delta_{b}^a\right), 
\end{eqnarray}
where $\epsilon^{abc}$ is the totally antisymmetric tensor with $\epsilon^{123}=1$.

Using these generators, our defined quantities $A^{a}_{MN} T^{a} = - i [T^c,T^{b}] A^c_M A^{b}_N$ become, where $abc$ is an even permutation of $123$,
\begin{eqnarray}
	A_{MN}^{a_L} &=& A_M^{b_L}A_N^{c_L} -A_M^{c_L}A_N^{b_L}
	+\frac{1}{2}\left(A_M^{\hat{b}}A_N^{\hat{c}} -A_M^{\hat{c}}A_N^{\hat{b}} 
	+A_M^{\hat{a}}A_N^{\hat{4}} -A_M^{\hat{4}}A_N^{\hat{a}}\right), 
	\label{AaL} \\
	A_{MN}^{a_R} &=& A_M^{b_R}A_N^{c_R} -A_M^{c_R}A_N^{b_R}
	+\frac{1}{2}\left(A_M^{\hat{b}}A_N^{\hat{c}} -A_M^{\hat{c}}A_N^{\hat{b}} 
	-A_M^{\hat{a}}A_N^{\hat{4}} +A_M^{\hat{4}}A_N^{\hat{a}}\right), 
	\label{AaR} \\
	A^{\hat{a}}_{MN} &=& \frac{1}{2}\left[\left(A_M^{b_L}+A_M^{b_R}\right)A_N ^{\hat{c}}
	-A_M ^{\hat{c}}\left(A_N^{b_L}+A_N^{b_R}\right)
	-\left(A_M^{c_L}+A_M^{c_R}\right)A_N ^{\hat{b}} \right.\nonumber \\ 
	& & \quad + \left. A_M ^{\hat{b}}\left(A_N^{c_L}+A_N^{c_R}\right)
	- \left(A_M^{a_L}-A_M^{a_R}\right)A_N ^{\hat{4}}  
	+ A_M ^{\hat{4}} \left(A_N^{a_L}-A_N^{a_R}\right)\right].
	\label{Ahata}
\end{eqnarray}
$A^{\hat{4}}_{MN}$ is given by,
\begin{equation}
	A^{\hat{4}}_{MN} = \frac{1}{2}\sum_{a=1}^3 
	\left[\left(A_M^{a_L}-A_M^{a_R}\right)A_N ^{\hat{a}}
	-A_M ^{\hat{a}}\left(A_N^{a_L}-A_N^{a_R}\right)\right].
\label{A4hat}
\end{equation}

 \section{Bessel functions \label{AppA}}

Bessel functions of the first and second kind, $J_n(x)$ and $Y_n(x)$, respectively, are solutions to Bessel's differential equation~\cite{Bessel,Wolfram,Maple},
\begin{equation}
	x^2\frac{d^2y}{dx^2}+x\frac{dy}{dx}+(x^2-n^2)y=0,
	\label{bessel}
\end{equation}
where $n$ is a non-negative real number. The two solutions are independent only if $n$ is not an integer; otherwise they are related through
\begin{equation}
	Y_n(z)=\frac{J_n(z)\cos(n\pi)-J_{-n}(z)}{\sin(n\pi)}.
\end{equation}

The Bessel functions obey the orthogonality relation,
\begin{equation}
	\int_0^a J_\nu\left(\alpha_n \frac{\rho}{a}\right)
	J_\nu\left(\alpha_m \frac{\rho}{a}\right)\rho \, d\rho=\frac{1}{2} \,a^2
	[J_{\nu+1}({\alpha_m})]^2 \, \delta_{mn}\,,
\end{equation}
where $\alpha_m$ is the $m$th zero of $J_\nu$, and similarly for $Y_\nu$. Their asymptotic approximations are,
\begin{eqnarray}
	J_n(x) &\approx& \left\{
	{
	\renewcommand{\arraystretch}{1}
	\begin{array}{l l}
	\frac{1}{\Gamma(n+1)}\left(\frac{z}{2}\right)^n & {\rm for }\,x \ll 1\,  \\
	\sqrt{\frac{2}{\pi x}}\cos\left(x-\frac{n\pi}{2}-\frac{\pi}{4}\right) \ \ & {\rm for }\,x\gg \left|n^2-\frac{1}	{4}\right|\,\\
	\end{array}
	}
	\right. \\
	Y_n(x) &\approx& \left\{
	{
	\renewcommand{\arraystretch}{1}
	\begin{array}{l l}
	 \frac{2}{\pi}\left[\ln\left(\frac{1}{2}x\right)+\gamma\right] &{\rm for }\,m=0,\,x \ll 1\, \\
	 \frac{\Gamma(n)}{\pi}\left(\frac{2}{x}\right)^n &{\rm for }\,m\neq 0,\,x \ll 1\, \\
	\sqrt{\frac{2}{\pi x}}\sin\left(x-\frac{n\pi}{2}-\frac{\pi}{4}\right) \ \ & {\rm for }\,x \gg 1.  \\
	 \end{array}
	 }
	 \right.
\end{eqnarray}
They obey the recurrence relations
\begin{eqnarray}
	&&J_{n+1}(x) = \frac{2n}{x}J_n(x)-J_{n-1}(x)\,, \\
	&&\frac{d}{dx}J_n(x) = \frac{1}{2}\left[J_{n-1}(x)-J_{n+1}(x)\right]
	=J_{n-1}(x)-\frac{n}{x}J_n(x)
	=\frac{n}{x}J_n(x)-J_{n+1}(x)\,,\\
	&&\frac{d}{dx}[x^m\,J_m(x)] = x^m\,J_{m-1}(x)\, \,,
	\label{recurrence}
\end{eqnarray}
and similarly for $Y_n(x)$.

The modified Bessel functions of the first and second kind, $I_n(x)$ and $K_n(x)$, are solutions to Eq.~(\ref{bessel}) under the transformation $x\rightarrow i x$~\cite{Bessel,Wolfram,Maple}. The Bessel function and modified Bessel function of the first kind are related through 
\begin{equation}
	I_n(x) = i^{-n} \,J_n(i x).
	\label{ItoJ}
\end{equation}
If $n$ is an integer, the Bessel function and modified Bessel function of the second kind are related as follows:
\begin{eqnarray}
	K_n(x) &=& \frac{\pi}{2}\frac{I_{-n}(x)-I_n(x)}{\sin(n\pi)} 
	=\frac{\pi}{2}\frac{i^{n}J_{-n}(ix)-i^{-n}J_n(ix)}{\sin(n\pi)} \nonumber \\
	&=& -i^{n}\frac{\pi}{2}\frac{J_{n}(ix)\cos(n\pi)-J_{-n}(ix)}{\sin(n\pi)}
	=-i^{n}\frac{\pi}{2}\,Y_n(i x).
	\label{KtoY}
\end{eqnarray}

The modified Bessel functions obey the same recurrence relations as the original Bessel function, and have the following asymptotic approximations, 
\begin{eqnarray}
	I_n(x) &\approx& \left\{
	{
	\renewcommand{\arraystretch}{1}
	\begin{array}{l l}
	\frac{1}{n!}\left(\frac{x}{2}\right)^n & {\rm for }\,x\ll 1\,  \\
	\frac{\exp(x)}{\sqrt{2\pi x}}& {\rm for }\,x\gg 1\,\\
	\end{array}
	}
	\right. \label{asympI}\\
	K_n(x) &\approx& \left\{
	{
	\renewcommand{\arraystretch}{1}
	\begin{array}{l l}
	 \sqrt{\frac{\pi}{2\,x}}\exp(-x) &{\rm for }\,\,x\gg 1\, \\
	 -\ln\left(\frac{x}{2}\right)-\gamma &{\rm for }\,\,x\ll 1, n=0\, \\
	 \frac{\Gamma(n)}{2}\left(\frac{2}{x}\right)^n&{\rm for }\,\,x \ll 1, n\neq 0\,, \\
	 \end{array}
	 }
	 \right. 
	 \label{asympK}
\end{eqnarray}
where $\gamma \simeq 0.5772$ is the Euler-Mascheroni constant.

\section{Profiles after EWSB-induced mixing\label{MixProfiles}}

Here we provide details of the mixing of the gauge boson and fermion modes after EWSB. Explicit calculations and summaries can also be found in Refs.~\cite{MSW,CMSW,Neutrinos,Lijun}.  Note however that these sources define the rotation angle $\theta_H$ differently from us; furthermore, in some cases their explicit solutions for the profile coefficients take a different form from ours due to differences in the solution method.  We have checked that our results are nevertheless equivalent to theirs.

To avoid notational clutter, we will omit the superscript $n$ denoting the KK mode number on all profiles $f^{(n)}$ and coefficients $C_G^{(n)}$ throughout this section; the KK mode number dependence will be expressed solely through the subscript on the mass parameters $m_n$.

\subsection{Gauge bosons}

The mixed particle profiles $f(m_n,z;v)$, where $v$ is the nonzero Higgs vev after EWSB, can be obtained from the pre-EWSB basis profiles through the gauge transformation in Eq.~(\ref{gaugetrans}). The base profiles $f(m_n,z;0)$ of the bosons are defined by 
\begin{eqnarray}
	f_{a_L}(m_n,z;0) &=& C_{a_L} C_A(m_n,z), \qquad\qquad 
	f_{B}(m_n,z;0) = C_{B} C_A(m_n,z), \nonumber \\
	f_{1_R}(m_n,z;0) &=& C_{1_R} S_A(m_n,z), \qquad\qquad 
	f_{2_R}(m_n,z;0) = C_{2_R} S_A(m_n,z), \nonumber \\
	f_{X}(m_n,z;0) &=& C_{X} S_A(m_n,z), \qquad\qquad 
	f_{\hat{a}}(m_n,z;0) = C_{\hat{a}} S_A(m_n,z),  
\end{eqnarray}
where $C_A(m_n,z)$ and $S_A(m_n,z)$ are the basis functions defined in Eqs.~(\ref{Cgaugeb}-\ref{Sgaugeb}), and the $C_G$ are normalization coefficients~\cite{MSW}. Note that it is $f_X(m_n,z;0)$ and $f_B(m_n,z;0)$ that are defined in terms of the basis functions, rather than $f_{3_R}(m_n,z;0)$ and $f_{U}(m_n,z;0)$. We will define $f_{3_R}(m_n,z;0)$ and $f_{U}(m_n,z;0)$ in terms of the $X$ and $B$ profiles below.

In the {\bf 5}-representation of the SO(5) generators (see Appendix~\ref{generators}), the transformation matrix $\Omega(z,v)$ of Eq.~(\ref{Omega}) can be rewritten as
\begin{equation}
	\Omega(z,v) = {\bf 1} - i \sqrt{2} \, T^{\hat{4}}\sin\theta_G
	+ {\bf 1}_{45} \left(\cos\theta_G-1\right),
\end{equation}
where ${\bf 1}$ is the unit matrix, $\theta_G \equiv \theta_G(z,v)$ is defined in Eq.~(\ref{Theta}), and ${\bf 1}_{45} \equiv {\rm diag}(0,0,0,1,1)$.  The gauge boson profiles after EWSB are determined by the equations,
\begin{eqnarray}
	f_{1_{L,R}}(v) &=& \frac{1}{2}\left(1\pm\cos\theta_G\right)C_{1_{L}}C_A
	+ \frac{1}{2}\left(1\mp\cos\theta_G\right)C_{1_{R}}S_A
	\pm \frac{\sqrt{2}}{2}\sin\theta_G\, C_{\hat{1}} S_A, 
	\nonumber \\
	f_{2_{L,R}}(v) &=& \frac{1}{2}\left(1\pm\cos\theta_G\right)C_{2_{L}}C_A
	+ \frac{1}{2}\left(1\mp\cos\theta_G\right)C_{2_{R}}S_A
	\pm \frac{\sqrt{2}}{2}\sin\theta_G\, C_{\hat{2}} S_A,  \nonumber \\
	f_{3_{L,R}}(v) &=& \frac{1}{2}\left(1\pm\cos\theta_G\right)C_{3_{L}}C_A 
	\pm\frac{\sqrt{2}}{2}\sin\theta_G\, C_{\hat{3}} S_A \nonumber \\ 
	&& +\frac{1}{2}\left(1\mp\cos\theta_G\right)
	\left[\cos\theta_H\,C_B C_A-\sin\theta_H\,C_{X} S_A\right], \nonumber \\
	f_{\hat{1}}(v) &=& \cos\theta_G \, C_{\hat{1}}S_A 
	+ \frac{1}{\sqrt{2}}\,\sin\theta_G\,\left(C_{1_R} S_A-C_{1_L} C_A\right), \nonumber \\
	f_{\hat{2}}(v) &=& \cos\theta_G \, C_{\hat{2}}S_A 
	+ \frac{1}{\sqrt{2}}\,\sin\theta_G\,\left(C_{2_R} S_A-C_{2_L} C_A\right), \nonumber \\
	f_{\hat{3}}(v) &=& \cos\theta_G \, C_{\hat{3}}S_A 
	+ \frac{1}{\sqrt{2}} \sin\theta_G
	\left(\cos\theta_H\,C_B C_A-\sin\theta_H\,C_{X} S_A-C_{3_L} C_A\right), \nonumber \\
	f_{\hat{4}}(v) &=& C_{\hat{4}} S_A, \nonumber \\
	f_U(v) &=& \sin\theta_H\,C_B C_A+\cos\theta_H\,C_{X} S_A,
	\label{fv}
\end{eqnarray}
where we suppress the $m_n$ and $z$ dependence of $f_G(v) \equiv f_G(m_n,z,v)$, $C_A \equiv C_A(m_n,z)$, and $S_A \equiv S_A(m_n,z)$ for compactness~\cite{Lijun}. Recall that $\cos\theta_H = \tan\theta_W$.

As discussed in Sec.~\ref{Mixing}, the TeV boundary conditions provide a system of equations by which the coefficients $C_G$ and mass eigenvalues may be determined. The boson masses are determined by setting the determinant of this system to zero, a requirement for the system to be solvable. The system of equations can be broken into subgroups corresponding to the charged and neutral gauge bosons, which mix independently.

\subsubsection{Charged gauge bosons}

The two sets $W^{1_L}$, $W^{1_R}$ and $A^{\hat{1}}$, and $W^{2_L}$, $W^{2_R}$ and $A^{\hat{2}}$, of electrically charged gauge bosons mix independently but with mathematically identical forms. We will represent these two systems by $W^{i_L}$, $W^{i_R}$ and $A^{\hat{i}}$ ($i=1,2$).  The TeV-brane boundary conditions are,
\begin{eqnarray}
	\left.\partial_z f_{i_{L}}^{(n)}(m_n,z;v)\right|_{z=L_1} &=& 0, \nonumber \\
	\left.\partial_z f_{i_{R}}^{(n)}(m_n,z;v)\right|_{z=L_1} &=& 0, \nonumber \\
	\left. f_{\hat{i}}^{(n)}(m_n,z;v)\right|_{z=L_1} &=& 0,
\end{eqnarray}
which we will use to solve for the coefficients $C_{i_L}$, $C_{i_R}$ and $C_{\hat{i}}$ that determine the composition of each mass eigenstate~\cite{Lijun}. Setting the determinant of this system to zero, we obtain two conditions, one of which must be satisfied for there to be a mass eigenstate:
\begin{eqnarray}
	&&S^{\prime}_A(m_n,L_1) = 0, \label{WRmass} \ {\rm or} \\
	&& 2 C^{\prime}_A(m_n,L_1)\,S_A(m_n,L_1)+ m_n k L_1 \sin^2\theta_G(L_1, v) = 0,
	\label{WLmass} 
\end{eqnarray}
where a prime represents the derivative with respect to the fifth coordinate $z$ and we have used the Wronskian relation~\cite{MSW,Wolfram} (also known as Abel's identity),
\begin{equation}
	S^{\prime}_A(m_n, z)\,C_A(m_n, z) - S_A(m_n, z)\,C^{\prime}_A(m_n, z) = m_n k z.
\end{equation}
The solutions $m_n$ of Eq.~(\ref{WRmass}) correspond to the masses of the $W^{i_R}$ KK modes, and are the same as the original $W^{i_R}$ masses before EWSB.  
The solutions $m_n$ of Eq.~(\ref{WLmass}) correspond to the masses of both the $W^{i_L}$ and $W^{\hat{i}}$ KK modes---compared to the corresponding masses before EWSB, these mass eigenvalues are shifted by a small amount.  Note that for $v\rightarrow 0$, the left-hand side of Eq.~(\ref{WLmass}) simplifies to the product of the left-hand sides of the original mass conditions for $W^{i_L}$ and $W^{\hat i}$.

These two sets of solutions for the mass eigenvalues correspond to two sets of solutions for the coefficients $C_G$.  The solution corresponding to Eq.~(\ref{WLmass}), which yields gauge KK modes that are mostly $W^{i_L}$ or $W^{\hat i}$, is given by (the superscript $W_L$ identifies this coefficient set)
\begin{eqnarray}
	C_{i_L}^{W_L} &=& \sqrt{2}\,C_{\hat{i}} 
	\frac{S_A(m_n,L_1) \left(1+\cos^2\theta_G(L_1,v)\right)}
	{C_A(m_n,L_1)\sin 2\theta_G(L_1,v)}, \nonumber\\
 	C^{W_L}_{i_R} &=& \frac{\sqrt{2}}{2} C_{\hat{i}} \tan\theta_G(L_1,v), \qquad 
	C^{W_L}_{\hat{i}} = C_{\hat{i}},
	\label{Wcoef1}
\end{eqnarray}
where the remaining coefficient $C_{\hat{i}}$ is fixed by the normalization condition with $\alpha = W_L$,
\begin{equation}
	\int_{L_0}^{L_1} \frac{dz}{kz}\left(\left[ f^\alpha_{i_L}(m,z;v) \right]^2
	+ \left[ f^\alpha_{i_R}(m,z;v) \right]^2 + \left[ f^\alpha_{\hat{i}}(m,z;v) \right]^2 \right) = 1.
	\label{WLnorm}
\end{equation}
When the coefficients given in Eq.~(\ref{Wcoef1}) are substituted into the profiles $f_{i_L}(m,z;v)$, $f_{i_R}(m,z;v)$ and $f_{\hat{i}}(m,z;v)$ of Eq.~(\ref{fv}), all of these profiles are non-zero. This indicates that the full $W^{i_L}$ mass eigenstate profile is a superposition of these mixed profiles, weighted by their associated generators,
\begin{equation}
	F_{i_L}(m,z;v)\,T^{F_{iL}} = f^{W_L}_{i_L}(m,z;v) T^{i_L} + f^{W_L}_{i_R}(m,z;v) T^{i_R} 
	+ f^{W_L}_{\hat{i}}(m,z;v) T^{\hat{i}},
	\label{WLfull}
\end{equation}
where the superscript $W_L$ on the profiles indicates that coefficients $C_{i_L}$, $C_{i_L}$, and $C_{\hat{i}}$ have been replaced by those in Eq.~(\ref{Wcoef1}). Here $T^{F_{iL}}$ is meant to denote the generator that would be associated with the profile $F_{iL}(m,z;v)$; however, in practice the two cannot be separated because the combination of generators on the right-hand side varies with $z$.  

The solution corresponding to Eq.~(\ref{WRmass}), which yields gauge KK modes that are mostly $W^{i_R}$, is given by 
\begin{equation}
	C^{W_R}_{i_L} =0, \qquad
	C^{W_R}_{i_R} = -\sqrt{2}\,C_{\hat{i}} \tan\theta_G(L_1,v),  \qquad
	C^{W_R}_{\hat{i}}=C_{\hat{i}}.
\end{equation}
The corresponding full profile $F_{i_R}(m,z;v)\,T^{F_{iR}}$ and its normalization condition are given by replacing $W_L\rightarrow W_R$ in Eq.~(\ref{WLfull}) and (\ref{WLnorm}).

\subsubsection{Neutral gauge bosons}

There are five neutral gauge boson degrees of freedom: $W^{3_L}$, $B$, $X$, $A^{\hat{3}}$, and $A^{\hat 4}$.  Their TeV-brane boundary conditions are,
\begin{eqnarray}
	&&\left.\partial_z f_{3_{L}}^{(n)}(m_n,z;v)\right|_{z=L_1} = 0, \nonumber \\
	&&\left.\partial_z f_{B}^{(n)}(m_n,z;v)\right|_{z=L_1} 
	= \left.\partial_z \left[\cos\theta_W f_{3_R}^{(n)}(m_n,z;v) 
	+ \sin\theta_W f_{U}^{(n)}(m_n,z;v)\right]\right|_{z=L_1} = 0, \nonumber \\
	&&\left.\partial_z f_{X}^{(n)}(m_n,z;v)\right|_{z=L_1} 
	= \left.\partial_z \left[\cos\theta_W f_{U}^{(n)}(m_n,z;v)
	- \sin\theta_W f_{3_R}^{(n)}(m_n,z;v)\right]\right|_{z=L_1}=0, \nonumber \\
	&&\left. f_{\hat{3}}^{(n)}(m_n,z;v)\right|_{z=L_1} 
	= \left. f_{\hat{4}}^{(n)}(m_n,z;v)\right|_{z=L_1} = 0,
\end{eqnarray}
which we use to solve for the coefficients $C_{3_L}$, $C_{B}$, $C_{X}$, $C_{\hat{3}}$, and $C_{\hat 4}$~\cite{Lijun}. 

The $A^{\hat{4}}_\mu$ boson does not mix with any other bosons via EWSB; as such, its profile after EWSB is simply 
\begin{equation}
	F_{\hat{4}}(m,z;v)= f_{\hat{4}}(m,z;v)= f_{\hat{4}}(m,z;0)=C_{\hat{4}}\, S_A(m,z).
\end{equation}
The single coefficient $C_{\hat 4}$ is then determined by applying a normalization condition analogous to Eq.~(\ref{WLnorm}).

Setting the determinant of the remaining system of equations to zero, we obtain three conditions, one of which must be satisfied for there to be a mass eigenstate:
\begin{eqnarray}
	&&C^{\prime}_A(m_n,L_1)=0, \ {\rm or} \label{photonmass}\\
	&&S^{\prime}_A(m_n,L_1)=0, \ {\rm or} \label{Xmass}\\
	&&2 C^{\prime}_A(m_n,L_1)\,S_A(m_n,L_1) 
	+ m_n k L_1 \sec^2\theta_W\,\sin^2\theta_G(L_1, v)=0.
	\label{Zmass}
\end{eqnarray}
The solutions $m_n$ of Eqs.~(\ref{photonmass}) and (\ref{Xmass}) correspond to the masses of the photon and $X$ boson KK modes, respectively; both are the same as the original $V$ and $X$ masses before EWSB.  In particular, the zero-mode photon remains massless, as it should. The solutions $m_n$ of Eq.~(\ref{Zmass}) correspond to the masses of both the $Z$ and $A^{\hat{3}}$ KK modes---compared to the corresponding masses before EWSB, these mass eigenvalues are shifted by a small amount as illustrated in Fig.~\ref{MCHMmass}.  Again, note that for $v\rightarrow 0$, the left-hand side of Eq.~(\ref{Zmass}) simplifies to the product of the left-hand sides of the original mass conditions for $Z$ and $A^{\hat 3}$.

These three sets of solutions for the mass eigenvalues correspond to three sets of solutions for the coefficients $C_G$.  The solution corresponding to Eq.~(\ref{photonmass}), which yields the photon KK modes, is given by 
\begin{equation}
	C^{V}_{3_L} = C_{B} \tan\theta_W, \qquad
	C^{V}_{B} = C_{B}, \qquad
	C^{V}_{X} = C^{V}_{\hat{3}}= 0,
	\label{photoncoef}
\end{equation}
where as usual the superscript $V$ identifies this coefficient set.
Notice in particular that the photon KK modes remain the usual mixtures of $W^{3_L}$ and $B$ and are unaffected by EWSB; as a result, the photon mass eigenstate profiles are given by
\begin{equation}
	F_{V}(m,z;v) = \cos\theta_W f^V_{B}(m,z;v) + \sin\theta_W f^V_{3_L}(m,z;v).
\end{equation}

The solution corresponding to Eq.~(\ref{Xmass}), which yields gauge KK modes that are mostly $X$, is given by
\begin{equation}
	C^{X}_{3_L} = C^{X}_{B} = 0, \qquad 
	C^{X}_{X} = \sqrt{2} C_{\hat{3}} \sin\theta_H \tan\theta_G(L_1,v), \qquad 
	C^{X}_{\hat{3}} = C_{\hat{3}}.
	\label{Xcoef}
\end{equation}
When these coefficients are substituted into the profiles $f_{3_L}(m,z;v)$, $f_{X}(m,z;v)$, $f_{B}(m,z;v)$ and $f_{\hat{3}}(m,z;v)$, all four of the profiles are non-zero; however, the linear combination 
\begin{equation}
	f_{V}(m,z;v)=  \sin\theta_W f_{3_L}(m,z;v) + \cos\theta_W f_{B}(m,z;v),
\end{equation}
corresponding to the photon, vanishes.  This means that EWSB does not mix the photon KK modes into $X$.  The full $X$ mass eigenstate profile can then be written in the same form as Eq.~(\ref{WLfull}), weighted by the appropriate generators, as
\begin{equation}
	F_{X}(m,z;v)\,T^{F_X} 
	= f^X_{Z} T^{Z} + f^X_{X}(m,z;v) T^{X} + f^X_{\hat{3}}(m,z;v) T^{\hat{3}},
\end{equation}
where the superscript $X$ on the profiles indicates that the relevant coefficients have been replaced by those in Eq.~(\ref{Xcoef}) and we define
\begin{equation}
	f_{Z}(m,z;v)= \cos\theta_W f_{3_L}(m,z;v) - \sin\theta_W f_{B}(m,z;v),
\end{equation}
with $T^Z$ being the corresponding generator.

The solution corresponding to Eq.~(\ref{Zmass}), which yields gauge KK modes that are mostly $Z$ or $A^{\hat 3}$, is given by 
\begin{eqnarray}
	C^{Z}_{B} &=& C^{Z}_{3_L} \cos\theta_H 
	= C_{X}\,\frac{S_A(m_n,L_1)}{C_A(m_n,L_1)}
	\frac{\left[1-\cos^2\theta_H+\left(1+\cos^2\theta_H\right)\cos^2\theta_G(L_1,v)\right]}
	{\sin\theta_H \left(1+\cos^2\theta_H\right) \left[-1+\cos^2\theta_G(L_1,v)\right]}, \nonumber\\
	C^{Z}_{X} &=& C_{X}, \qquad
	C^{Z}_{\hat{3}} = \frac{\sqrt{2}\,C_{X}}{\sin\theta_H \tan\theta_G(L_1,v)}.
	\label{Zcoef}
\end{eqnarray}
When these coefficients are substituted into the profiles $f_{3_L}(m,z;v)$, $f_{X}(m,z;v)$, $f_{B}(m,z;v)$ and $f_{\hat{3}}(m,z;v)$, the linear combination corresponding to the photon again vanishes.  The $Z$ and $A^{\hat 3}$ mass eigenstate profiles, weighted by their associated generators, are then given by
\begin{equation}
	F_{Z}(m,z;v)\,T^{F_Z} 
	= f^Z_{Z}(m,z;v) T^{Z} + f^Z_{X}(m,z;v) T^{X} + f^Z_{\hat{3}}(m,z;v) T^{\hat{3}},
\end{equation}
where the superscript $Z$ on the profiles indicates that the relevant coefficients have been replaced by those in Eq.~(\ref{Zcoef}).  In all cases, the last overall coefficient in the profiles is determined by a normalization condition analogous to Eq.~(\ref{WLnorm}).

\subsection{Fermions}

The mixing of the fermions after EWSB is implemented analogously to that of the gauge bosons.  
The quarks and leptons are embedded into SO(5) multiplets in the same way; we thus use generic notation for both sectors.  This is valid in the absence of Majorana boundary masses for the neutrinos, which we are in any case ignoring.

The fermions of a single quark [Eq.~(\ref{quark123})] or lepton [Eq.~(\ref{lep123})] generation can be written in the spinorial representation of SO(5) according to, 
\begin{equation}
\begin{array}{l l}
\begin{array}{l}
\xi_{1_L}^{q_i} = \left(
{
\begin{array}{l}
\chi_{1_L}^i (-,+)_{\frac{5}{3}} \\
\tilde t_{1_L}^i (-,+)_{\frac{2}{3}} \\
t_{1_L}^i (+,+)_{\frac{2}{3}} \\
b_{1_L}^i (+,+)_{-\frac{1}{3}} \\
\hat t_{1_L}^i (-,+)_{\frac{2}{3}} \\
\end{array}
}
\right),
\\
\\
\xi_{2_R}^{q_i} = \left(
{
\begin{array}{l}
\chi_{2_R}^i (-,+)_{\frac{5}{3}} \\
\tilde t_{2_R}^i (-,+)_{\frac{2}{3}} \\
t_{2_R}^i (-,+)_{\frac{2}{3}} \\
b_{2_R}^i (-,+)_{-\frac{1}{3}} \\
\hat t_{2_R}^i (+,+)_{\frac{2}{3}} \\
\end{array}
}
\right),
\end{array}
&
\xi_{3_R}^{q_i} = \left(
{
\begin{array}{l}
\chi_{3_R}^i (-,+)_{\frac{5}{3}} \\
\tilde t_{3_R}^i (-,+)_{\frac{2}{3}} \\
t_{3_R}^i (-,+)_{\frac{2}{3}} \\
b_{3_R}^i (-,+)_{-\frac{1}{3}} \\
\Xi_{3_R}^i (-,+)_{\frac{5}{3}} \\
T_{3_R}^i (-,+)_{\frac{2}{3}} \\
B_{3_R}^i (-,+)_{-\frac{1}{3}} \\
\Xi_{3_R}^{\prime i} (-,+)_{\frac{5}{2}} \\
T_{3_R}^{\prime i} (-,+)_{\frac{2}{3}} \\
B_{3_R}^{\prime i} (+,+)_{-\frac{1}{3}} \\
\end{array}
}
\right),
\end{array}
\end{equation}
for the quarks of generation $i$, and analogously for the leptons.  
The corresponding 5D profiles are then given by,
\begin{equation}
\begin{array}{l l}
\begin{array}{l}
f_{1_L}(m_n,z;0)= \left(
{
\begin{array}{l}
C_{ 1}\,S_{c_1}^{+}(m_n,z)  \\
C_{ 2}\,S_{c_1}^{+}(m_n,z) \\
C_{ 3}\,\dot{S}_{c_1}^{-}(m_n,z) \\
C_{ 4}\,\dot{S}_{c_1}^{-}(m_n,z) \\
C_{ 5}\,S_{c_1}^{+}(m_n,z) \\
\end{array}
}
\right),
\\
\\
f_{ {2_R}}(m_n,z;0)= \left(
{
\begin{array}{l}
C_{ 6}\,S_{c_2}^{-}(m_n,z)  \\
C_{ 7}\,S_{c_2}^{-}(m_n,z) \\
C_{ 8}\,{S}_{c_2}^{-}(m_n,z) \\
C_{ 9}\,{S}_{c_2}^{-}(m_n,z) \\
C_{ 10}\,\dot{S}_{c_2}^{+}(m_n,z)  \\
\end{array}
}
\right),
\end{array}
&
f_{ {3_R}}(m_n,z;0)= \left(
{
\begin{array}{l}
C_{ {11}}\,S_{c_3}^{-}(m_n,z)  \\
C_{ {12}}\,S_{c_3}^{-}(m_n,z) \\
C_{ {13}}\,{S}_{c_3}^{-}(m_n,z) \\
C_{ {14}}\,{S}_{c_3}^{-}(m_n,z) \\
C_{ {15}}\,S_{c_3}^{-}(m_n,z) \\
C_{ {16}}\,S_{c_3}^{-}(m_n,z)  \\
C_{ {17}}\,S_{c_3}^{-}(m_n,z) \\
C_{ {18}}\,{S}_{c_3}^{-}(m_n,z) \\
C_{ {19}}\,{S}_{c_3}^{-}(m_n,z) \\
C_{ {20}}\,\dot{S}_{c_3}^{+}(m_n,z) \\
\end{array}
}
\right),
\end{array}
\label{BaseFermVec}
\end{equation}
where $S_{c}^{\pm}$ and $\dot{S}_{c}^{\pm}$ are the fermion basis functions defined in Eqs.~(\ref{Sfermion}-\ref{dotSFermion})~\cite{MSW,Neutrinos,Lijun}.
The fermions of opposite chirality can be read from these by replacing $S^\pm_c \leftrightarrow \dot{S}^\pm_c$.

The SO(5) generators for the spinorial representation are obtained from those given in Appendix~\ref{generators} through a change of basis.  The gauge transformation that ``turns on'' the Higgs vev then takes the form~\cite{MSW,CMSW},
\begin{eqnarray}
	f_{ {1_L}}(m_n,z;v) &=& A \Omega(z,v) A^{-1} f_{ {1_L}}(m_n,z;0), \nonumber \\
	f_{ {2_R}}(m_n,z;v) &=& A \Omega(z,v) A^{-1} f_{ {2_R}}(m_n,z;0), \nonumber \\
	f_{ {3_R}}(m_n,z;v) &=& B \Omega(z,v) B^{-1} f_{ {3_R}}(m_n,z;0),
\end{eqnarray}
where the $T^{\hat{4}}$ generator in $\Omega(z,v)$ is in the {\bf 5}-representation for the first two equations and the {\bf 10}-representation for the third, and the change-of-basis matrices $A$ and $B$ are given by~\cite{MSW,CMSW}
\begin{equation}
A=\frac{1}{\sqrt{2}}\left(
\begin{array}{ccccc}
-i & -1 & 0 & 0 & 0 \\
0 & 0 & -i & 1 & 0 \\
0 & 0 & i & 1 & 0 \\
-i & 1 & 0 & 0 & 0 \\
0 & 0 & 0 & 0 & \sqrt{2} \\
\end{array}
\right),  \qquad
B=\frac{1}{\sqrt{2}}\left(
\begin{array}{cccccccccc}
i & 1 & 0 & 0 & 0 & 0 & 0 & 0 & 0 & 0\\
0 & 0 & -i & 1 & 0 & 0 & 0 & 0 & 0 & 0 \\
0 & 0 & -i & -1 & 0 & 0 & 0 & 0 & 0 & 0 \\
-i & 1 & 0 & 0 & 0 & 0 & 0 & 0 & 0 & 0 \\
0 & 0 & 0 & 0 & -1 & i & 0 & 0 & 0 & 0 \\
0 & 0 & 0 & 0 & 0 & 0 & \sqrt{2} & 0 & 0 & 0 \\
0 & 0 & 0 & 0 & 1 & i & 0 & 0 & 0 & 0 \\
0 & 0 & 0 & 0 & 0 & 0 & 0 & -1 & i & 0 \\
0 & 0 & 0 & 0 & 0 & 0 & 0 & 0 & 0 & \sqrt{2} \\
0 & 0 & 0 & 0 & 0 & 0 & 0 & 1 & i & 0 \\
\end{array}
\right).
\label{AB}
\end{equation}

Performing this transformation, we obtain the following mixed profiles:
\begin{eqnarray}
f_{ {1_L}}(m_n,z;v) &=& \left(
{
\begin{array}{c}
C_{ 1}\,S_{c_1}^{+}
\\
\cos^2\!\left(\frac{\theta_G}{2}\right) C_{ 2}\, S_{c_1}^{+}
-\sin^2\!\left(\frac{\theta_G}{2}\right) C_{ 3}\,\dot{S}_{c_1}^{-}
-\frac{\sqrt{2}}{2}\,\sin\!\theta_G\, C_{ 5}\,S_{c_1}^{+}
\\
-\sin^2\!\left(\frac{\theta_G}{2}\right) C_{ 2} \,S_{c_1}^{+}
+\cos^2\!\left(\frac{\theta_G}{2}\right) C_{ 3}\,\dot{S}_{c_1}^{-}
-\frac{\sqrt{2}}{2}\,\sin\!\theta_G \,C_{ 5}\,S_{c_1}^{+}
\\
C_{ 4}\,\dot{S}_{c_1}^{-}
\\
\frac{\sqrt{2}}{2}\,\sin\!\theta_G \,C_{ 2}\, S_{c_1}^{+}
+\frac{\sqrt{2}}{2}\sin\!\theta_G \,C_{ 3}\, \dot{S}_{c_1}^{-}
+\cos\!\theta_G \,C_{ 5}\, S_{c_1}^{+}
\\
\end{array}
}
\right), \nonumber \\
f_{ {2_R}}(m_n,z; v) &=& \left(
{
\begin{array}{c}
C_{ 6}\,S_{c_2}^{-}
\\
\cos^2\!\left(\frac{\theta_G}{2}\right) C_{ 7} \,S_{c_2}^{-}
-\sin^2\!\left(\frac{\theta_G}{2}\right) C_{ 8}\, S_{c_2}^{-}
-\frac{\sqrt{2}}{2}\,\sin\!\theta_G\, C_{ 10}\,\dot{S}_{c_2}^{+}
\\
-\sin^2\!\left(\frac{\theta_G}{2}\right) C_{ 7} \,S_{c_2}^{-}
+\cos^2\!\left(\frac{\theta_G}{2}\right) C_{ 8}\, S_{c_2}^{-}
-\frac{\sqrt{2}}{2}\,\sin\!\theta_G \,C_{ 10}\,\dot{S}_{c_2}^{+}
\\
C_{ 9} \,S_{c_2}^{-}
\\
\frac{\sqrt{2}}{2}\,\sin\!\theta_G \,C_{ 7} \,S_{c_2}^{-}
+\frac{\sqrt{2}}{2}\,\sin\!\theta_G \,C_{ 8}\, S_{c_2}^{-}
+\cos\!\theta_G \,C_{ 10}\,\dot{S}_{c_2}^{+}
\\\end{array}
}
\right),  
\label{fermionmixing} \\
f_{ {3_R}}(m_n,z;v) &=& \left(
{
\begin{array}{c}
\cos\!\theta_G \, C_{ {11}}\, S_{c_3}^{-}  
+\frac{i\sqrt{2}}{2}\,\sin\!\theta_G \left(C_{ {18}}-C_{ {15}}\right) S_{c_3}^{-}  
\\ 
\cos^2\!\left(\frac{\theta_G}{2}\right) C_{ {12}} \,  S_{c_3}^{-} 
-\sin^2\!\left(\frac{\theta_G}{2}\right) C_{ {13}} \, S_{c_3}^{-} 
+\frac{i}{2}\,\sin\!\theta_G \left(C_{ {19}}-C_{ {16}}\right) S_{c_3}^{-} 
\\
\cos^2\!\left(\frac{\theta_G}{2}\right) C_{ {13}} \,S_{c_3}^{-} 
-\sin^2\!\left(\frac{\theta_G}{2}\right) C_{ {12}}\,  S_{c_3}^{-} 
+\frac{i}{2}\,\sin\!\theta_G \left(C_{ {19}}-C_{ {16}}\right) S_{c_3}^{-} 
\\
\cos\!\theta_G\, C_{ {14}}\,  S_{c_3}^{-} 
 -\frac{i\sqrt{2}}{2}\,\sin\!\theta_G \,C_{ {17}} \,S_{c_3}^{-}
+\frac{i\sqrt{2}}{2}\,\sin\!\theta_G \,C_{ {20}}\, \dot{S}_{c_3}^{+}
\\
-\frac{i\sqrt{2}}{2}\,\sin\!\theta_G \,C_{ {11}} \, S_{c_3}^{-} 
+\cos^2\!\left(\frac{\theta_G}{2}\right) C_{ {15}} \, S_{c_3}^{-} 
-\sin^2\!\left(\frac{\theta_G}{2}\right) C_{ {18}} \,S_{c_3}^{-} 
\\
\cos^2\!\left(\frac{\theta_G}{2}\right) C_{ {16}} \, S_{c_3}^{-} 
-\sin^2\!\left(\frac{\theta_G}{2}\right) C_{ {19}} \, S_{c_3}^{-} 
-\frac{i}{2}\,\sin\!\theta_G \left(C_{ {12}}+C_{ {13}}\right) S_{c_3}^{-} 
\\
-\frac{i\sqrt{2}}{2}\,\sin\!\theta_G \,C_{ {14}} \, S_{c_3}^{-} 
+\cos^2\!\left(\frac{\theta_G}{2}\right)C_{ {17}} \, S_{c_3}^{-} 
-\sin^2\!\left(\frac{\theta_G}{2}\right) C_{ {20}}\, \dot{S}_{c_3}^{+} 
\\
\frac{i\sqrt{2}}{2}\,\sin\!\theta_G \,C_{ {11}}\, S_{c_3}^{-} 
+\cos^2\!\left(\frac{\theta_G}{2}\right) C_{ {18}} \, S_{c_3}^{-} 
-\sin^2\!\left(\frac{\theta_G}{2}\right) C_{ {15}}\, S_{c_3}^{-} 
\\
\frac{i}{2}\,\sin\!\theta_G \left(C_{ 12}+C_{ {13}}\right)  S_{c_3}^{-} 
-\sin^2\!\left(\frac{\theta_G}{2}\right) C_{ {16}} \, S_{c_3}^{-} 
+\cos^2\!\left(\frac{\theta_G}{2}\right) C_{ {19}} \, S_{c_3}^{-} 
\\
\frac{i\sqrt{2}}{2}\,\sin\!\theta_G \,C_{ {14}}\,  S_{c_3}^{-} 
-\sin^2\!\left(\frac{\theta_G}{2}\right) C_{ {17}} \, S_{c_3}^{-} 
+\cos^2\!\left(\frac{\theta_G}{2}\right) C_{ {20}} \,\dot{S}_{c_3}^{+} 
\\
\end{array}
}
\right), \nonumber
\end{eqnarray}
where we have dropped the $(m_n,z)$ dependence of the basis functions $S^{\pm}$ and $\dot S^{\pm}$ for compactness.
In terms of this spinorial form, the fermion boundary conditions at the TeV brane ($z = L_1$) of Eq.~(\ref{FermBC}) can be rewritten as 
\begin{eqnarray}
	&&f_{ {1_L}}^{1,..,4}+M_2 f_{ {3_R}}^{1,..,4}=0, \qquad 
	f_{ {1_L}}^{5}+M_2 f_{ {2_R}}^{5}=0, \qquad 
	f_{ {2_L}}^{1,..,4}=0, \nonumber \\
	&&f_{ {3_L}}^{1,..,4}-M_2 f_{ {1_L}}^{1,..,4}=0, \qquad 
	f_{ {2_L}}^{5}-M_1 f_{ {1_L}}^{5}=0, \qquad 
	f_{ {3_L}}^{5,..,10}=0,
\end{eqnarray}
where the superscript denotes the appropriate row of the fermion multiplet~\cite{MSW,CMSW,Neutrinos}.  As usual, the fermions of opposite chirality ($f_{1_R}$, $f_{2_L}$, and $f_{3_L}$) can be read from Eq.~(\ref{fermionmixing}) by replacing $S^\pm_c \leftrightarrow \dot{S}^\pm_c$.

A single generation of quarks contains ten fields with charge $+2/3$, five with charge $-1/3$, and five with charge $+5/3$.  The mixing among these states after EWSB is summarized in Table~\ref{MixingGroupsQ}.  The fermion structure in the lepton sector is the same, but with charges $0$, $-1$, and $+1$, respectively.  This mixing has a very small effect on the zero-mode electron profiles that enter our calculation of $e^+e^- \to ZH$.  Its main effect on our calculation is instead to affect the total widths of the neutral KK gauge bosons exchanged in the $s$-channel through its effect on the fermion couplings to gauge bosons.  

\begin{table}
\begin{center}
\begin{tabular}{| >{\centering\arraybackslash}m{1.2in} || >{\centering\arraybackslash}m{1.2in}  | >{\centering\arraybackslash}m{1in}  | >{\centering\arraybackslash}m{1in} |}
\hline \hline
Charge & $2/3$ & $-1/3$ & $5/3$ \\
\hline
Mixed quarks & 
$\tilde{t}_1$, $t_1$, $\hat{t}_1$, $\tilde{t}_2$, $t_2$,  $\hat{t}_2$, $\tilde{t}_3$, $t_3$, $T_3$, $T'_3$ & 
$b_1$, $b_3$, $B_3$, $B'_3$ & 
$\chi_1$, $\chi_3$, $\Xi_3$, $\Xi'_3$
 \\
\hline
Unmixed quark & 
-- & 
$b_2$ & 
$\chi_2$ \\
\hline \hline
Mass conditions combined into single equation & 
$\left(\tilde{t}_1,\tilde{t}_3\right)$, $\left(t_1, \hat{t}_1, t_2,  \hat{t}_2, t_3\right)$ & 
$\left(b_1, b_3, B'_3\right)$ & 
$\left(\chi_1, \chi_3\right)$
 \\
\hline
Mass eigenvalues unaffected by EWSB & 
$\tilde{t}_2$, $T_3$, $T'_3$  & 
$b_2$, $B_3$ & 
$\chi_2$, $\Xi_3$, $\Xi'_3$
 \\
\hline \hline
\end{tabular}
\end{center}
\caption{Quark mixing induced by EWSB. Recall that, before EWSB, $t_{1_L}$, $b_{1_L}$, $\hat{t}_{2_R}$, and $B^{\prime}_{3_R}$ are the only quarks that have zero modes and correspond to the SM quarks of one generation. Lepton mixing is analogous. \label{MixingGroupsQ}}
\end{table}

The post-EWSB fermion mass eigenstates are determined by applying the appropriate TeV-brane boundary conditions on Eq.~(\ref{fermionmixing}) and solving the resulting system of equations for the coefficients $C_i$.  The mass eigenvalues are found by requiring that a solution for the coefficients exists, i.e., by setting the determinant of the system to zero.  We summarize the details below for completeness.

\subsubsection{Up-type fermions}

Each up-type fermion of the SM is accompanied by nine additional 5D fields, all of which mix after EWSB.  Imposing the TeV-brane boundary conditions on Eq.~(\ref{fermionmixing}) allows a solution for the coefficients $C_{ {2}}$, $C_{ {3}}$, $C_{ {5}}$, $C_{ {7}}$, $C_{ {8}}$, $C_{ {10}}$ $C_{ {12}}$, $C_{ {13}}$, $C_{ {16}}$, and $C_{ {19}}$.  Requiring that a solution exists results in the mass conditions, one of which must be satisfied for there to be a fermion mass eigenstate:
\begin{eqnarray}
	&& \dot{S}_{c_3}^{-\,2}=0, \ {\rm or} \nonumber \\
	&& \dot{S}_{c_2}^{-}=0, \ {\rm or} \nonumber \\
	&& \dot{S}_{c_1}^{+}\,\dot{S}_{c_3}^{-} + M_2^2\,S_{c_1}^{+}\,S_{c_3}^{-}=0, \ {\rm or}
	\nonumber \\
	&& 2\left[M_1^2 S^+_{c_1} \dot{S}^-_{c_2} \dot{S}^+_{c_2}\left(M_2^2 S^-_{c_3} \dot{S}^-_{c_1} + S^-_{c_1} \dot{S}^-_{c_3}\right)+ S^+_{c_2} \dot{S}^-_{c_2} \left(M_2^2 \dot{S}^-_{c_1} \dot{S}^+_{c_1} S^-_{c_3} + S^-_{c_1}\dot{S}^+_{c_1} \dot{S}^-_{c_3}\right)\right] \nonumber \\
	&& + \left[M_2^2 S^+_{c_2} S^-_{c_3} \dot{S}^-_{c_2} + M_1^2 \left(2 M_2^2 S^+_{c_1} S^-_{c_3} \dot{S}^-_{c_1} + 2 S^+_{c_1} S^-_{c_1} \dot{S}^-_{c_3} -   \dot{S}^+_{c_2} \dot{S}^-_{c_2} \dot{S}^-_{c_3}\right)\right](k L_1)^4\sin^2\theta_G \nonumber\\
	&& -M_1^2 \dot{S}^{-}_{c_3}(k L_1)^8\sin^4\theta_G = 0,
\end{eqnarray}
where the functions for the 5D profiles are to be evaluated at $z = L_1$ and $\theta_G \equiv \theta_G(L_1,v)$.  The last of these conditions contains the zero-mode solution.

The first mass condition sets the masses of the two degenerate states that are mostly $T_3$ and $T^{\prime}_3$.  It yields two coefficient sets, which we denote by $\rho$ and $\sigma$,
\begin{eqnarray}
	C^{\rho,\sigma}_{ {2}} &=& C^{\rho,\sigma}_{ {3}}= C^{\rho,\sigma}_{ {5}}= C^{\rho,\sigma}_{ {7}}= C^{\rho,\sigma}_{ {8}}= C^{\rho,\sigma}_{ {10}}=0, \nonumber\\
	C^{\rho,\sigma}_{ {12}} &=& C^{\rho,\sigma}_{ {13}}= C_{ {13}}, \qquad
	C^{\rho,\sigma}_{ {16}} = -\frac{2\,i}{\tan\!\theta_G\,}\,C_{ {13}}+C_{ {19}}, \qquad
	C^{\rho,\sigma}_{ {19}}= C_{ {19}}.
\end{eqnarray}
Here we choose $(C_{ {13}},C_{ {19}})=(0,1),(1,0)$ to obtain the two independent solutions $\rho$ and $\sigma$, respectively.

The second mass condition sets the mass of the state that is mostly $\tilde t_2$.  It results in the coefficient set,
\begin{eqnarray}
	C^\tau_{ {2}} &=& C^\tau_{ {3}}= C^\tau_{ {5}}= C^\tau_{ {10}}= C^\tau_{ {12}}= C^\tau_{ {13}}= C^\tau_{ {16}}= C^\tau_{ {19}}= 0, \nonumber\\
	C^\tau_{ {7}} &=& -C_{ {8}}, \qquad
	C^\tau_{ {8}}= C_{ {8}},
\end{eqnarray}
which we denote by $\tau$.

The third mass condition yields the masses of the states that are mostly $\tilde t_1$ and $\tilde t_3$.  The resulting coefficient set is,
\begin{eqnarray}
	C^{\omega}_{ {2}} &=& C_{ {2}},\qquad 
	C^{\omega}_{ {3}}= C^{\omega}_{ {7}}= C^{\omega}_{ {8}}= C^{\omega}_{ {10}}=C^{\omega}_{ {13}}= 0, \qquad
	C^{\omega}_{ {5}}= -\frac{\sqrt{2}}{2}\,\tan\!\theta_G \,C_{ {2}}, \nonumber \\
	C^{\omega}_{ {16}} &=& -C^{\omega}_{ {19}} = \frac{i}{2}\,\tan\!\theta_G \,C^{\omega}_{ {12}} = \frac{i}{2}\,M_2\,\tan\!\theta_G \,C_{ {2}}\,\frac{S_{c_1}^{+}}{\dot{S}_{c_3}^{-}}, 
\end{eqnarray}
which we will denote by $\omega$. 

Finally, the fourth mass condition yields the masses of the states that are mostly $t_1$, $\hat t_1$, $t_2$, $\hat t_2$, and $t_3$.  The resulting coefficient set, which we call $u$, is
\begin{eqnarray}
	C^u_{ {2}} &=& \frac{1}{\sqrt{2}}\tan\theta_G \, C^u_{ {5}}= C_3\,\frac{\dot{S}_{c_3}^{-} S_{c_1}^{-}+M_2^2 S_{c_1}^{-} \dot{S}_{c_1}^{-}}{\dot{S}_{c_1}^{+} \dot{S}_{c_3}^{-}+M_2^2 S_{c_1}^{+} S_{c_3}^{-}}, \qquad
	C^u_{ {3}}= C_3, \nonumber \\
	-C^u_{ {12}} &=& \frac{\sin^2\theta_G}{1+\cos^2\theta_G} C^u_{ {13}} 
= i\,\tan\theta_G C^u_{ {16}} = -i\,\tan\theta_G C^u_{ {19}}
= \frac{1}{2} \frac{M_2 \,C_3 (k z)^4\sin^2\theta_G}{\dot{S}_{c_1}^{+} \dot{S}_{c_3}^{-}+M_2^2 S_{c_1}^{+} S_{c_3}^{-}}, \nonumber \\
	C^u_{ {7}} &=& C^u_{ {8}} = \frac{1}{\sqrt{2}}\tan\theta_G \frac{S^{+}_{c_2}}{\dot{S}^{-}_{c_2}}C^u_{ {10}}= -\frac{M_1}{M_2} \frac{\dot{S}_{c_3}^{-}}{\dot{S}_{c_2}^{-}} C^u_{ {12}} + M_1\, \frac{S_{c_1}^{+}}{\dot{S}_{c_2}^{-}} C^u_{ {2}}.
\end{eqnarray}

In the end, the profiles of our original ten 5D up-type fermion fields have been re-expressed after EWSB in terms of five functions, which we can write schematically as $F_{u_{L,R}}^{\frac{2}{3}}$, $F_{\rho_{L,R}}^{\frac{2}{3}}$, $F_{\sigma_{L,R}}^{\frac{2}{3}}$, $F_{\tau_{L,R}}^{\frac{2}{3}}$ and $F_{\omega_{L,R}}^{\frac{2}{3}}$. The full profiles are defined by their action on an arbitrary matrix $M$ (e.g., a gauge generator) according to
\begin{equation}
	F_{a_{L,R}}^{Q\,\dag} \,M\, F_{a_{L,R}}^Q =\sum_{i=1}^3 f_{ {iL,iR}}^{a\,\dag}\,M\,f_{ {iL,iR}}^{a} \,,
\label{F53}
\end{equation} 
where $a$ denotes the appropriate coefficient solution.  The profiles are normalized according to
\begin{equation}
	\int\frac{dz}{(kz)^4}\sum_{i=1}^3 f_{ {iL,iR}}^{a\,\dag}\,f_{ {iL,iR}}^{a} =1,
\end{equation} 
where all profiles $f_\ell$ are evaluated at the same mass eigenvalue.
 The notation $f_{ i}^{a}$ indicates the vector $f_{ i}$ with its coefficients evaluated according to coefficient set $a$ (any coefficients not explicitly defined by the solution are set to zero). 
 
The 4D coupling between the mass eigenstates of a gauge boson $G$ and two fermions $a$ and $b$ therefore takes the form,
\begin{equation}
	g_{G\bar{a} b} = g_5 \int_{L_0}^{L_1} \frac{dz}{(kz)^4} \sum_{\alpha,i} f^{a\,\dag}_{iL,iR}(m_n,z;v) T^\alpha f^G_\alpha(m_n,z;v) f_{iL,iR}^b(m_n,z;v). \label{fermmixcoup}
\end{equation}

\subsubsection{Down-type fermions}

Each down-type fermion of the SM is accompanied by four additional 5D fields; one of the new fields remains unmixed after EWSB while the other four down-type states mix with each other.  Imposing the TeV-brane boundary conditions again allows a solution for the coefficients $C_{ {4}}$, $C_{ {14}}$, $C_{ {17}}$ and $C_{ {20}}$.  
(The coefficient $C_9$, corresponding to the unmixed down-type fermion $b_2$, is set by the profile normalization condition; the masses of the corresponding fermion's KK modes are not affected by EWSB.)

Requiring that a coefficient solution exist results in the following mass conditions, one of which must be satisfied for there to be a fermion mass eigenstate:
\begin{eqnarray}
	&& \dot{S}_{c_3}^{-} = 0, \ {\rm or} \label{down1}\\
	&& 2\,S_{c_3}^{+}\left(M_2^2\,S_{c_3}^{-}\,\dot{S}_{c_1}^{-}+S_{c_1}^{-}\,\dot{S}_{c_3}^{-}\right)-(k L_1)^4 M_2^2\,\dot{S}_{c_1}^{-}\,\sin^2\!\theta_G = 0.\label{down2}
\end{eqnarray}
The second of these conditions contains the zero-mode solution.

The first mass condition sets the mass of the state that is mostly $B_3$.  It yields the coefficient set, 
\begin{equation}
	C^{\eta}_{ {4}} =  C^{\eta}_{ {20}} = 0, \qquad
	C^{\eta}_{ {14}}=\frac{i\sqrt{2}}{2}\tan\theta_G\,C_{ {17}}, \qquad
	C^{\eta}_{ {17}} = C_{ {17}},
\end{equation}
which we denote by $\eta$.

The second mass condition sets the masses of the states that are mostly $b_1$, $b_3$, and $B_3^{\prime}$.  It results in the coefficient set,
\begin{eqnarray}
	C^d_{ {4}} &=& \frac{i\sqrt{2}}{M_2\,\sin\!\theta_G}\,C_{ {20}}\, \frac{S_{c_3}^{+}}{\dot{S}_{c_1}^{-}}, \qquad
	C^d_{ {20}} =C_{ {20}}, \nonumber\\
	C^d_{ {14}} &=& -\frac{i\sqrt{2}}{\tan\!\theta_G}\, C^d_{ {17}}  = \frac{i\sqrt{2}}{\tan\!\theta_G}\,C_{ {20}}\, \frac{S_{c_3}^{+}}{\dot{S}_{c_3}^{-}}, 
\end{eqnarray}
which we denote by $d$. 

In the end, the profiles of our original five 5D down-type fermion fields have been re-expressed after EWSB in terms of three functions, which we can write schematically as $F_{d_{L,R}}^{-\frac{1}{3}}$, $F_{\eta_{L,R}}^{-\frac{1}{3}}$ and $F_{b_{2L,R}}^{-\frac{1}{3}}$. The full profiles are defined by Eq.~(\ref{F53}), where $a=d,\eta,b_2$.

\subsubsection{Exotic fermions}

The fermion sector also contains five exotic 5D quarks (with $Q = 5/3$) and five exotic 5D leptons (with $Q = 1$) per generation.  We discuss here the quark sector; all details carry over to the lepton sector.  One of these exotic fermions remains unaffected by EWSB, while the remaining four mix with each other.  Imposing the TeV-brane boundary conditions again allows a solution for the coefficients $C_{ {1}}$, $C_{ {11}}$, $C_{ {15}}$ and $C_{ {18}}$.  (The coefficient $C_{ {6}}$, corresponding to the unmixed exotic fermion $\chi_2$, is set by the profile normalization condition; the masses of the corresponding fermion's KK modes are not affected by EWSB.)

Requiring that a coefficient solution exist results in the following mass conditions, one of which must be satisfied for there to be a fermion mass eigenstate:
\begin{eqnarray}
	&&\dot{S}_{c_3}^{-\,2} = 0, \ {\rm or} \nonumber \\
	&&\dot{S}_{c_1}^{+}\,\dot{S}_{c_3}^{-} + M_2^2\,S_{c_1}^{+}\,S_{c_3}^{-}=0. 
\end{eqnarray}

The first mass condition sets the masses of the two degenerate states that are mostly $\Xi_3$ and $\Xi_3^{\prime}$.  It yields two coefficient sets, which we denote by $\alpha$ and $\beta$,
\begin{eqnarray}
	C^{\alpha,\beta}_{ {1}} &=& 0, \qquad
	C^{\alpha,\beta}_{ {11}}=\frac{i\sqrt{2}}{2}\tan\theta_G\left(C_{ {15}}-C_{ {18}}\right), \nonumber \\
	C^{\alpha,\beta}_{ {15}} &=& C_{ {15}}, \qquad C^{\alpha,\beta}_{ {18}} =C_{ {18}}.
\end{eqnarray}
Here we choose $(C_{ {15}},C_{ {18}})=(0,1)$, $(1,0)$ to obtain the two independent solutions $\alpha$ and $\beta$, respectively.

The second mass condition yields the masses of the states that are mostly $\chi_1$ and $\chi_3$.  The resulting coefficient set is,
\begin{equation}
	C^\gamma_{ {1}} = C_{ {11}} \,\frac{1}{M_2\cos\!\theta_G}\, \frac{\dot{S}_{c_3}^{-}}{S_{c_1}^{+}}, \qquad 
	C^\gamma_{ {11}} = C_{ {11}}, \qquad
	C^\gamma_{ {15}} = -C^\gamma_{ {18}}=\frac{i\sqrt{2}}{2}\tan\!\theta_G\,C_{ {11}},
\end{equation}
which we denote by $\gamma$. 

In the end, the profiles of our original five 5D exotic fermion fields have been re-expressed after EWSB in terms of four functions, which we can write schematically as $F_{\alpha_{L,R}}^{\frac{5}{3}}$, $F_{\beta_{L,R}}^{\frac{5}{3}}$, $F_{\gamma_{L,R}}^{\frac{5}{3}}$ and $F_{\chi_{2L,R}}^{\frac{5}{3}}$.  The full profiles are again defined by Eq.~(\ref{F53}), where $a= \alpha,\beta, \gamma, \chi_2$.

\section{Interaction vertices \label{sec:vertices}}

In the following we summarize the relevant vertices for our calculation.  We quote the couplings with respect to the 4D fields; these couplings therefore contain the fifth-dimensional profiles of the associated particles.  The 4D couplings among specific KK modes can be found by integrating these expressions over the fifth dimension.  The 5D couplings can be read off by dropping the fifth-dimensional profiles.

The Feynman rule for triple-gauge-boson vertices takes the form,
\begin{eqnarray}
	G_{1\mu}^{(n)}(p_1)G_{2\nu}^{(m)}(p_2)G_{3 \lambda}^{(l)}(p_3) 
	&\rightarrow& 
	\frac{C_{G_1 G_2 G_3}}{k z} \frac{g_5}{g}
	\left[\eta^{\mu\lambda}\left(p_3 - p_1\right)^{\nu} 
	+ \eta^{\mu\nu}\left(p_1 - p_2\right)^{\lambda} \right. \nonumber \\
	&& + \left.\eta^{\nu\lambda}\left(p_2 - p_3\right)^{\mu} \right]\,
	f_{G_1}^{(n)}(z) f_{G_2}^{(m)}(z) f_{G_3}^{(l)}(z),
	\label{vertW123}
\end{eqnarray}
where we take all particle momenta to be incoming.  The couplings $C_{G_1 G_2 G_3}$ are given by
\begin{eqnarray}
	&&C_{Z W^{+}_L W^{-}_L} = i g_Z \cos^2{\theta_W}, \quad 
	C_{Z W_R W_R} = -i g_Z \sin^2{\theta_W}, \quad
	C_{Z A^{\widehat \pm} A^{\widehat \mp}} = \frac{i}{2} g_Z \cos{2\theta_W},
	\nonumber \\
	&&C_{Z A^{\hat{3}} A^{\hat{4}}} = \frac{g_Z}{2}, \quad
	C_{X W^{+}_R W^{-}_R} = -i g_X, \quad
	C_{X A^{\widehat \pm} A^{\widehat \mp}} = -\frac{i}{2} g_X, \quad
	C_{X A^{\hat{3}} A^{\hat{4}}} = \frac{g_X}{2},
	\nonumber \\
	&&C_{W_{L}^{\mp} A^{\widehat \pm} A^{\hat{3}}} 
	= C_{W_{R}^{\mp} A^{\widehat \pm} A^{\hat{3}}} = \pm \frac{i}{2} g, \qquad
	C_{W_{L}^{\mp} A^{\widehat \pm} A^{\hat{4}}} 
	= -C_{W_{R}^{\mp} A^{\widehat \pm} A^{\hat{4}}} = \pm \frac{1}{2} g,
\end{eqnarray}
where $g_Z$ and $g_X$ were given in Eqs.~(\ref{eq:gZ}) and (\ref{eq:gX}), respectively.

The Feynman rules for the gauge-gauge-Higgs vertices involving $Z$ or $X$ arise from quadruple-gauge interactions involving the fifth components of two $A^{\hat 4}$ bosons, with one of them replaced by the Higgs vev:
\begin{eqnarray}
	Z^{(n)}_\mu Z^{(m)}_\nu H &\rightarrow& 
	\frac{i g_Z^2}{2 k z}\,\frac{g_5}{g} \,v \left[f_H(z)\right]^2 f_Z^{(n)}(z) f_Z^{(m)}(z)\, 
	\eta_{\mu\nu}, 
	\nonumber \\
	X^{(n)}_\mu Z^{(m)}_\nu H &\rightarrow& 
	\frac{i g_Z g_X}{2 k z}\,\frac{g_5}{g}\, v \left[f_H(z)\right]^2 f_X^{(n)}(z) f_Z^{(m)}(z)\,
	\eta^{\mu\nu},
	\nonumber \\
	X^{(n)}_\mu X^{(m)}_\nu H &\rightarrow& 
	\frac{i g_X^2}{2 k z}\,\frac{g_5}{g}\, v \left[f_H(z)\right]^2 f_X^{(n)}(z) f_X^{(m)}(z) \,
	\eta^{\mu\nu}.
\end{eqnarray}

Feynman rules for gauge-gauge-Higgs vertices involving one $A^{\hat a}$ gauge KK mode arise from the 5D triple gauge vertex:
\begin{eqnarray}
	G_\mu^{(n)}  A^{\hat{a}\,(m)}_\nu   H &\rightarrow& 
	\frac{i}{2 k z} C_{G A^{\hat{a}} H} \frac{g_5}{g} f_H(z) 
	\left\{\left[\partial_z f_G^{(n)}(z)\right]f_A^{(m)}(z)\right. \nonumber \\ 
&& \left. -\left[\partial_z f_A^{(m)}(z)\right] f_G^{(n)}(z)\right\},
\end{eqnarray}
where $G=Z$, $X$, or $W^\pm_{L,R}$, $\hat{a}= \widehat +$, $\widehat -$, or $\hat{3}$, and the couplings are
\begin{equation}
	C_{Z A^{\hat{3}} H} = g_Z, \qquad
	C_{X A^{\hat{3}} H} = g_X, \qquad
	C_{W^\pm_L A^{\widehat \mp} H} = - C_{W^\pm_R A^{\widehat \mp} H} = g.
\end{equation}
Note that, in the absence of EWSB-induced mixing, the corresponding 4D couplings involving zero-modes of $Z$ or $W^{\pm}_L$ will be zero after integration over $z$.  This is because $\partial_z f^{(0)}_G(z) = 0$ (flat profile) and $\int_{L_0}^{L_1} \partial_z f_A^{(m)}(z) \, dz = f_A^{(m)}(L_1) - f_A^{(m)}(L_0) = 0$ (Dirichlet boundary conditions on both branes).

The Feynman rules for gauge-gauge-fermion vertices are
\begin{eqnarray}
	\bar{\Psi}_{L,R} Z_\mu \Psi_{L,R} &\rightarrow& 
	\frac{g_Z}{(k z)^4}\,\frac{g_5}{g} \left(\widetilde{T}^{3_L} - Q\sin^2\theta_W \right) 
	\gamma_\mu, 
	\nonumber \\
	\bar{\Psi}_{L,R} X_\mu \Psi_{L,R} &\rightarrow& 
	\frac{g_X}{(k z)^4}\,\frac{g_5}{g} \left[\frac{\left(Q - \widetilde{T}^{3_L}\right) \sin^2\theta_W
	- \widetilde{T}^{3_R}\cos^2\theta_W}{\cos{2\theta_W}}\right] \gamma_\mu,
	\nonumber \\
	\bar{\Psi}_{L,R} A^{\hat{3}}_\mu \Psi_{L,R} &\rightarrow& 
	\frac{g_5}{(k z)^4} T^{\hat{3}} \gamma_{\mu},
	\label{Gff}
\end{eqnarray}
where $Q = \widetilde{T}^{3_L} + Y$ is the electric charge operator in units of $e$, and we have defined 
\begin{equation}
	\widetilde{T}_5^{3_{L,R}} = A\,T_5^{3_{L,R}} A^{-1}, \qquad
	\tilde{T}_{10}^{3_{L,R}} = B\,T_{10}^{3_{L,R}} B^{-1},
\end{equation}
for fermions in the {\bf 5} and {\bf 10} representations of SO(5), respectively.
The relevant generators in the {\bf 5} and {\bf 10} representations were given in Appendix~\ref{generators} and the basis transformation matrices $A$ and $B$ were defined in Eq.~(\ref{AB}).
The relevant quantum numbers of the fermions that couple to $Z$, $X$ and $A_{\mu}^{\hat 3}$ are summarized in Tables~\ref{FermionVertices1} and \ref{FermionVertices2}.

\begin{table}
\begin{center}
\begin{tabular}{cccccc}
\hline\hline
Lepton pair & $Q_{\ell}$ & Quark pair & $Q_q$ & $\widetilde{T}^{3_L}$ & $\widetilde{T}^{3_R}$ \\
\hline
$\bar{\kappa}_{1}\kappa_{1}$, $\bar{\kappa}_{2}\kappa_{2}$, $\bar{\kappa}_{3}\kappa_{3}$ 
& $1$
& $\bar{\chi}_{1}\chi_{1}$, $\bar{\chi}_{2}\chi_{2}$, $\bar{\chi}_{3}\chi_{3}$ 
& $\frac{5}{3}$ 
& $\frac{1}{2}$ 
& $\frac{1}{2}$ \\
$\bar{\hat{\eta}}_{1}\hat{\eta}_{1}$, $\bar{\hat{\eta}}_{2}\hat{\eta}_{2}$, $\bar{\hat{\eta}}_{3}\hat{\eta}_{3}$ 
& $0$
&$\bar{\hat{t}}_{1}\hat{t}_{1}$, $\bar{\hat{t}}_{2}\hat{t}_{2}$, $\bar{\hat{t}}_{3}\hat{t}_{3}$ 
& $\frac{2}{3}$ 
& $-\frac{1}{2}$ 
& $\frac{1}{2}$ \\
$\bar{\eta}_{1}\eta_{1}$, $\bar{\eta}_{2}\eta_{2}$, $\bar{\eta}_{3}\eta_{3}$ 
& $0$
& $\bar{t}_{1}t_{1}$, $\bar{t}_{2}t_{2}$, $\bar{t}_{3}t_{3}$  
& $\frac{2}{3}$ 
& $\frac{1}{2}$ 
& $-\frac{1}{2}$ \\
$\bar{\ell}_{1}\ell_{1}$, $\bar{\ell}_{2}\ell_{2}$, $\bar{\ell}_{3}\ell_{3}$ 
& $-1$
& $\bar{b}_{1}b_{1}$, $\bar{b}_{2}b_{2}$, $\bar{b}_{3}b_{3}$ 
& $-\frac{1}{3}$ 
& $-\frac{1}{2}$ 
& $-\frac{1}{2}$  \\
$\bar{\tilde{\eta}}_{1}\tilde{\eta}_{1}$, $\bar{\tilde{\eta}}_{2}\tilde{\eta}_{2}$, $\bar{N}_{3}N_{3}$, $\bar{N}'_{3}N'_{3}$ 
& $0$
& $\bar{\tilde{t}}_{1}\tilde{t}_{1}$, $\bar{\tilde{t}}_{2}\tilde{t}_{2}$, $\bar{T}_{3}T_{3}$, $\bar{T}'_{3}T'_{3}$  
& $\frac{2}{3}$ & 0 & 0 \\
$\bar{K}_{3} K_{3}$
& $1$
& $\bar{\Xi}_{3} \Xi_{3}$  
& $\frac{5}{3}$ & 1 & 0 \\
$\bar{L}_{3} L_{3}$
& $-1$
& $\bar{B}_{3} B_{3}$  
& $-\frac{1}{3}$ & $-1$ & 0 \\
$\bar{K}'_{3} K'_{3}$
& $1$
& $\bar{\Xi}'_{3} \Xi'_{3}$  
& $\frac{5}{3}$ & 0 & 1 \\
$\bar{L}'_{3} L'_{3}$  
& $-1$
& $\bar{B}'_{3} B'_{3}$  
& $-\frac{1}{3}$ & 0 & $-1$ \\
\hline\hline
\end{tabular}
\caption{Quantum numbers of the fermions that couple to $Z$ and $X$.  The fermion $L,R$ subscripts have been suppressed.  The exotic fermions are defined in Eq.~(\ref{quark123}). \label{FermionVertices1}}
\end{center}
\end{table}

\begin{table}
\begin{center}
\begin{tabular}{ccc}
\hline\hline
Lepton pair & Quark pair & $\widetilde T^{\hat 3}$ \\
\hline
$\bar{\tilde{\eta}}_3 N'_3$, $\bar{N}_3 {\eta}_3$ + h.c. 
& $\bar{\tilde{t}}_3 T'_3$, $\bar{T}_3 {t}_3$ + h.c.
& $-\frac{i}{\sqrt{2}}$ \\
$\bar{\tilde{\eta}}_3 N_3$, $\bar{N}'_3 {\eta}_3$ + h.c. 
& $\bar{\tilde{t}}_3 T_3$, $\bar{T}'_3 {t}_3$ + h.c.
& $\frac{i}{\sqrt{2}}$ \\
$\bar{\ell}_3 L'_3$, $\bar{\ell}_3 {L}_3$ + h.c.
& $\bar{b}_3 B'_3$, $\bar{b}_3 {B}_3$ + h.c. 
& $-\frac{i}{2}$ \\
$\bar{\kappa}_3 K'_3$, $\bar{\kappa}_3 {K}_3$ + h.c.
& $\bar{\chi}_3 \Xi'_3$, $\bar{\chi}_3 {\Xi}_3$ + h.c. 
& $\frac{i}{2}$ \\
$\bar{\tilde{\eta}}_1 \hat{\eta}_1$, $\bar{\tilde{\eta}}_2 \hat{\eta}_2$ + h.c.
& $\bar{\tilde{t}}_1 \hat{t}_1$, $\bar{\tilde{t}}_2 \hat{t}_2$ + h.c.  
& $-\frac{1}{2}$ \\
$\bar{{\eta}}_1 \hat{\eta}_1$, $\bar{{\eta}}_2 \hat{\eta}_2$ + h.c.
& $\bar{{t}}_1 \hat{t}_1$, $\bar{{t}}_2 \hat{t}_2$ + h.c.
& $\frac{1}{2}$ \\
\hline\hline
\end{tabular}
\caption{Quantum numbers of the fermions that couple to $A^{\hat 3}$.  The fermion $L,R$ subscripts have been suppressed.  The exotic fermions are defined in Eq.~(\ref{quark123}). \label{FermionVertices2}}
\end{center}
\end{table}

\section{Gauge boson decay widths \label{AppDecay}}

In this section we summarize the formulas for the decay widths of gauge KK modes to pairs of fermions or gauge bosons, to a gauge boson plus a Higgs boson, and to pairs of gauge bosons.
In the following, all coupling constants and 5D integral factors from the interaction vertices will be expressed as overall couplings $C_i^{L,R}$.  Note that after EWSB-induced particle mixing is implemented, the coupling constants and 5D integrals cannot be separated. The relevant couplings are given in Appendix~\ref{sec:vertices}.

The decay width for a gauge boson $G$ of mass $M_G$ to two fermions of mass $m_1$ and $m_2$ is given by 
\begin{eqnarray}
	\Gamma(G \to f \bar f^{\prime}) 
	&=& \frac{\lambda^{1/2}(M_G^{2},m_1^2,m_2^2)}{48\pi M_G^5} 
	\left\{ \left[ \left(C_{Gff^{\prime}}^L \right)^2 + \left(C_{Gff^{\prime}}^R \right)^2 \right] 
	\beta(M_G^{2},m_1^2,m_2^2) \right. 
	\nonumber\\
	&& + \left. 12 C_{Gff^{\prime}}^{L} C_{Gff^{\prime}}^{R} M_G^{2} m_1 m_2 \right],
\end{eqnarray}
where $C_{Gff^{\prime}}^{L,R}$ are the appropriate overall left- and right-handed couplings (including 5D integral factors), $\lambda$ is defined in Eq.~(\ref{eq:lambda}), and 
\begin{equation}
	\beta(x,y,z) = 2 x^2 - y^2 - z^2 - x y - x z + 2 y z.
\end{equation}
For $M_G \gg m_1, m_2$, this decay width grows proportional to $M_G$.

The decay width for a gauge boson $G_1$ of mass $M_1$ to a lighter gauge boson $G_2$ of mass $M_2$ and a Higgs boson is given by
\begin{equation}
	\Gamma(G_1 \to G_2H) = \frac{\lambda^{1/2}(M_1^2, M_2^2,M_H^2)}{48 \pi M_1^3} 
	C_{G_1G_2H}
	\left[2+ \frac{\left(M_1^2 + M_2^2 - M_H^2 \right)^2}{4 M_1^2 M_2^2} \right],
\end{equation}
where $M_H$ is the mass of the Higgs boson and $C_{G_1G_2H}$ is the appropriate coupling of the two gauge bosons to the Higgs.  For $M_1 \gg M_2$, this decay width grows proportional to $M_1$.

The decay width for a gauge boson $G_1$ of mass $M_1$ to two other gauge bosons $G_2$ and $G_3$ with masses $M_2$ and $M_3$, respectively, is given by
\begin{equation}
	\Gamma(G_1 \to G_2G_3) = 
	\frac{\lambda^{1/2}\left(M_1^2, M_2^2, M_3^2 \right)}{48\pi M_1^3}
	C_{G_1G_2G_3} N(M_1,M_2,M_3),
	\label{GzwwL}
\end{equation}
where $C_{G_1G_2G_3}$ is the appropriate coupling and $N$ is a kinematic function,
\begin{eqnarray}
	N(x,y,z) &=& - 8(x^2+y^2+z^2)
	+ 2\left(\frac{y^4 + z^4}{x^2} + \frac{x^4 + z^4}{y^2} + \frac{x^4 + y^4}{z^2} \right)
	\nonumber \\
	&& + \frac{1}{4} \left(\frac{x^6}{y^2 z^2} + \frac{y^6}{x^2 z^2} + \frac{z^6}{x^2 y^2} \right)
	- \frac{9}{2} \left(\frac{y^2 z^2}{x^2} + \frac{x^2 z^2}{y^2} + \frac{x^2 y^2}{z^2} \right).
\end{eqnarray}
For $M_1 \gg M_2 \sim M_3$, this decay width grows proportional to $M_1^5/M_2^2 M_3^2$.

\providecommand{\href}[2]{#2}\begingroup\raggedright\endgroup


\begin{thebibliography}{10}

\bibitem{LEP}
  R.~Barate {\em et~al.} [LEP Working Group for Higgs Boson Searches], {\it {Search for the standard model Higgs boson at
  LEP}},  {\em Phys.\ Lett.} {\bf B565} (2003) 61--75
  [\href{http://xxx.lanl.gov/abs/hep-ex/0306033}{{\tt hep-ex/0306033}}].
  
  \bibitem{LEPEWWG}
  LEP Electroweak Working Group, \href{http://lepewwg.web.cern.ch/LEPEWWG/}{{http://lepewwg.web.cern.ch/LEPEWWG/}}, March 2012.

\bibitem{LHCJuly4}
ATLAS Collaboration, {\it Observation of an excess of events in the search for the Standard Model Higgs boson with the ATLAS detector at the LHC},
\href{http://cdsweb.cern.ch/record/1460439}{ATLAS-CONF-2012-093} (2012);
CMS Collaboration, {\it Observation of a new boson with a mass near 125 GeV},
\href{http://cdsweb.cern.ch/record/1460438?ln=en}{CMS-PAS-HIG-12-020} (2012).


\bibitem{SUSY1}
P.~Fayet, {\it Supersymmetry and weak, electromagnetic and strong
  interactions},  {\em Phys.\ Lett.} {\bf B64} (1976) 159--162.

\bibitem{SUSY2}
P.~Fayet, {\it Spontaneously broken supersymmetric theories of weak,
  electromagnetic and strong interactions},  {\em Physics Letters B} {\bf 69}
  (1977) 489--494.

\bibitem{SUSY3}
P.~Fayet, {\it Weak interactions of a light gravitino: A lower limit on the
  gravitino mass from the decay psi $\to$ gravitino + antiphotino},  {\em Physics
  Letters B} {\bf 84} (1979) 421--426.

\bibitem{SUSY4}
G.~R.~Farrar and P.~Fayet, {\it Phenomenology of the production, decay, and detection of
new hadronic states associated with supersymmetry}, {\em Physics Letters B} {\bf 76} (1978) 575.

\bibitem{SUSY5}
S.~P.~Martin, {\it {A Supersymmetry Primer}},
  \href{http://xxx.lanl.gov/abs/hep-ph/9709356}{{\tt hep-ph/9709356}}.

\bibitem{SUSY6}
J.~Kalinowski, {\it {SUSY Theory Review}},  {\em Acta Physica Polonica B} {\bf
  38} (2007) 0531.

\bibitem{LittleHiggs1}
N.~Arkani-Hamed, A.~G.~Cohen and H.~Georgi, {\it Electroweak symmetry breaking
  from dimensional deconstruction},  {\em Physics Letters B} {\bf 513} (2001)
  232--240.

\bibitem{LittleHiggs2}
N.~Arkani-Hamed, A.~Cohen, E.~Katz and A.~Nelson, {\it {The Littlest Higgs}},
  {\em JHEP} {\bf 0207} (2002) 034
  [\href{http://xxx.lanl.gov/abs/hep-ph/0206021}{{\tt hep-ph/0206021}}].

\bibitem{LittleHiggs3}
N.~Arkani-Hamed, A.~Cohen, E.~Katz, A.~Nelson, T.~Gregoire and J.~G.~Wacker, {\it
  {The Minimal moose for a little Higgs}},  {\em JHEP} {\bf 0208} (2002) 021
  [\href{http://xxx.lanl.gov/abs/hep-ph/0206020}{{\tt hep-ph/0206020}}].

\bibitem{LittleHiggs4}
M.~Schmaltz and D.~Tucker-Smith, {\it {Little Higgs Review}},  {\em Ann. Rev.
  Nucl. Part. Sci.} {\bf 55} (2005) 229--270
  [\href{http://xxx.lanl.gov/abs/hep-ph/0502182}{{\tt hep-ph/0502182}}].

\bibitem{TQC}
V.~A.~Miransky, M.~Tanabashi and K.~Yamawaki, {\it Dynamical electroweak
  symmetry breaking with large anomalous dimension and t quark condensate},
  {\em Phys.\ Lett.\ B} {\bf 221} (1989) 177--183.

\bibitem{TC}
C.~T.~Hill, {\it Topcolor: top quark condensation in a gauge extension of the
  standard model},  {\em Phys.\ Lett.\ B} {\bf 266} (1991) 419--424.

\bibitem{FG}
C.~T.~Hill, M.~A.~Luty and E.~A.~Paschos, {\it Electroweak symmetry breaking
  by fourth-generation condensates and the neutrino spectrum},  {\em Phys.\ Rev.\
  D} {\bf 43} (1991) 3011--3025.

\bibitem{CS}
W.~J.~Marciano, {\it Exotic new quarks and dynamical symmetry breaking},  {\em
  Phys.\ Rev.\ D} {\bf 21} (1980) 2425--2428.

\bibitem{Technicolor}
L.~Susskind, {\it Dynamics of spontaneous symmetry breaking in the
  Weinberg-Salam theory},  {\em Phys.\ Rev.\ D} {\bf 20} (1979) 2619--2625.

\bibitem{Technicolor2}
S.~Weinberg, {\it Implications of dynamical symmetry breaking: An addendum},
  {\em Phys.\ Rev.\ D} {\bf 19} (1979) 1277--1280.

\bibitem{Technicolor3}
S.~Weinberg, {\it Implications of dynamical symmetry breaking},  {\em Phys.\
  Rev.\ D} {\bf 13} (1976) 974--996.

\bibitem{ExTech}
S.~Dimopoulos and L.~Susskind, {\it Mass without scalars},  {\em Nucl.\
  Phys.\ B} {\bf 155} (1979) 237--252.

\bibitem{WTech}
T.~Appelquist and L.~C.~R.~Wijewardhana, {\it Chiral hierarchies from slowly
  running couplings in technicolor theories},  {\em Phys.\ Rev.\ D} {\bf 36}
  (1987) 568--580.

\bibitem{Technicolor4}
M.~Piai, {\it {Lectures on walking technicolor, holography and gauge/gravity
  dualities}},  {\em Adv.\ High Energy Phys.} {\bf 2010} (2010) 464302
  [\href{http://xxx.lanl.gov/abs/1004.0176}{{\tt arXiv:1004.0176}}].

\bibitem{RS1}
L.~Randall and R.~Sundrum, {\it {A large mass hierarchy from a small extra
  dimension}},  {\em Phys.\ Rev.\ Lett.} {\bf 83} (1999) 3370--3373
  [\href{http://xxx.lanl.gov/abs/hep-ph/9905221}{{\tt hep-ph/9905221}}].
  
  \bibitem{0508279}
H.~Davoudiasl, B.~Lillie, T.~G.~Rizzo, {\it Off-the-Wall Higgs in the Universal Randall-Sundrum Model},  {\em
  JHEP} {\bf 0608} (2006) 042 [\href{http://arxiv.org/abs/hep-ph/0508279}{{\tt arXiv:hep-ph/0508279}}].
  
\bibitem{0611358}
G.~Cacciapaglia, C.~Cs{\'a}ki, G.~Marandella, J.~Terning, {\it The Gaugephobic Higgs},  {\em
  JHEP} {\bf 0702} (2007) 036 [\href{http://arxiv.org/abs/hep-ph/0611358}{{\tt arXiv:hep-ph/0611358}}].





\bibitem{Maldacena}
J.~M.~Maldacena,
  {\it The Large N limit of superconformal field theories and supergravity},
 {\em Adv.\ Theor.\ Math.\ Phys.} {\bf 2} (1998) 231
  [{\em Int.\ J.\ Theor.\ Phys.} {\bf 38}, (1999) 1113]
  [\href{http://arxiv.org/abs/hep-th/9711200}{\tt hep-th/9711200}].

\bibitem{MCHM}
K.~Agashe, R.~Contino and A.~Pomarol, {\it {The Minimal Composite Higgs
  Model}},  {\em Nucl.\ Phys.} {\bf B719} (2005) 165--187
  [\href{http://xxx.lanl.gov/abs/hep-ph/0412089}{{\tt hep-ph/0412089}}].

\bibitem{Zbb}
K.~Agashe, R.~Contino, L.~Da~Rold and A.~Pomarol, {\it {A custodial symmetry
  for Z b anti-b}},  {\em Phys.\ Lett.} {\bf B641} (2006) 62--66
  [\href{http://xxx.lanl.gov/abs/hep-ph/0605341}{{\tt hep-ph/0605341}}].
  
  \bibitem{0308036}
K.~Agashe, A.~Delgado, M.~J.~May, R.~Sundrum,{\it RS1, Custodial Isospin and Precision Tests},  {\em
  JHEP} {\bf 0308} (2003) 050 [\href{http://arxiv.org/abs/hep-ph/0308036}{{\tt arXiv:hep-ph/0308036}}].
  
\bibitem{0612048}
R.~Contino, L.~Da~Rold, A.~Pomarol, {\it Light custodians in natural composite Higgs models},  {\em
 Phys.~Rev.~D} {\bf 75} (2007) 055014 [\href{http://arxiv.org/abs/hep-ph/0612048}{{\tt arXiv:hep-ph/0612048}}].
  
\bibitem{MSW}
A.~D.~Medina, N.~R.~Shah and C.~E.~M.~Wagner, {\it Gauge-higgs unification and
  radiative electroweak symmetry breaking in warped extra dimensions},  {\em
  Phys.\ Rev.\ D} {\bf 76} (2007) 095010.
  
\bibitem{MCHMEWPT}
K.~Agashe and R.~Contino, {\it {The minimal composite Higgs model and
  electroweak precision tests}},  {\em Nucl.\ Phys.} {\bf B742} (2006) 59--85
  [\href{http://xxx.lanl.gov/abs/hep-ph/0510164}{{\tt hep-ph/0510164}}].

\bibitem{EWCs}
M.~S.~Carena, E.~Ponton, J.~Santiago and C.~E.~M.~Wagner, {\it {Electroweak
  constraints on warped models with custodial symmetry}},  {\em Phys.\ Rev.\ D}
  {\bf 76} (2007) 035006 [\href{http://xxx.lanl.gov/abs/hep-ph/0701055}{{\tt
  hep-ph/0701055}}].

\bibitem{LightKK}
M.~S.~Carena, E.~Ponton, J.~Santiago and C.~E.~M.~Wagner, {\it {Light
  Kaluza-Klein states in Randall-Sundrum models with custodial SU(2)}},  {\em
  Nucl.\ Phys.} {\bf B759} (2006) 202--227
  [\href{http://xxx.lanl.gov/abs/hep-ph/0607106}{{\tt hep-ph/0607106}}].

\bibitem{CMSW}
M.~Carena, A.~D.~Medina, B.~Panes, N.~R.~Shah and C.~E.~M.~Wagner, {\it
  Collider phenomenology of gauge-higgs unification scenarios in warped extra
  dimensions},  {\em Phys.\ Rev.\ D} {\bf 77} (2008) 076003.

\bibitem{Neutrinos}
M.~Carena, A.~D.~Medina, N.~R.~Shah and C.~E.~M.~Wagner, {\it Gauge-higgs
  unification, neutrino masses, and dark matter in warped extra dimensions},
  {\em Phys.\ Rev.\ D} {\bf 79} (2009) 096010.
  
\bibitem{GHU}
  N.~S.~Manton,
  {\it A New Six-Dimensional Approach to the Weinberg-Salam Model},
 {\em Nucl.\ Phys.} {\bf B158} (1979) 141;
  D.~B.~Fairlie,
  {\it Two Consistent Calculations Of The Weinberg Angle},
 {\em J.\ Phys.\ G} {\bf 5} (1979) L55;
  {\it Higgs' Fields and the Determination of the Weinberg Angle},
 {\em Phys.\ Lett.}  {\bf B 82} (1979) 97;
  P.~Forgacs and N.~S.~Manton,
  {\it Space-Time Symmetries in Gauge Theories},
 {\em Commun.\ Math.\ Phys.}  {\bf 72} (1980) 15;
  D.~Kapetanakis and G.~Zoupanos,
  {\it Coset space dimensional reduction of gauge theories},
 {\em Phys.\ Rept.}  {\bf 219} (1992) 1.

\bibitem{Coleman:1973jx} 
  S.~R.~Coleman and E.~J.~Weinberg,
  {\it Radiative Corrections as the Origin of Spontaneous Symmetry Breaking},
 {\em Phys.\ Rev.\ D} {\bf 7} (1973) 1888.

 \bibitem{Hosotani}
Y.~Hosotani, {\it Dynamical mass generation by compact extra dimensions},  {\em
 Phys.~Lett.~B} {\bf 126} (1983) 309.   
 
\bibitem{Falkowski}
A.~Falkowski, {\it {About the holographic pseudo-Goldstone boson}},  {\em Phys.\
  Rev.\ D} {\bf 75} (2007) 025017
  [\href{http://xxx.lanl.gov/abs/hep-ph/0610336}{{\tt hep-ph/0610336}}].
  
\bibitem{HiggsPGB}
R.~Contino, Y.~Nomura and A.~Pomarol, {\it {Higgs as a holographic
  pseudo-Goldstone boson}},  {\em Nucl.\ Phys.} {\bf B671} (2003) 148--174
  [\href{http://xxx.lanl.gov/abs/hep-ph/0306259}{{\tt hep-ph/0306259}}].
  
\bibitem{0302001}
T.~Gherghetta, A.~Pomarol, {\it The Standard Model Partly Supersymmetric},  {\em
 Phys.~Rev.~D} {\bf 67} (2003) 085018 [\href{http://arxiv.org/abs/hep-ph/0302001}{{\tt arXiv:hep-ph/0302001}}].
  
\bibitem{9909411}
N.~Arkani-Hamed, Y.~Grossman, M.~Schmaltz, {\it Split Fermions in Extra Dimensions and Exponentially Small Cross-Sections at Future Colliders},  {\em Phys.~Rev.~D} {\bf 61} (2000) 115004 [\href{http://arxiv.org/abs/hep-ph/9909411}{{\tt arXiv:hep-ph/9909411}}].  
  
  
\bibitem{WarpedReview}
H.~Davoudiasl, S.~Gopalakrishna, E.~Ponton and J.~Santiago, {\it {Warped
  5-Dimensional Models: Phenomenological Status and Experimental Prospects}},
  {\em New J.\ Phys.} {\bf 12} (2010) 075011
  [\href{http://xxx.lanl.gov/abs/0908.1968}{{\tt arXiv:0908.1968}}].

 \bibitem{RandallSchwartz}
L.~Randall and M.~D.~Schwartz, {\it {Quantum field theory and unification in
  AdS5}},  {\em JHEP} {\bf 11} (2001) 003
  [\href{http://xxx.lanl.gov/abs/hep-th/0108114}{{\tt hep-th/0108114}}].

\bibitem{Houches}
T.~Gherghetta, {\it {Warped models and holography}},
  \href{http://xxx.lanl.gov/abs/hep-ph/0601213}{{\tt hep-ph/0601213}}.

\bibitem{FifthDim}
R.~Sundrum, {\it {To the fifth dimension and back (TASI 2004)}},
  \href{http://xxx.lanl.gov/abs/hep-th/0508134}{{\tt hep-th/0508134}}.

 \bibitem{Davoudiasl:1999tf} 
  H.~Davoudiasl, J.~L.~Hewett and T.~G.~Rizzo,
  {\it Bulk gauge fields in the Randall-Sundrum model},
  {\em Phys.\ Lett.\ B} {\bf 473}, 43 (2000)
  [\href{http://arxiv.org/abs/hep-ph/9911262}{\tt hep-ph/9911262}];
  A.~Pomarol,
  {\it Gauge bosons in a five-dimensional theory with localized gravity},
  {\em Phys.\ Lett.\ B} {\bf 486}, 153 (2000)
  [\href{http://arxiv.org/abs/hep-ph/9911294}{\tt hep-ph/9911294}].

\bibitem{HOOS1}
Y.~Hosotani, {\it {Dynamical Electroweak Symmetry Breaking in $SO(5) \times
  U(1)$ Gauge-Higgs Unification in the Randall-Sundrum Warped Space}},
  \href{http://xxx.lanl.gov/abs/0901.2415}{{\tt arXiv:0901.2415}}.

\bibitem{HOOS2}
Y.~Hosotani, M.~Tanaka and N.~Uekusa, {\it {Collider signatures of the SO(5)$\times$U(1) gauge-Higgs unification}},
  \href{http://xxx.lanl.gov/abs/1103.6076}{{\tt arXiv:1103.6076}}.
  
 \bibitem{PDG}
 K.~Nakamura {\it et al.}  [Particle Data Group Collaboration],
  {\it Review of particle physics},
 {\em  J.\ Phys.\ G} {\bf 37}, 075021 (2010).

\bibitem{Espinosa:2010vn} 
  J.~R.~Espinosa, C.~Grojean and M.~Muhlleitner,
  {\it Composite Higgs Search at the LHC},
  {\em JHEP} {\bf 1005} (2010) 065
  [\href{http://arxiv.org/abs/arXiv:1003.3251}{\tt arXiv:1003.3251 [hep-ph]}].


\bibitem{McDonald:2010fe} 
  K.~L.~McDonald and D.~E.~Morrissey,
  JHEP {\bf 1102}, 087 (2011)
  [arXiv:1010.5999 [hep-ph]].

\bibitem{ILC2}
J.~Brau {\em et~al.} [ILC Collaboration],
  {\it {ILC Reference Design Report: ILC Global Design Effort and World Wide
  Study}},  \href{http://xxx.lanl.gov/abs/0712.1950}{{\tt arXiv:0712.1950}};
  {\it International Linear Collider Reference Design Report Volume 2: Physics at the ILC},
  \href{http://arxiv.org/abs/arXiv:0709.1893}{\tt arXiv:0709.1893}.

\bibitem{CLIC}
CLIC Study Team Collaboration, {\it {A 3 TeV $e^+\,e^-$
  Linear Collider Based on CLIC Technology}}, CERN Report CERN-2000-008 (2000),
 available from \verb+http://cdsweb.cern.ch+.

\bibitem{Bessel}
J.~R.~Culham, {\it Bessel Functions of the First and Second Kind},
  \verb+http://www.mhtl.uwaterloo.ca/courses/me755/web_chap4.pdf+,
  August 2011.

\bibitem{Wolfram}
E.~W.~Weisstein, {\it Wolfram MathWorld}, \verb+http://mathworld.wolfram.com+,
 August 2011.

\bibitem{Maple}
Maplesoft, {\it Maple Help Center},
  \verb+http://www.maplesoft.com/support/help/Maple+, August 2011.

\bibitem{Lijun}
K.~Agashe {\em et~al.}, {\it {LHC Signals for Coset Electroweak Gauge Bosons
  in Warped/Composite PGB Higgs Models}},  {\em Phys.\ Rev.\ D} {\bf 81} (2010)
  096002 [\href{http://xxx.lanl.gov/abs/0911.0059}{{\tt arXiv:0911.0059}}].
  
  
\end{thebibliography}
\end{document}